\documentclass[dvipsnames]{aa} 

\usepackage{graphicx}
\usepackage{txfonts}
\usepackage{stfloats}

\usepackage[colorlinks=true,linkcolor=blue,citecolor=blue,urlcolor=blue]{hyperref}
\usepackage{newtxtext,newtxmath}
\usepackage{float}
\usepackage[T1]{fontenc}
\usepackage{graphicx}	
\usepackage{amsmath}	

\usepackage{ulem}

\usepackage{lastpage}
\usepackage{caption}
\usepackage{subcaption}

\newcommand{\edit}[1]{\textcolor{black}{#1}}

\begin{document}

    \title{The $\textsc{[C ii]}$ line emission as an interstellar medium probe in the \textsc{Marigold} galaxies}

   \author{
        Prachi Khatri 
        \and Emilio Romano-Díaz
        \and Cristiano Porciani
        }

   \institute{Argelander Institute f\"ur Astronomie, Auf dem H\"ugel 71, D-53121 Bonn, Germany \\
             \email{pkhatri@astro.uni-bonn.de }
             }

   \date{Received 18 November 2024 / Accepted 25 February 2025}

    \titlerunning{$\textsc{[C ii]}$ emission in the \textsc{Marigold} galaxies}
    \authorrunning{P. Khatri et al.}
 
  \abstract
   {
   The [\ion{C}{II}] fine-structure line at 157.74 $\mu$m is one of the brightest far-infrared emission lines in galaxies and an important probe of galaxy properties such as the star formation rate (SFR) and the molecular gas mass ($M_{\mathrm{mol}}$). 
   }
   {Using high-resolution numerical simulations, we tested the reliability of the [\ion{C}{II}] line as a tracer of $M_{\mathrm{mol}}$ in high-redshift 
   galaxies and investigated secondary dependences of the $[\ion{C}{II}]-M_{\mathrm{mol}}$ relation on the SFR and metallicity. We also investigated the time evolution of the [\ion{C}{II}] luminosity function (LF) and the relative spatial extent of [\ion{C}{II}] emission and star formation.
   }
   {
   We post-processed galaxies from the \textsc{Marigold} cosmological simulations at redshifts $3 \le z \leq 7$ to obtain their [\ion{C}{II}] emission. These simulations were performed with the sub-grid chemistry model, \textsc{Hyacinth}, to track the non-equilibrium abundances of $\mathrm{H_2}$, $\mathrm{CO}$, $\rm C$ and $\mathrm{C^+}$ on the fly. Based on a statistical sample of galaxies at these redshifts, we investigated correlations between the [\ion{C}{II}] line luminosity ($L_{[\ion{C}{II}]}$) and the SFR, the $M_{\mathrm{mol}}$, the total gas mass
   and the metal mass in gas phase ($M_{\mathrm{metal}}$). 
   }
   {We find that accounting for secondary dependencies in the $L_{[\ion{C}{II}]}-M_{\mathrm{mol}}$ relation improves the $M_{\mathrm{mol}}$ prediction by a factor of 2.3 at all redshifts. 
   Our simulations predict a mild evolution in the slope of the $L_{[\ion{C}{II}]}-\mathrm{SFR}$ relation ($\lesssim 0.15$ dex) and an increase in the intercept by 0.5 dex in the above redshift range. 
   Among the various galaxy properties we explore, the [\ion{C}{II}] emission in our simulated galaxies shows the tightest correlation with $M_{\mathrm{metal}}$, indicating the potential of this line to constrain the metallicity of high-redshift galaxies. 
   About 20\% (10\%) of our simulated galaxies at $z=5$ ($z=4$) have [\ion{C}{II}] emission extending $\geq 2$ times farther than the star formation activity. The [\ion{C}{II}] LF evolves rapidly and is always well approximated by a double power law that does not show an exponential cut-off at the bright end. We record a  600-fold increase in the number density of $L_{[\ion{C}{II}]} \sim 10^9 \, \mathrm{L_{\odot}}$ emitters in 1.4 Gyr.}
   {}
 

   \maketitle
%


\section{Introduction}
\label{sec:intro}
The fine-structure line of singly ionised carbon ($\mathrm{C^+}$) at a wavelength of  $157.74 \, \mu \rm m$ (hereafter $[\ion{C}{II}]$), is an important coolant in the interstellar medium (ISM) of galaxies near and far \edit{\citep[][also see \citealt{carilli-walter13}  and \citealt{hodge-daCunha20} for a review]{crawford85, tielens_hollenback85, wolfire95}}. Being one of the brightest lines in the infrared, accounting for $\sim 0.1-1 \%$ of the total infrared luminosity in star-forming galaxies \citep{diaz-santos13}, it is particularly useful for observing high-redshift ($z \gtrsim 4$) galaxies, where it is conveniently redshifted to the transparent atmospheric window at millimetre wavelengths and is accessible from the ground with the Atacama Large Millimeter/submillimeter Array (ALMA) and the Northern Extended Millimeter Array (NOEMA), among others. \edit{Some recent surveys include the ALMA Large Program to Investigate $\mathrm{C^+}$ at Early Times 
\citep[ALPINE;][]{lefevre20} at $4.4 \leq z \leq 5.9$,  the Reionization Era Bright Emission Line Survey
\citep[REBELS;][]{bouwens22} at $6 \leq z \leq 9$, and the  [\ion{C}{II}] Resolved ISM in STar-forming galaxies with ALMA (CRISTAL).}

The strength of this line has been shown to correlate well with both the galaxy-integrated star formation rate \citep[SFR;][]{stacey10, delooze11, delooze14, carniani18, matthee19, schaerer20} and the spatially resolved SFR \citep[the $\Sigma_{[\ion{C}{II}]} - \Sigma_{\mathrm{SFR}}$ relation;][]{pineda14, delooze14, herrera-camus15}. Additionally, in recent years, the line strength has been used as a tracer of other galaxy-integrated quantities such as the molecular gas mass \citep{hughes17, madden20, zanella18}, particularly of CO-dark molecular gas \citep{madden20, accurso17}, the $\rm H \, \textsc{i}$ mass \citep{heintz21, heintz22}, the total gas mass \citep{eugenio23}, as well as the metal content \citep{heintz23}. 
However, it is known from observations of [\ion{C}{II}] emission from galactic centres and luminous infrared galaxies, with infrared (IR) luminosities $L_{\rm IR}\gtrsim 10^{11} \, \rm L_{\odot}$, that the $L_{[\ion{C}{II}]}/L_{\rm IR}$ ratio decreases with increasing $L_{\rm IR}$ \citep[e.g.,][]{malhotra01, gracia-carpio11, diaz-santos13}, thereby hinting at a possible breakdown of the [\ion{C}{II}]-SFR relation at high SFRs or high SFR surface densities \citep[see for example,][]{gracia-carpio11}, \edit{commonly referred to as the ``[\ion{C}{II}] deficit''}. 

From the theoretical point of view, the observed correlations between $[\ion{C}{II}]$ line strength and the galaxy properties arise naturally as $[\ion{C}{II}]$ is a metal cooling line and is linked to both the metal content and the heating via star formation in regions where cooling is dominated by this line such as photon-dominated regions (PDRs) and the cold neutral medium. Moreover, the various galaxy properties are themselves correlated; for example, the Kennicutt-Schmidt relation connects the gas surface density and the SFR surface density \citep{kennicutt98, leroy08, bigiel08}, while the mass-metallicity relation \citep{tremonti04} connects the stellar mass and gas metallicity. This implies that any correlation of the $[\ion{C}{II}]$ line with another galaxy property will have secondary dependencies, often manifested as a scatter, that must be quantified to provide robust calibrations. 

In this regard, the $[\ion{C}{II}]$-SFR correlation has garnered a lot of attention on the theoretical front. Several studies have meticulously tested this correlation and its redshift evolution using chemical and radiative transfer modelling in individual galaxies \citep{vallini15, katz19, katz22} or for entire simulation suite at targeted redshifts \citep{olsen16, olsen17, pallottini17, lagache18, lupi18, popping19, leung20, casavecchia24a}. 

However, unlike the $[\ion{C}{II}]$-SFR relation, the $[\ion{C}{II}]-M_{\mathrm{mol}}$ relation has received limited attention in simulations so far \citep[see for example,][]{vizgan22, casavecchia24b}, partly because current state-of-the-art cosmological simulations do not self-consistently follow the evolution of the molecular gas component in galaxies, and they often rely on analytical relations to be used in post-processing \citep[e.g.,][]{lagos15, lagos16}, which might not hold at high redshifts. For instance, \cite{vizgan22} found a shallower $L_{[\ion{C}{II}]}-M_{\mathrm{mol}}$ relation at $z \sim 6$ compared to the one obtained by \cite{zanella18} using a compilation of  $M_* \gtrsim 10^{10} \, \mathrm{M_{\odot}}$ galaxies at $z=0-5.5$, highlighting the need for robust testing of this calibration. 

Therefore, while observations of the [\ion{C}{II}] line have opened up an interesting avenue for probing the high-$z$ ISM, several open questions still remain: does the [\ion{C}{II}]-SFR relation evolve with redshift? Does the [\ion{C}{II}]-$M_{\mathrm{mol}}$ relation show secondary dependencies on other galaxy properties? 
What is faint-end slope of the [\ion{C}{II}] luminosity function and how does it evolve with redshift? To provide a theoretical insight on these, we have performed a suite of cosmological simulations, called the \textsc{Marigold} suite, wherein we follow the non-equilibrium abundance of $\mathrm{H_2}$, CO, C, and $\mathrm{C^+}$ on the fly using the sub-grid model \textsc{Hyacinth} \citep{khatri24}. The $[\rm \ion{C}{II}]$ emission from the simulated galaxies is calculated by solving the radiative-transfer problem in post-processing. In this paper, we use this simulation suite to investigate the usefulness of this line as a probe of the interstellar medium (ISM) in high-redshift ($3 \leq z \leq 7$) galaxies. In particular, we provide a calibration for inferring the molecular gas mass of a galaxy from its [\ion{C}{II}] luminosity, accounting for secondary dependencies in this relation across redshift. 

In the past few years, observations of high-redshift galaxies have detected $[\ion{C}{II}]$ emission extending farther than the UV continuum emission from these galaxies \citep{fujimoto19, fujimoto20, ginolfi20a, akins22, lambert23, posses24}, hinting at an extended gas reservoir rich in ionised carbon. 
This extended $[\ion{C}{II}]$ emission is often referred to as a $[\ion{C}{II}]$ halo, despite the misleading term.  Satellites galaxies, galactic outflows, and gas stripped due to galaxy interactions are all plausible sources of an extended [\ion{C}{II}] halo. Reproducing extended [\ion{C}{II}] in numerical simulations has proven challenging so far \citep{fujimoto19, munoz-elgueta24}, making it difficult to pinpoint its exact origin(s). This is further complicated by the faint nature of the extended emission and the limited spatial resolution of high-$z$ observations. In this study, we further explore the existence of extended [\ion{C}{II}] emission using our simulations.

The rest of the paper is organised as follows: In Sect.~\ref{sec:sims}, we describe the simulation suite and detail the modelling of the $[\ion{C}{II}]$ emission in Sect.~\ref{sec:cii_calc}. In Sect.~\ref{sec:cii_lf}, we investigate the redshift evolution of the $[\ion{C}{II}]$ luminosity function from the simulations. We investigate the $[\ion{C}{II}]$-star formation rate correlation in our simulated galaxies on global and spatially resolved scales in Sect.~\ref{sec:cii_sfr_all}. In Sect.~\ref{sec:cii_mmol}, we examine the reliability of the $[\ion{C}{II}]$ line as a molecular gas tracer and quantify secondary dependencies of the $L_{[\ion{C}{II}]}-M_{\mathrm{mol}}$ relation on the star formation rate and the gas metallicity. Then, we explore the extended $[\ion{C}{II}]$ emission in Sect.~\ref{sec:extended_cii}. We compare our [\ion{C}{II}]-SFR relation and [\ion{C}{II}] luminosity function with those from previous numerical studies in Sect.~\ref{sec:lit_comparison} and conclude with a summary of our main results in Sect.~\ref{sec:discussion}.

\renewcommand{\arraystretch}{1.1} 
\begin{table*}
    \caption{Specifications of the \textsc{Marigold} simulation suite.} 
    \centering
    \begin{tabular}{ c c c c c c c c c}
    \hline
    \hline
    Simulation & $L_{\rm box} \, \rm $& $N_{\rm DM}$ & $l_{\rm initial}$ & $l_{\rm final}$ & $\Delta x^{\rm min} \, \rm $  &  $m_{\rm DM}$ & $m_*$ & $m_{\mathrm{gas}}^{\rm ini}$\\
    & (cMpc) & & & & (pc)  & $(\mathrm{M_{\odot}})$ & $(\mathrm{M_{\odot}})$ & $(\mathrm{M_{\odot}})$\\
    \hline
    {\tt M25} & 25 & $1024^3$ & 10 & 17 & 32  & $5.0 \times 10^5$ & $7.2 \times 10^3$ & $9.3 \times 10^4$\\
    {\tt M50} & 50 &$1024^3$ & 10 & 17 & 64  & $4.0 \times 10^6$  & $5.8 \times 10^4$ & $7.4\times 10^5$\\
    \hline
    \end{tabular}
    \tablefoot{From left to right, the columns list: the name of the simulation, the comoving box size, the number of dark-matter (DM) particles, the initial and final refinement levels, the minimum cell size achieved in the simulation in physical units, the DM and stellar particle masses, and the average gas mass per grid cell in the initial conditions.}
    \label{tab:sims}
\end{table*}

\section{Simulations}
\label{sec:sims}

We use the \textsc{Marigold} suite of cosmological simulations (\textcolor{blue}{Khatri et al., in prep.}), which comprises two hydrodynamical simulations -- a $(25 \, \rm Mpc)^3$ comoving volume ({\tt M25}) and a $(50 \, \rm Mpc)^3$ comoving volume ({\tt M50}).  The simulations and the statistical properties of the simulated galaxies are described in full detail in \textcolor{blue}{Khatri et al. (in prep.)}. Here we briefly summarize some key details.

The simulations adopt the Planck cosmology \citep{planck18} 
with $\Omega_{\Lambda}=0.6847$, $\Omega_{\rm m}=0.3153$, $\Omega_{\rm b}=0.0493$, $\sigma_{8}=0.8111$, $n_{\rm s}=0.9649$, and $h=0.6736$. The simulations are started from uni-grid initial conditions (ICs) set at $z = 99$ generated with the code \textsc{Music} \citep{music}. The ICs have an initial refinement level $l_{\rm initial}=10$ corresponding to $1024^3$ grid cells and an equal number of dark matter particles. We allowed the grid to refine naturally down to $z=3$, which results in a maximal spatial resolution (i.e., minimum grid cell size $\Delta x^{\rm min}$ in physical units) achieved during the course of the  simulations of $\Delta x^{\rm min} =  32 \, \rm pc$ for {\tt M25} and $\Delta x^{\rm min} =  64 \, \rm pc$ for {\tt M50}. The simulation volumes have periodic boundary conditions and the dynamical evolution of dark matter, gas, and stars is tracked with the adaptive mesh refinement code \textsc{Ramses} \citep{ramses} down to $z= 3$. The simulation specification are given in Table~\ref{tab:sims}.

These simulations were performed using the sub-grid model \textsc{Hyacinth} \citep{khatri24} to evolve the non-equilibrium abundances of $\mathrm{H_2}$, CO, C, and $\mathrm{C^+}$. \textsc{Hyacinth} comprises a sub-grid chemical network based on our modified version of the widely used \cite{NL99} chemical network. It accounts for the unresolved density structure of the ISM by assuming a probability distribution function (PDF) of sub-grid densities. The PDF is designed to vary with the state of star formation. A metallicity-dependent temperature-density relation based on high-resolution molecular cloud simulations \citep{hu21} assigns a (sub-grid) temperature to each sub-grid density. The chemical rate equations are solved at each sub-grid density and the cell-averaged chemical abundances are obtained by integrating over the density PDF. The chemical abundances from \textsc{Hyacinth} are consistent with PDR codes \citep{wolfire10}. The model is described in full detail in \cite{khatri24}. These simulations adopted an $\mathrm{H_2}$-based prescription for star formation.

We used the Amiga Halo Finder \edit{\citep[AHF;][]{ahf}} to identify halos and subhalos in each snapshot. AHF identifies halos by locating density peaks within the simulation and then iteratively determining the gravitationally bound particles that constitute each peak. Each resulting halo is a spherical region with virial radius $R_{\mathrm{vir}}$ and a mean matter density (i.e., including dark matter, gas, and stars) equal to 200 times the critical density $\rho_{\rm crit}$. The virial mass of the halo can be written as $M_{\mathrm{vir}} = \frac{4}{3} \pi R_{\mathrm{vir}}^3 \; 200 \, \rho_{\rm crit}$, where the masses and sizes of halos are calculated accounting for unbinding. These are referred to as `main halos' in the following. Subhalos are defined as gravitationally bound objects within main halos and lying within common isodensity contours of the host halo. We imposed that every halo be resolved with at least 100 particles, unless otherwise specified (see for example, Sect.~\ref{sec:cii_lf}). Galaxies were defined in terms of their parent halo. For main halos, the stellar concentration at their centre is referred to as the main galaxy. For each main galaxy, we started with a spherical region of size 0.1 $R_{\mathrm{vir}}$ and calculated the stellar half-mass radius $r_{1/2, *}$ (that is, the radius containing half of the stellar mass within 0.1$R_{\mathrm{vir}}$). We adopted $2 r_{1/2, *}$ as the size of the galaxy and all (galaxy-integrated) quantities were measured within this radius. Conversely, the stellar concentration residing at the centre of a subhalo was called a satellite galaxy, whose size is defined by the radius corresponding to the maximum of the subhalo rotation curve, $R_{V_{\rm max}}$ \citep{klypin11, prada12}. In other words, $R_{\rm V_{\rm max}}$ sets the boundary of a satellite galaxy. 
To ensure that the stellar component of the galaxies was well resolved, we imposed a cut of 100 stellar particles, which resulted in stellar mass cuts of $7.2\times 10^5 \, \mathrm{M_{\odot}}$ and $5.8 \times 10^6 \, \mathrm{M_{\odot}}$ for {\tt M25} and {\tt M50}, respectively.
The median mass of a simulated galaxy at $z=7$ is $8.7 \times 10^7 \, \mathrm{M_{\odot}}$ ({\tt M25}) and $2.6 \times 10^8 \, \mathrm{M_{\odot}}$ ({\tt M50}). At $z=3$, these are $2.2 \times 10^8 \, \mathrm{M_{\odot}}$ and $1.8 \times 10^9 \, \mathrm{M_{\odot}}$, respectively.

\section{Modelling $[\ion{C}{II}]$ emission}
\label{sec:cii_calc}
The first ionisation potential of atomic carbon (11.3 eV) is lower than that of hydrogen (13.6 eV). Consequently, singly ionised carbon ($\mathrm{C^+}$) exists in different ISM phases including molecular clouds, neutral atomic gas, and \ion{H}{II} regions and emits the [\ion{C}{II}] fine-structure line in a variety of conditions. Disentangling the contribution of the different phases to the total [\ion{C}{II}] emission of a galaxy is therefore challenging and requires using other emission lines. 

\begin{figure*}
    \centering
    \begin{subfigure}[b]{0.95\textwidth}
    \includegraphics[width=0.95\textwidth, trim={0 0 0 0},clip]{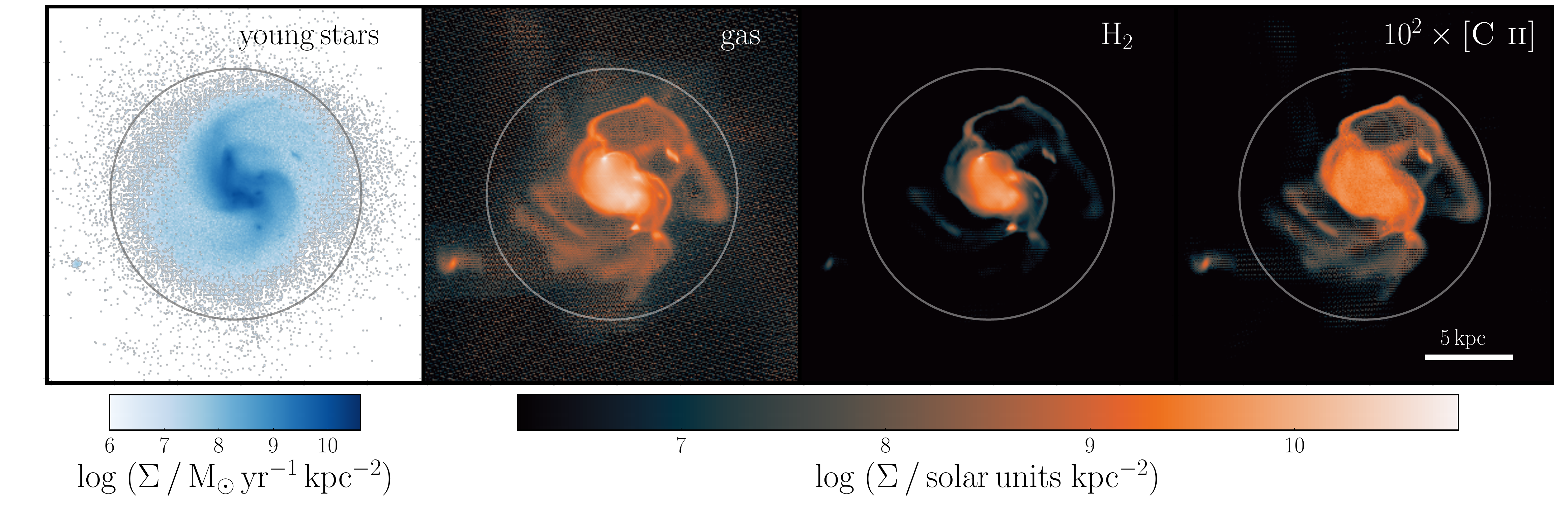}
    \end{subfigure}
    \caption{Face-on view of a simulated galaxy at $z=4$. From left to right, the columns show the surface density of young stars (with ages $\leq 200$ Myr), total gas \edit{(including H$_2$)} surface density, $\mathrm{H_2}$ surface density, and [\ion{C}{II}] surface brightness. In each panel, the circle indicates 0.1 times the virial radius of the parent DM halo.}
    \label{fig:image}
\end{figure*}

Observational studies of [\ion{C}{II}] emission in $z \gtrsim 1$ galaxies have shown that the bulk of the [\ion{C}{II}] emission originates from PDRs that represent a transition region between the \ion{H}{II} region around a young massive star and a molecular cloud. 
For instance, \cite{stacey10} found that in a sample of 12 galaxies at $1 \lesssim z \lesssim 2$, more than $70\%$ of the [\ion{C}{II}] emission arises from PDRs.
Also, \cite{gullberg15} found that in their sample of 20 dusty star-forming galaxies at $2.1 < z < 5.7$, the [\ion{C}{II}] emission is consistent with arising from PDRs.
This finding is further supported by theoretical studies that calculate the [\ion{C}{II}] emission from simulated galaxies in post-processing, accounting for emission arising from different phases, for example, \cite{olsen15}, who employ a multi-phase ISM model for each gas particle in the simulation and use the photoionisation code \textsc{Cloudy} \citep{ferland92} to calculate the emission arising from each phase (also see \citealt{pallottini17} and \citealt{casavecchia24b}). 
\edit{
These numerical studies show} that the bulk ($\gtrsim 70-90 \%$) of the [\ion{C}{II}] emission arises from neutral atomic and molecular gas. Moreover, based on a sample of low-metallicity dwarf galaxies from the \textit{Herschel} Dwarf Galaxy Survey \citep{madden13}, \cite{cormier19} found that $\gtrsim 70\%$ [\ion{C}{II}] emission arises from PDRs. As such galaxies are expected to be similar to high-redshift galaxies in terms of their ISM structure, it is reasonable to assume that a similar fraction of the [\ion{C}{II}] emission arises from PDRs in high-redshift galaxies as well.  


We used \textsc{Hyacinth} to obtain the abundances of $\mathrm{H_2}$, $\mathrm{CO}$, $\rm C$, and $\mathrm{C^+}$ in our simulations. 
This approach allowed us to model the [\ion{C}{II}] emission without relying on assumptions regarding the $\mathrm{C^+} $ abundance that might not hold across the entire galaxy population at all redshifts.

The [\ion{C}{II}] line arises from the $^2P_{3/2} \, \rightarrow \, ^2P_{1/2}$ fine-structure transition of $\mathrm{C^+}$, that can be excited by collisions with hydrogen molecules ($\mathrm{H_2}$), hydrogen atoms ($\rm H$), and electrons ($e^-$). The transition can also be excited by an external radiation field such as the cosmic microwave background (CMB), which becomes particularly important at higher redshifts \citep{daCunha13}. De-excitation can either happen spontaneously or be stimulated by the external radiation field. 

When calculating the $[\ion{C}{II}]$ emission from a simulated galaxy, we accounted for optical depth effects within individual cells. For this, we assumed a plane-parallel configuration and divided each cell into $N$ slices. The density in each slice was obtained from the underlying density PDF \citep[same as in \textsc{Hyacinth}; see also][]{vallini15}.
The level populations in each slice were computed accounting for the emission from all other slices. Conversely, the fraction of emission from a given slice that manages to reach the edge of the cell (where the optical depth $\tau=0$) depends on its location within the cell. Solving this radiative transfer problem requires an iterative approach, which is detailed in Appendix~\ref{sec:appA}. We validate our model against \textsc{Cloudy} in Appendix~\ref{sec:cloudy}. 

However, when calculating the total [\ion{C}{II}] luminosity of a galaxy, we assumed that the cells are radiatively decoupled from each other. This means that the $[\ion{C}{II}]$ emission that escapes the cell of origin, travels unattenuated to the edge of the galaxy. This assumption is valid whenever the velocity difference between neighbouring cells is larger than the intrinsic line width due to the gas motions within a cell: 
the emitted spectra is shifted out of the frequency range where it could be absorbed by another cell. This is a common approximation in the literature \citep[see e.g.][]{olsen15, vallini15}. The total $[\ion{C}{II}]$ luminosity of the galaxy is then calculated as the sum of the luminosities of each cell within the galaxy.

Fig.~\ref{fig:image} shows the different stellar and gas components and the [\ion{C}{II}] surface brightness of a simulated galaxy at $z=4$. \edit{From left to right, the panels show the surface density of young stars with ages $\leq 200$ Myr, the surface density of the total gas in the galaxy (including molecular hydrogen), the $\mathrm{H_2}$ surface density, and the surface brightness of the [\ion{C}{II}] line.}
\section{[C II] luminosity function}
\label{sec:cii_lf}

\begin{figure*}
    \centering
    \includegraphics[width=0.80 \textwidth, trim={1cm 0 4cm 2cm},clip]{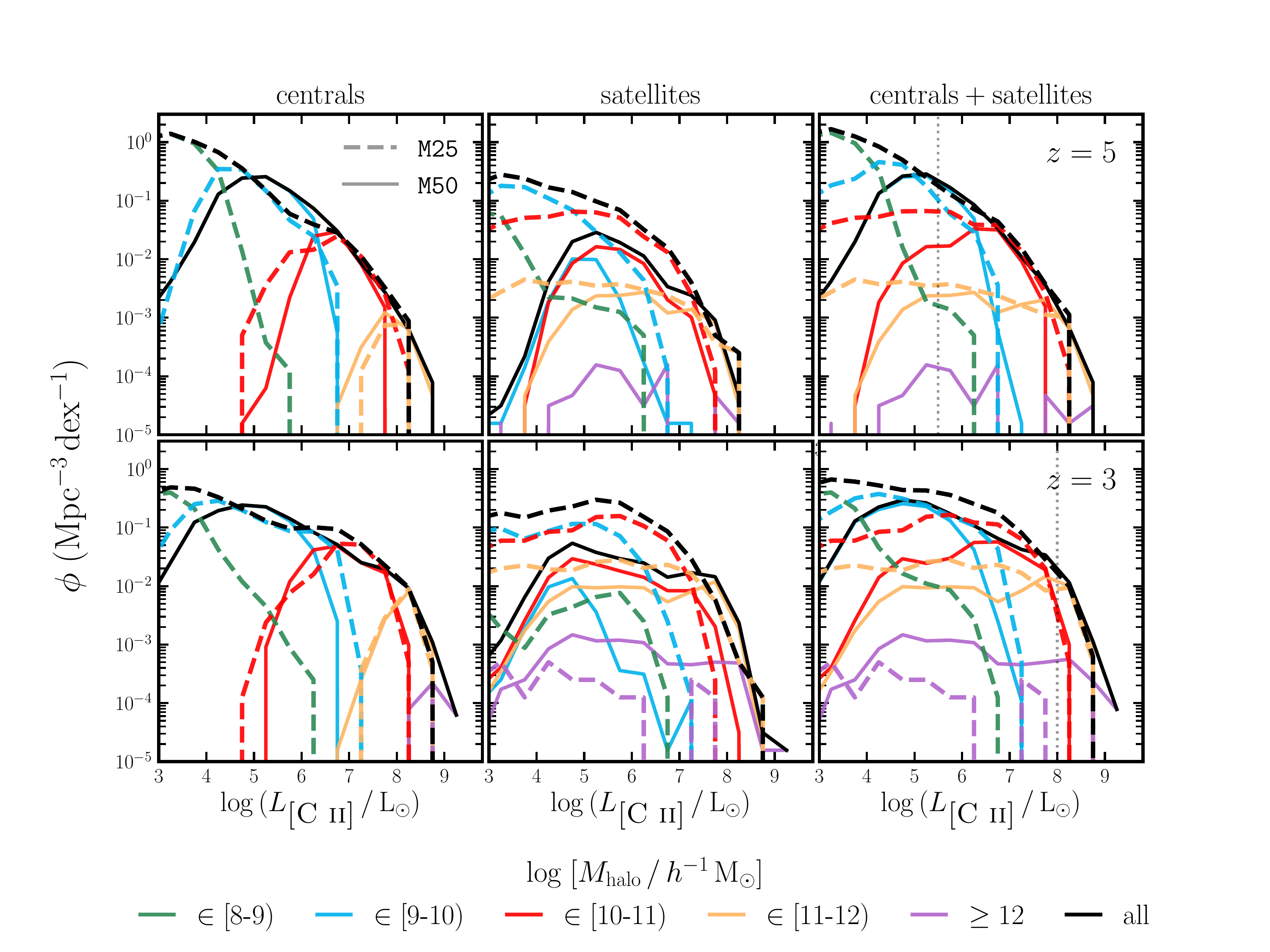}
    \caption{Conditional [\ion{C}{II}] LF from the M25 (dashed lines) and M50 (solid lines) simulations at redshifts $z=5$ and 3 for central galaxies (left panels), satellites galaxies (middle panels), and all galaxies (right panels). The coloured lines show the CLF of emitters residing in DM halos in different $M_{\mathrm{halo}}$ bins and the black lines show the total LFs. The dotted grey line in the right panels denotes the luminosity threshold, $L_{\mathrm{thr}}$, below which the total LFs from two simulations differ significantly due to resolution effects. }
    \label{fig:conditional_cii_lf}
\end{figure*}

In this section, we examine the [\ion{C}{II}] luminosity function (LF) from the \textsc{Marigold} simulations and investigate how it evolves with redshift. 

\subsection{Conditional luminosity function and resolution effects}
The difficulty we have to face is to combine simulations with different spatial and mass resolutions in
a consistent way while accounting for sample variance given the relatively small computational volumes. For these reasons, we first performed a consistency check between the outputs of ${\tt M25}$ and ${\tt M50}$ by
measuring the conditional luminosity function \cite[CLF,][]{yang03}, i.e. the luminosity function of emitters hosted
by DM halos within a narrow mass bin. \edit{Numerical simulation typically adopt a threshold of $\sim 100$ particles to define resolved halos. Here, we have adopted a more conservative threshold of 300 DM particles} for our (main) halos and examined three different cases: (i) central galaxies only, (ii) satellite galaxies only, and (iii) centrals+satellites.
Our results for $z=5$ and 3 are shown in 
Fig.~\ref{fig:conditional_cii_lf}. These cases are representative of what happens in the redshift ranges
$5\leq z\leq 7$ and $3\leq z<5$, respectively.  



For all halo masses, the CLFs of the central galaxies in the two simulations are in excellent agreement. The CLF has a characteristic bell shape and its width grows with decreasing halo mass.
Conversely, satellites have a much broader CLF and
{\tt M25} presents many more faint satellites than {\tt M50} at fixed halo mass. This discrepancy reflects the different mass and spatial resolutions in the simulations that regulate the abundance of DM satellites and the [\ion{C}{II}] emission from their gas, respectively.
Inspection of the total LF reveals that the two simulations show small differences (which are compatible with sample variance) above a threshold luminosity $L_\mathrm{thr}$ and substantial systematic differences for $L_{[\ion{C}{II}]}<L_\mathrm{thr}$. 
We find that $L_\mathrm{thr}\simeq 10^{5.5}$ L$_\sun$ for $5\leq z\leq 7$ and
$L_\mathrm{thr}\simeq 10^{8}$ L$_\sun$ for $3\leq z< 5$
(dotted grey lines in the right panels of Fig.~\ref{fig:conditional_cii_lf}).
This confirms that, as we stated before, our high-resolution ${\tt M25}$ simulation is excellent for probing the faint end of the LF while the ${\tt M50}$ simulation is ideally suited for probing the bright end because of its larger volume. 

Finally, we note that at $L_{[\ion{C}{II}]} \geq 10^5 \, \mathrm{L_{\odot}}$, the total luminosity function is fully represented by halos with masses $M_\mathrm{halo}\geq 10^9 \, h^{-1} \, \mathrm{M_{\odot}}$. Therefore, in the following, we restrict our analysis to these ranges of luminosities and masses. 



\subsection{Bayesian curve fitting}

At each redshift, we fitted the simulated LF, $\phi(L_{[\ion{C}{II}]})=\mathrm{d}n/\mathrm{d} \log L_{[\ion{C}{II}]}$\footnote{Note that here and throughout the text, we use `log' to denote $\rm log_{10}$; for the natural logarithm $\rm log_{e}$, we use `ln' instead.}, with different
functional forms. Based on the discussion above,
we included all galaxies with $L_{[\ion{C}{II}]} \geq 10^5 \, \mathrm{L_{\odot}}$ hosted by halos with $M_\mathrm{halo}\geq 10^9 \, h^{-1} \, \mathrm{M_{\odot}}$ from {\tt M25} but only those with $L_{[\ion{C}{II}]}>L_\mathrm{thr}$ from {\tt M50}. 
For the functional forms, we considered a Schechter function,
\begin{equation}
\label{eq:schechter}
    \phi(L_{[\ion{C}{II}]}) = 
    \ln(10) \; \phi_* \left( \frac{L_{[\ion{C}{II}]}}{L_*} \right)^{\alpha + 1} \exp \left( - \frac{L_{[\ion{C}{II}]}}{L_*} \right) \, ,
\end{equation}
and a double power-law (DPL),
\begin{align}
\label{eq:double_pl}
    \phi(L_{[\ion{C}{II}]}) 
    =&  
    \displaystyle  \ln(10) \, \phi_* \; 
    \left[ \left(\frac{L_{[\ion{C}{II}]}}{L_*} \right)^{-(\alpha+1) } \!\!\!\!\!+\left( \frac{L_{[\ion{C}{II}]}}{L_*} \right)^{-(\beta+1)} \right]^{-1} \,. 
\end{align}
where all parameters have the usual meaning and $\beta<\alpha$.


We followed a Bayesian approach and sampled the parameter space using a Markov-chain Monte-Carlo (MCMC) method implemented with the python package {\tt emcee} \citep{emcee}. We assumed that the counts $N_i$ in 
each logarithmic bin of luminosity follow a Poisson distribution and write the log-likelihood function for each simulation as
\begin{equation}
    \label{eq:likelihood}
    \ln \mathcal{L} = \sum_{i=1}^{N_\mathrm{bins}} N_i \, \ln (N_{\mathrm{model},i}) 
    - N_{\mathrm{model},i} + \mathrm{constant}\,.
\end{equation}
Eventually, we summed the results obtained for {\tt M25} and {\tt M50}.

To account for the sample variance of the simulated volumes, we followed an approach that builds upon the method originally proposed by \cite{trenti08} to obtain the LF from the combination of observational data with different depths. Briefly, we fitted the LF data from the two simulations
with exactly the same shape but (slightly) different normalisation factors that can be written as
$\mathrm{log} \, \phi_{*, j} = \mathrm{log} \, \phi_{*} + \Delta_{j}$, where $\phi_*$ represents the `cosmic' normalisation and 
$\Delta_{j}$ is the correction due to sample variance in the $j^\mathrm{th}$ simulation.
We imposed independent Gaussian priors on each $\Delta_j$. In other words, $\Delta_{j} \sim \mathcal{N}(0,(\sigma_{\mathrm{v}, \, j} / \ln 10 )^2)$, where 
$\sigma^2_{\mathrm{v}, \, j}$ is the sample variance of the overdensity within the respective simulation volume. The latter was estimated from the calculations presented in \cite{sommerville04}, considering the halo mass bin that gives the dominant contribution to the counts of emitters around $L_*$. We adopted (broad) flat priors on all other parameters. 

As an example, we show in Appendix~\ref{sec:schechter}, the resulting posterior distribution for the DPL fit at $z=3$. In what follows, we present results obtained after marginalising the posterior distributions over the parameters, $\Delta_j$.

Since computing the model evidence from the Markov chains is problematic, for simplicity, we used the deviance information criterion \citep[DIC,][]{spiegelhalter02} to identify whether the Schechter function or the DPL
provide a better fit to the simulated data. 
Deviance is a measure of goodness of fit defined as $D=-2 \ln \mathcal{L}(\boldsymbol{\theta})$, where $\boldsymbol{\theta}$ indicates the parameters of a model. The DIC is obtained correcting the deviance with a penalty, $p_{\mathrm{D}}$, for the complexity of the model which
is quantified in terms of the effective number of fit parameters. There are two common approaches to estimate $p_{\mathrm{D}}$ from a Markov chain: $p_{\mathrm{D}}^{(a)}=\overline{D(\boldsymbol{\theta})}-D(\overline{\boldsymbol{\theta}})$ and
$p_{\mathrm{D}}^{(b)}=\mathrm{Var}(D)/2$, where an overbar denotes the average over the posterior distribution (i.e. over the sampled chain) and the symbol Var denotes the corresponding variance.
The DIC is then defined as DIC$=D(\overline{\boldsymbol{\theta}})+2p_{\mathrm D}$.
Models with smaller DIC are better supported by the data.
Typically, differences ($\Delta$DIC) above 5 are considered substantial and the model with the higher DIC is ruled out if the difference grows above 10. Based on this test, we find that the LF from our simulations is better represented by a DPL. The corresponding parameters are listed in Table~\ref{tab:cii_lf} along with the $\Delta$DIC values computed with both estimators for $p_{\mathrm{D}}$. We provide in Appendix~\ref{sec:schechter} a comparison of the two functional forms with the simulation data.

\begin{figure*}
    \centering
    \includegraphics[width=0.78\linewidth, scale=0.45, trim={2cm 0.5cm 5cm 3.5cm},clip]{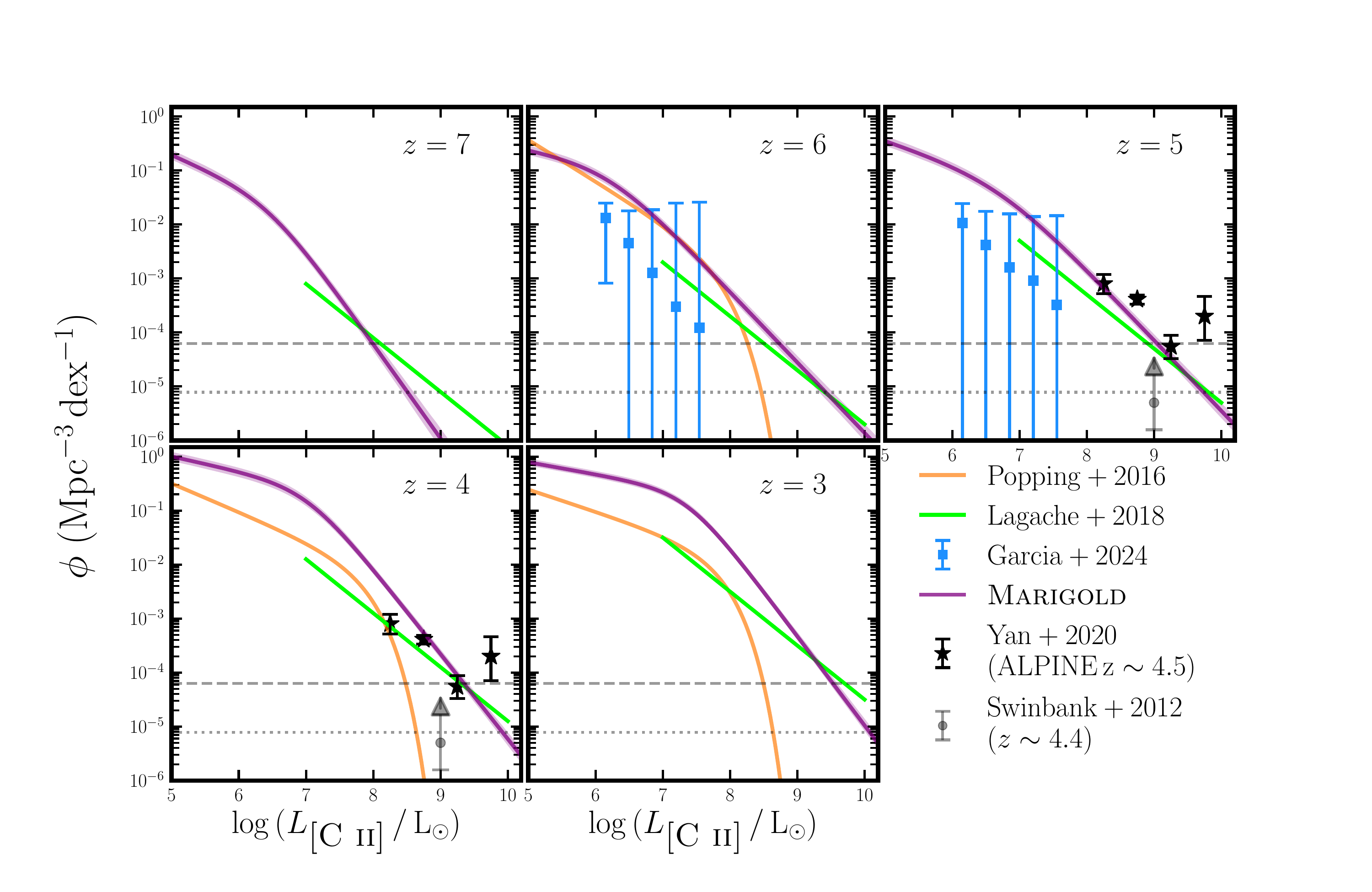}
    \caption{Simulated $[\ion{C}{II}]$ LF compared with observational estimates. The coloured lines represent the best-fit DPL -- Eq.~(\ref{eq:double_pl}) -- to the simulated LF and the shaded area represents the central 68\% credibility range obtained using the MCMC chains. Black stars represent the observational estimates at $z \sim 4.5$ from the ALPINE survey \citep{yan20} and the grey arrow shows the lower limit from \cite{swinbank12} based on observations of two galaxies at $z \sim 4.4$. The dashed and dotted horizontal lines represent a number count of 1 per dex in the entire simulation volume of {\tt M25} and {\tt M50}, respectively.
     The [\ion{C}{II}] LF from \cite{popping16} and \cite{lagache18} both based on a semi-analytical galaxy formation model, and from \cite{garcia24} obtained by post-processing the \textsc{Simba} simulations at $z=6$ and 5 are included in the respective panels.     }
    \label{fig:cii_lf}
\end{figure*}

Fig.~\ref{fig:cii_lf} shows the resulting $[\ion{C}{II}]$ luminosity function obtained using a DPL at different redshifts. The shaded band indicates the central 68\% credibility region obtained with the MCMC method. The solid line represents the best fit (evaluated at the posterior mean $\overline{\boldsymbol{\theta}}$). We see a clear redshift evolution in the LF. The turnover luminosity ($L_*$) increases (almost) monotonically with time and the faint-end slope ($\alpha$) tends to flatten at late times. In contrast, the bright-end slope ($\beta$) does not show a clear evolutionary trend. The large jump in the LF at $L_{[\ion{C}{II}]} \lesssim 10^{8} \, \mathrm{L_{\odot}}$ is partially due to the spatial refinement that happens shortly after $z=5$ in both simulations, that allows them, especially {\tt M25}, to resolve more emitters. We also see a significant evolution in the number density of bright emitters. The number density of galaxies at $L_{[\ion{C}{II}]} \sim 10^9 \, \mathrm{L_{\odot}}$ increases from $\sim 10^{-6} \, \mathrm{dex^{-1} \, Mpc^{-3}}$ at $z=7$ to  $6 \times 10^{-4} \, \mathrm{dex^{-1} \, Mpc^{-3}}$ at $z=3$, indicating a 600-fold increase in the number of $L_{[\ion{C}{II}]} \sim 10^9 \, \mathrm{L_{\odot}}$ galaxies in a less than 1.5 Gyr timespan.

Observational constraints on the [\ion{C}{II}] luminosity function at these redshifts come from \cite{yan20} for targeted detections at $4.5\lesssim z \lesssim 5.9$ in ALPINE 
and a lower limit at $z \sim 4.4$ reported by \cite{swinbank12} based on [\ion{C}{II}] detections in two observed galaxies. We find that our LFs at $z=5$ and 4 are in excellent agreement with both estimates.

\edit{In Fig.~\ref{fig:cii_lf}, we also compare our [\ion{C}{II}] LF against the results 
from \cite{popping16} and \cite{lagache18}, both based a semi-analytical galactic formation models. \cite{popping16} fit a Schechter function to their predicted LFs at different redshifts. At $z=6$, our predicted LF is very similar to \cite{popping16}, despite the differences in our methods. However, deviations start to appear at late times. For instance, despite similar faint-end slopes at redshifts $z=4$, our simulations predict a higher number density of emitters at the all luminosities. Moreover, our LFs extend out to higher luminosities, similar to \cite{lagache18}, who fit their LFs with a single power law with a slope of $-1.0$ at all redshifts. 
In this regard, the underabundance of bright galaxies in semi-analytical models (SAMs) that track the $\mathrm{H_2}$ abundance is a well-known problem and is related to the underabundance of cold gas in the galaxies simulated with SAMs \citep[][]{popping15}. At all redshifts, we find a steeper bright-end slope in comparison to \cite{lagache18}.}
\edit{We further compare with the LF from \cite{garcia24} obtained by post-processing the \textsc{Simba} simulations at $z=5$ and 6. We find that although consistent within 1 $\sigma$ uncertainty, their predictions are consistently below ours and can be up to an order of magnitude lower at the bright end.}

\edit{Discrepancies in the predicted LFs from different studies can have important consequences for predicting the power spectrum of the line intensity mapping (LIM) signal. 
LIM is an emerging technique that measures the integrated line emission from galaxies and the intergalactic medium without resolving individual sources \citep[see][for a recent review]{bernal22}. The first moment of the LF governs the overall amplitude of the LIM power spectrum, while the second moment influences the shot noise component. 
As bright emitters are expected to have a dominant contribution to the [\ion{C}{II}] LIM power spectrum at $z \gtrsim 4$ (e.g. \textcolor{blue}{Marcuzzo et al., in prep.}), current and upcoming LIM surveys can prove extremely useful in constraining the different numerical models.}

\renewcommand{\arraystretch}{1.4} 
\begin{table*}[h]
\centering
\caption{Best-fit parameters to the LF for the DPL function given in Eq.~(\ref{eq:double_pl}).}
\begin{tabular}{c|cccc|cc}
\hline
\hline
$z$  & $\log (\phi_{*} \, / \, \rm Mpc^{-3} \, dex^{-1}$) & $\log (L_{*} \, / \, \rm L_{\odot}$) & $\alpha$ & $\beta$ & $\Delta\mathrm{DIC}^{(a)}$ & $\Delta\mathrm{DIC}^{(b)}$  \\ \hline
7 &
$-1.84_{-0.15}^{+0.15}$ &
$6.44_{-0.11}^{+0.10}$ &
$-1.54_{-0.05}^{+0.06}$ &
$-2.75_{-0.09}^{+0.08}$ &
105 & 107
\\
6 &
$-1.26_{-0.12}^{+0.12}$ &
$6.20_{-0.10}^{+0.10}$ &
$-1.24_{-0.06}^{+0.06}$ &
$-2.30_{-0.05}^{+0.04}$ &
222 & 223
\\
5 &
$-1.52_{-0.13}^{+0.13}$ &
$6.72_{-0.11}^{+0.11}$ &
$-1.42_{-0.03}^{+0.03}$ &
$-2.31_{-0.05}^{+0.05}$ &
159 & 161
\\
4 &
$-0.91_{-0.07}^{+0.08}$ &
$7.03_{-0.04}^{+0.04}$ &
$-1.28_{-0.01}^{+0.01}$ &
$-2.57_{-0.05}^{+0.05}$ &
449 & 451
\\
3 &
$-1.00_{-0.08}^{+0.08}$ &
$7.37_{-0.04}^{+0.04}$ &
$-1.22_{-0.01}^{+0.01}$ &
$-2.65_{-0.05}^{+0.05}$ &
174 & 176
\\
\hline
\end{tabular}
\tablefoot{The last two columns show the values of $\Delta$DIC$^a$ and $\Delta$DIC$^b$ between the Schechter function and the DPL (see text).}
\label{tab:cii_lf}
\end{table*}

\subsection{[\ion{C}{II}] luminosity density}
\label{sec:rho_cii}
\begin{figure}
    \centering
    \includegraphics[width=0.45\textwidth, trim={0 1.5cm 0 0.8cm},clip]{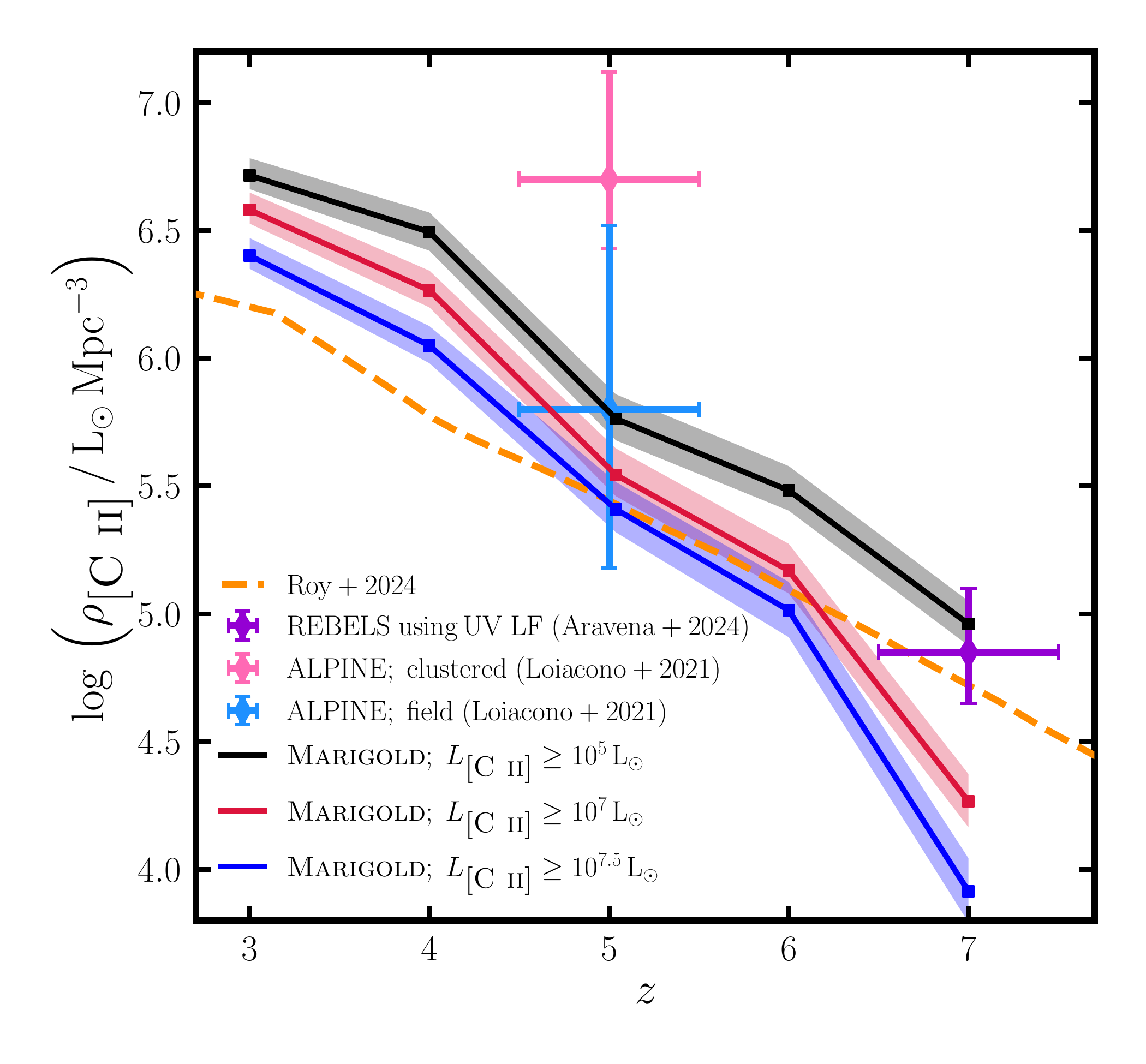}
    \caption{Comparison of the cosmic [\ion{C}{II}] luminosity density ($\rho_{[\ion{C}{II}]}$) for different luminosity cuts in the \textsc{Marigold} simulations with observational estimates from ALPINE \citep{loiacono21} -- clustered and field estimates in pink and blue, respectively; from REBELS \citep{aravena24} in purple and from a semi-empirical model by \cite{roy24} shown as an orange dashed line. The $\rho_{[\ion{C}{II}]}$ from the simulations is obtained by integrating the LFs shown in Fig.~\ref{fig:cii_lf} down to $\mathrm{log} (L_{\mathrm{min}} \, /\, \mathrm{L_{\odot}})= 5$ (black), 7 (red), and 7.5 (blue). The shaded regions represent the 16-84 percentile range of $\rho_{[\ion{C}{II}]}$ obtained from the LFs.
    }
    \label{fig:rho_cii}
\end{figure}

Measuring the cosmic star formation rate density (SFRD) at different epochs has been the subject of several studies \citep[see][for a complete review]{madau-dickinson14}. Similarly, estimating the cosmic molecular gas density from blind and targeted surveys of molecular gas tracers such as CO and dust continuum \citep[see e.g.][among others]{aspecs, riechers19, scoville17, magnelli20} has also gained significant interest over the last decade. Owing to the correlation between the [\ion{C}{II}] luminosity with both the SFR and molecular gas mass across redshift, estimates  of the cosmic [\ion{C}{II}] luminosity density ($\rho_{[\ion{C}{II}]}$) can be used to infer the cosmic SFRD and the cosmic molecular gas density (see for example, \citealt{yan20, loiacono21} for the ALPINE survey and \citealt{aravena24} for \edit{REBELS}). 

In Fig.~\ref{fig:rho_cii}, we show the redshift evolution of $\rho_{[\ion{C}{II}]}$ predicted by our simulations and compare with observational estimates at these redshifts. For this, we integrated the LF in Fig.~\ref{fig:cii_lf} down to $\mathrm{log} (L_{\mathrm{min}} \, /\, \mathrm{L_{\odot}})= 5$. This is shown by a black line in Fig.~\ref{fig:rho_cii} and referred to as our fiducial estimate in the following. For a fair comparison with observational estimates of $\rho_{[\ion{C}{II}]}$ from the ALPINE \citep{loiacono21} and REBELS \citep{aravena24} surveys, we show using red and blue lines, respectively, the $\rho_{[\ion{C}{II}]}(z)$ for $\mathrm{log} (L_{\mathrm{min}} \, /\, \mathrm{L_{\odot}})= 7$ and $\mathrm{log} (L_{\mathrm{min}} \, /\, \mathrm{L_{\odot}})= 7.5$, 
which correspond to the luminosity cuts adopted in the two surveys for integrating the luminosity function. Note that \cite{loiacono21} provide two estimates for $\rho_{[\ion{C}{II}]}$ based on serendipitously detected galaxies in ALPINE at $z \sim 5$; one of these is obtained by integrating the [\ion{C}{II}] LF based on a sample of galaxies that seem to be part of a local overdensity and are therefore not representative of the galaxy population at the targeted redshift (this is referred to as the `clustered estimate'). The $\rho_{[\ion{C}{II}]}$ for the field population (the `field estimate') can be estimated by only considering emitters detected outside the aforementioned overdensity. Since only one such emitter is detected in their sample, \cite{loiacono21} obtain the field estimate by multiplying the clustered estimate by the ratio between the number of emitters in the field and clustered subsamples (=1/11), on account of the similar volumes estimated for the two subsamples. We find that the $\rho_{[\ion{C}{II}]}$ predicted by our simulations at $z =5$ shows an excellent agreement with that from ALPINE for their field sample, irrespective of the luminosity cut. 
Conversely, as expected, the $\rho_{[\ion{C}{II}]}^{\rm clustered}$ from ALPINE in almost an order of magnitude higher than our predicted $\rho_{[\ion{C}{II}]}$.
Our $\rho_{[\ion{C}{II}]}$ at $z=7$ is lower than the REBELS one when considering their luminosity cut of $10^{7.5} \, \rm L_{\odot}$ (blue line), but our $\rho_{[\ion{C}{II}]}$ with lower luminosity cuts (red and black lines) bracket the REBELS estimate. Overall, our estimates show a good agreement with observations. At $z \lesssim 4$, the impact of a luminosity cut is not as severe. 

We also include in Fig.~\ref{fig:rho_cii}, the $\rho_{[\ion{C}{II}]}(z)$ estimate from \cite{roy24}, based on a semi-empirical model that performs an abundance matching of the theoretical halo mass function and the observed stellar mass function, and assigns a [\ion{C}{II}] luminosity to every halo based on the empirical SFR-stellar-mass relation and the [\ion{C}{II}]-SFR relation from \cite{vallini15}. 
The resulting $\rho_{[\ion{C}{II}]}(z)$ from their approach is similar in shape to our fiducial estimate but consistently lower by a factor of $\sim 2$ in the redshift range shown here.

%
\begin{figure*}
    \centering
    \includegraphics[width=0.78\linewidth]{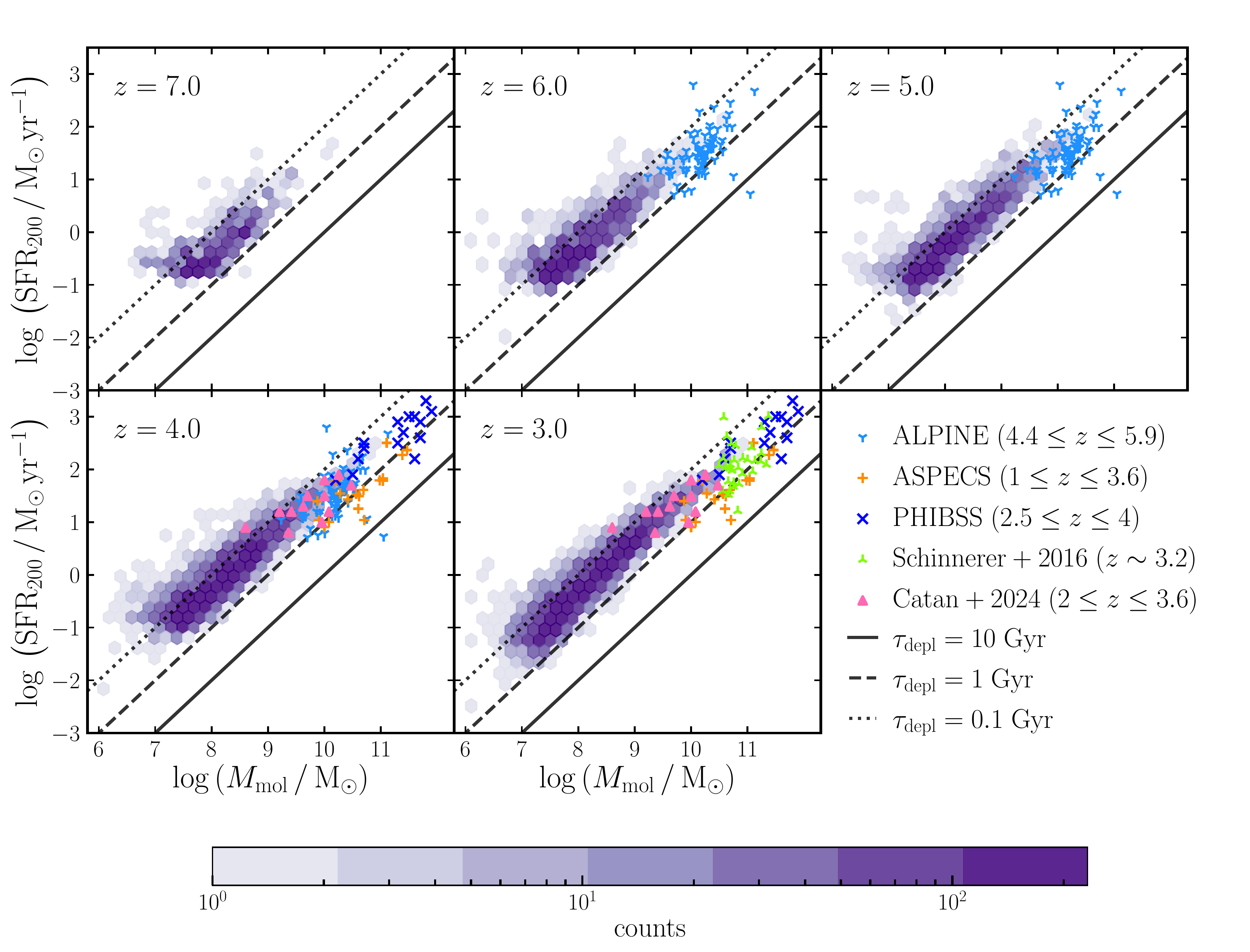}
    \caption{\edit{Distribution of the \textsc{Marigold} galaxies in the SFR-$M_{\mathrm{mol}}$ plane at different redshifts. These are shown as purple hexbins, where the counts are sampled logarithmically.}}
    \label{fig:ks}
\end{figure*}
\begin{figure*}
    \centering
    \includegraphics[width=0.78\linewidth]{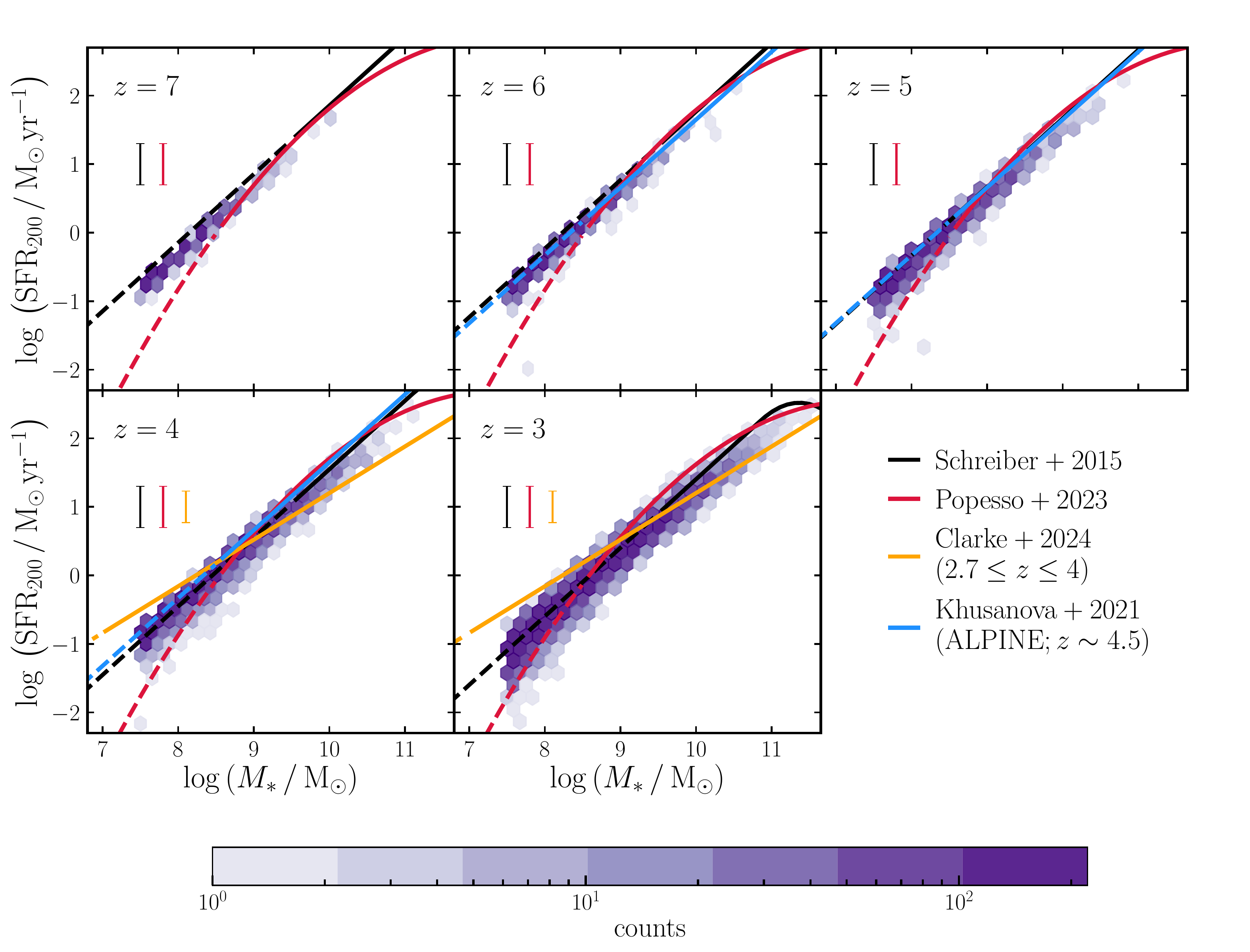}
    \caption{\edit{SFR versus $M_*$ in the \textsc{Marigold} galaxies at different redshifts compared with observational estimates of the main-sequence (MS) of star-forming galaxies. These include the MS relations from \citet[][black]{sch15} and \citet[][red]{popesso23} at the respective redshifts, and the MS relations from \citet[][orange]{clarke24} based on a sample of galaxies at $2.7 \leq z \leq 4$ and \citet[][blue]{khusanova21} based on ALPINE galaxies. The coloured errors bars on the left-hand side denote the scatter of the respective MS relations. All observed relations are extrapolated beyond the range constrained by the respective studies using a dashed line of the same colour.}}
    \label{fig:ms}
\end{figure*}
\section{Correlations between intrinsic galaxy properties}
\edit{Before we examine the relationship between the [\ion{C}{II}] luminosity and the other properties of our simulated galaxies such as the SFR and the molecular gas mass ($M_{\mathrm{mol}}$), it is instructive to inspect how these properties correlate with each other and compare them with the observed galaxy population.}

\edit{In Fig.~\ref{fig:ks}, we show the distribution of our simulated galaxies in the SFR-$M_{\mathrm{mol}}$ plane. The ratio of $M_{\mathrm{mol}}$ to SFR is used to define the depletion timescale $\tau_{\mathrm{depl}}$, that represents the time required to exhaust the molecular gas reservoir of the galaxy if star formation continues at the measured rate. 
We show lines of constant $\tau_{\mathrm{depl}}$ in the figure. The observed galaxies are shown using different coloured symbols. In particular, we include the galaxies from the ALPINE survey (in light blue), the ALMA Spectroscopic Survey in the Hubble Ultra Deep Field \citep[ASPECS,][in orange]{decarli19, aspecs}, from the  Plateau de Bure High-$z$ Blue Sequence Survey \citep[PHIBSS][in deep blue]{lenkic20}, as well as galaxies from studies by \citet[][in green]{schinnerer16} and \citet[][pink]{catan24}. All observations except ALPINE obtain the $M_{\mathrm{mol}}$ based on CO rotational lines. For ALPINE, the $M_{\mathrm{mol}}$ is obtained from the observed [\ion{C}{II}] luminosity by adopting a conversion factor $\alpha_{[\ion{C}{II}]}$ of 31 $\mathrm{M_{\odot} \, L_{\odot}^{-1}}$ from \cite{zanella18}. We see that a majority of the observed galaxies have $\tau_{\mathrm{depl}} \sim 0.1-1$ Gyr.}

Fig.~\ref{fig:ms} shows SFR versus $M_*$ for the simulated galaxies at different redshifts compared with various observational estimates of the SFR-$M_*$ relation, commonly known as the main-sequence (MS) of star-forming galaxies. Given the scatter in these relations, the distribution of our simulated galaxies in the SFR-$M_*$ plane is consistent with the observed galaxy population at the respective redshifts. 

\section{The $L_{[\ion{C}{II}]}-\mathrm{SFR}$ relation}
\label{sec:cii_sfr_all}
\begin{figure*}
    \centering
    \centering
     \includegraphics[width=0.81 \textwidth, trim={0 0 0 0},clip]{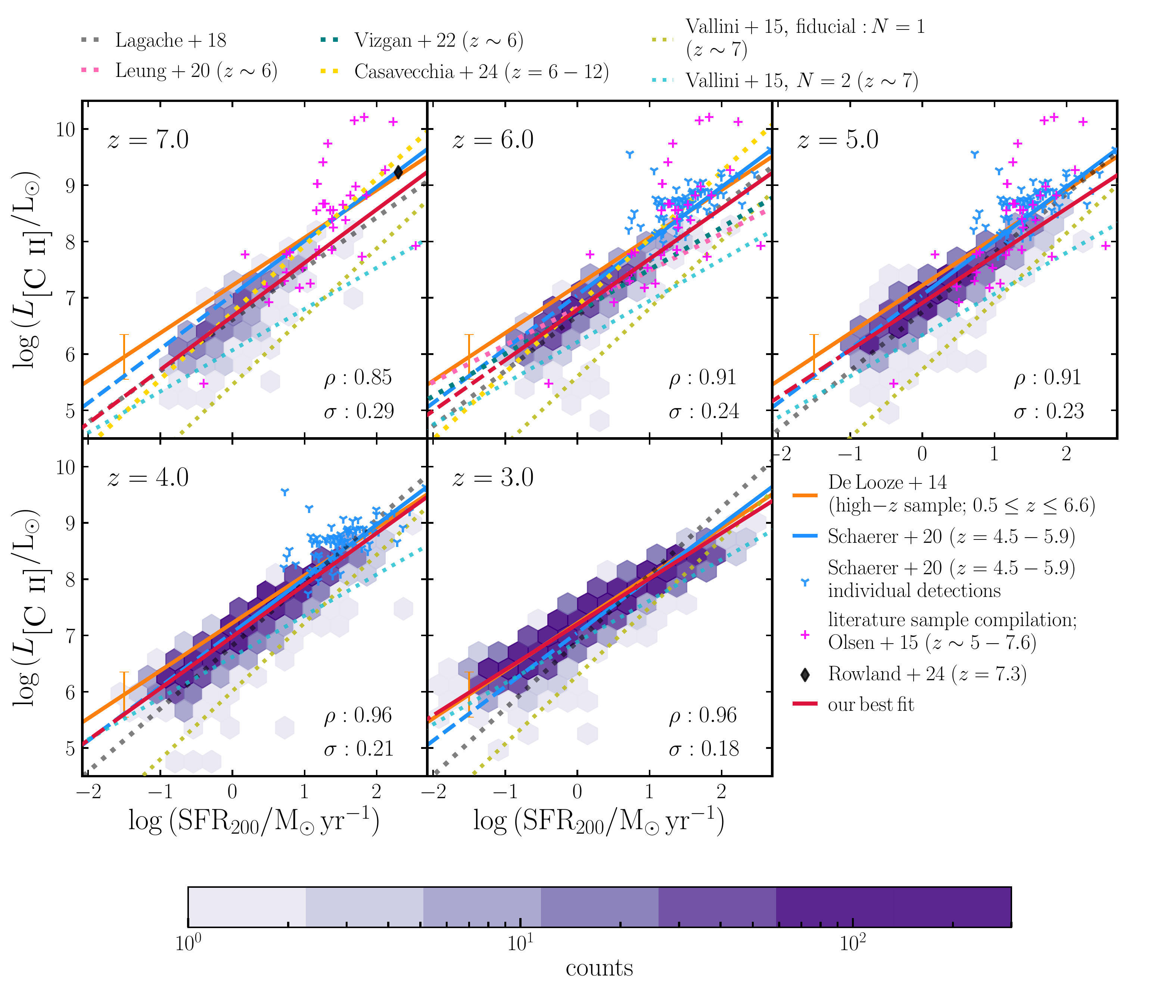}
     \caption{$[\ion{C}{II}]-\mathrm{SFR}$ relation from the \textsc{Marigold} simulations at $3 \le z \leq 7$ compared with observations. The simulated galaxy population is represented as purple hexbins, with the colour indicating the galaxy counts per bin. The red line showing the best-fit to these galaxies (see Table~\ref{tab:best-fits1} for the fit parameters). In each panel, we report the Spearman's rank correlation coefficient ($\rho$) and the $1 \sigma$ scatter around the best-fit relation. The best-fit relation from \cite{delooze14} for their high$-z$  ($0.5<z<6.6$) sample is shown in orange, with the shaded area representing the $1 \sigma$ scatter. The blue line indicates the best-fit relation for $4.5 \lesssim z \lesssim 5.9$ galaxies from the ALPINE survey \citep{schaerer20}. The best-fits mentioned above are extrapolated beyond the  range constrained by the respective studies using a dashed line of the same colour. The individual ALPINE galaxies (at $4.5 \lesssim z \lesssim 5.9$), the literature sample (at $5 \lesssim z \lesssim 7.6$) taken from \cite{olsen15}, and REBELS-25 from the REBELS survey \cite{rowland24} are shown with blue, pink, and black symbols, respectively. \edit{The relations from other simulations are shown as dotted lines. The relations from \cite{lagache18} at the respective redshift are shown in grey. The olive and cyan dotted lines along with the scatter represent the \cite{vallini15} relations with $N=1$ and $N=3$, respectively for the Kennicutt-Schmidt relation. In both cases we show the relation assuming the median $Z_{\mathrm{gas}}$ of our simulated galaxies at a given redshift. The pink and teal lines represent the relations from \cite{leung20} and \cite{vizgan22}, respectively, both obtained by post-processing the $z \sim 6$ snapshot of the \textsc{Simba} simulations \citep{dave20} with different versions of \textsc{S\'igame} \citep{olsen15, olsen17}. The grey line represents the relation from \cite{casavecchia24a}.}  
     }
    \label{fig:cii_sfr}
\end{figure*}
\begin{figure*}
    \centering    
    \includegraphics[width=0.79\textwidth, trim={0 0 0 0},clip]{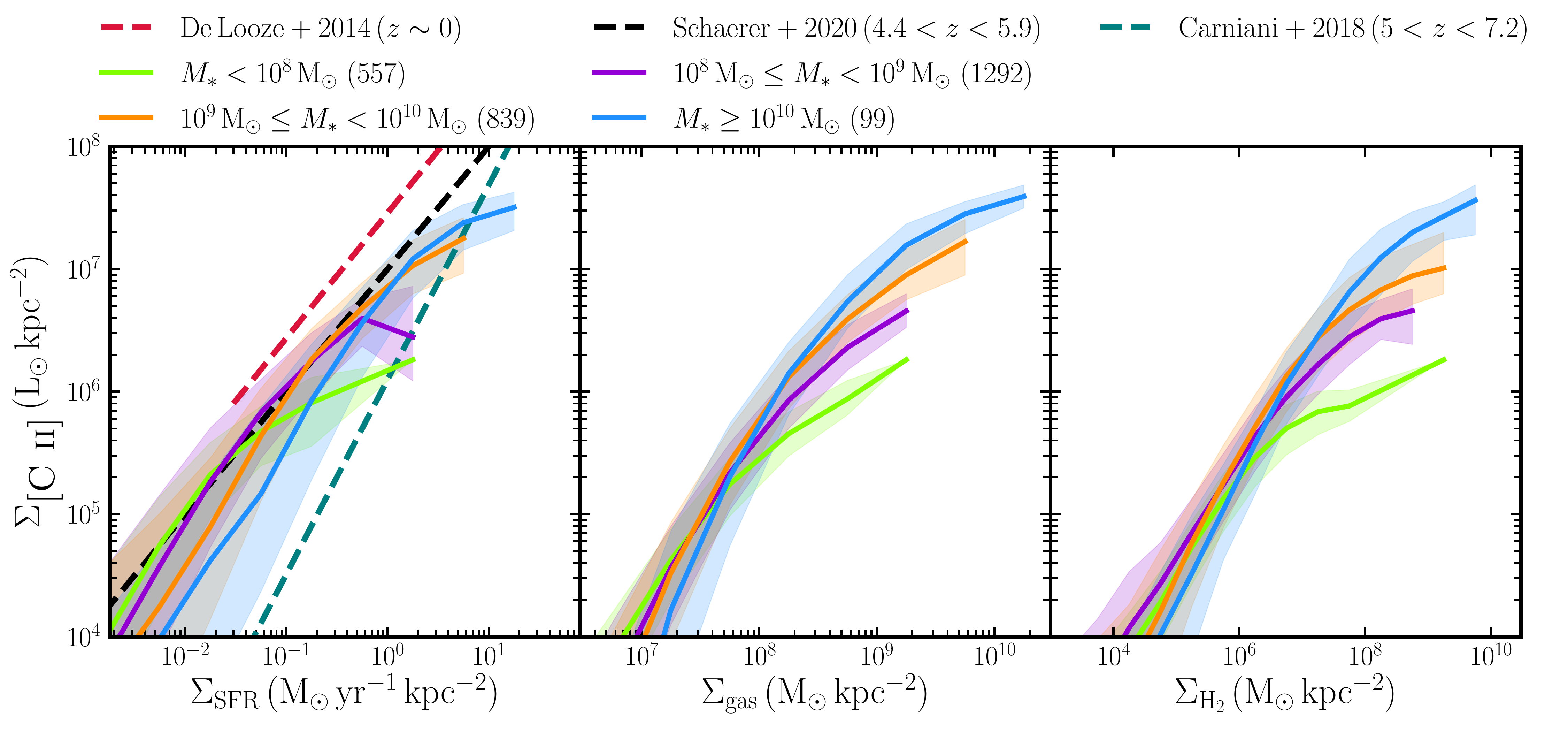}
    \caption{Spatially resolved $[\ion{C}{II}]-\mathrm{SFR} $ relation for the \textsc{Marigold} galaxies at $z=4$. The galaxies are divided into different stellar-mass bins (the number of galaxies in each bin is indicated in the legend). The solid lines show the median $\Sigma_{[\ion{C}{II}]}$ as a function of the SFR surface density (\textit{left}), of the gas surface density (\textit{middle}), and of the $\mathrm{H_2}$ surface density (\textit{right}). The shaded areas represent the 16-84 percentile range. In the left panel, dashed lines indicate the empirical relations from \citet[][red]{delooze14} based on local dwarf galaxies from the \textit{Herschel} Dwarf Galaxy Survey, for ALPINE galaxies in black \citep[based on global values only][black]{schaerer20}, and for a sample of galaxies at $5<z<7.2$ from \citet[][teal]{carniani18}. For the simulated galaxies, the [\ion{C}{II}] surface brightness, and the SFR, gas, and $\mathrm{H_2}$ surface densities were obtained from a face-on projection of a cube centred on the galaxy and of side length equal to twice the radius of the galaxy. 
    }
    \label{fig:sigma_cii_sfr}
\end{figure*}
We now turn our attention to investigating how the [\ion{C}{II}] luminosity correlates with the SFR (this section) and the molecular gas mass (Sect.~\ref{sec:cii_mmol}) using a statistical sample of simulated galaxies. For this, we only consider central galaxies as the [\ion{C}{II}] LFs of the centrals from the two simulations show an excellent agreement across redshift (see the left panels of  Fig.~\ref{fig:conditional_cii_lf}). 

The luminosity of the [\ion{C}{II}] line correlates strongly with the SFR in normal star-forming galaxies \citep{stacey10, delooze14}. It is one of the brightest emission lines in high-$z$ galaxies and is unaffected by dust obscuration. Thus, it can provide an estimate of the total (unobscured+obscured) SFR in distant galaxies. Moreover, if the [\ion{C}{II}]-SFR calibration does not evolve significantly with redshift, then it can be robustly employed across cosmic time. In this section, we aim to investigate the correlation between [\ion{C}{II}] luminosity ($L_{[\ion{C}{II}]}$) and the SFR in the \textsc{Marigold} galaxies on both galaxy-wide (Sects.~\ref{sec:cii_sfr}-~\ref{sec:lit_comparison}) and spatially resolved (Sect.~\ref{sec:sigma_cii_sfr}) scales. 

\subsection{Galaxy-integrated [\ion{C}{II}]-SFR relation}
\label{sec:cii_sfr}
Fig.~\ref{fig:cii_sfr} shows our simulated galaxies in the $L_{[\ion{C}{II}]}-\mathrm{SFR}$ plane for different redshifts.  We calculated the SFR averaged over the last 200 Myr to be consistent with observations that commonly use a combination of SF tracers to estimate the total obscured + unobscured SF \citep[see e.g.][]{kennicutt12, delooze14, herrera-camus15, lomaeva22}. For comparison, we show the best-fit relation from \citep[][]{delooze14} for their high-redshift sample ($0.5 \leq z \leq 6.6$, orange solid line). We further compare with $[\ion{C}{II}]$-detected galaxies at $4.4 \lesssim z \lesssim 5.9$ from the ALPINE survey \citep{bethermin20}. The best-fit $L_{[\ion{C}{II}]}-\mathrm{SFR}$ for these galaxies \citep{schaerer20} is shown using a blue solid line. 
We also include a literature sample of  [\ion{C}{II}]-detected galaxies compiled by \citep[][pink symbols]{olsen15} and the galaxy REBELS-25 \citep[black diamond][]{rowland24} observed as part of REBELS. Many of our galaxies occupy the same region in the [\ion{C}{II}]-SFR plane as the observed galaxies (shown as blue, pink, and black scatter points). 

At each redshift, we fitted a relation of the form $\mathrm{log}(L_{[\ion{C}{II}]}/{\rm L_{\odot}}) = a \, {\rm log (SFR_{200}/M_{\odot} \, yr^{-1}}) \,+\, b$ (where $\{a,b\}\in \mathbb{R}^2$) to our simulated galaxies using an ordinary least squares linear regression (shown as a solid red line in Fig.~\ref{fig:cii_sfr}). 
We report in Table~\ref{tab:best-fits1} the resulting parameters along with the $1 \sigma$ dispersion, where $\sigma$ is the standard deviation of the residuals around the best fit in each case. 

Firstly, from the distribution of simulated galaxies at different redshifts in the $[\ion{C}{II}]$-SFR plane, we immediately see an increase in $L_{[\ion{C}{II}]}$ at a given SFR from $z=7$ to $z=3$. At $z=3$, our galaxies are well-distributed around the \cite{delooze14} relation and our best fit is in excellent agreement with theirs, while at $z=4$, we obtain a similar relation as \cite{schaerer20} (considering conservative upper limits on [\ion{C}{II}] non-detections to the ALPINE galaxy sample).
The value of the Spearman's rank correlation coefficient, indicated in each panel in Fig.~\ref{fig:cii_sfr}, increases with time and remains high ($\gtrsim 0.86$) at all redshifts. We further find that the scatter $L_{[\ion{C}{II}]}-\mathrm{SFR}$ relation increases with increasing redshift, similar to previous findings by \cite{carniani18} for a sample of $\sim 50$ galaxies at $5 \lesssim z \lesssim 7$. In contrast, based on a semi-analytical galaxy formation model coupled to the spectral synthesis code \textsc{Cloudy} \citep{ferland92}, \cite{lagache18} reported a slight decrease in the scatter of the $L_{[\ion{C}{II}]}-\mathrm{SFR}$ relation with redshift. 



%
\subsection{Redshift evolution of the $[\ion{C}{II}]-\mathrm{SFR}$ relation}
\label{sec:cii_sfr_z_evol}
An important question concerning  the $[\ion{C}{II}]$-SFR relation is whether it shows any evolution with redshift. For instance, based on the ALPINE survey, \cite{schaerer20} found that star-forming galaxies at $4 \lesssim z \lesssim 6$ follow the same or a slightly steeper [\ion{C}{II}]-SFR relation compared to local galaxies \citep[][shown as an orange line in Fig.~\ref{fig:cii_sfr}]{delooze14}, depending on the treatment of non-detections in their galaxy sample. For the \textsc{Marigold} galaxies, the slope of the $L_{[\ion{C}{II}]}-\mathrm{SFR}$ relation shows little redshift evolution in the range $3 \leq z \leq 7$ ($ \lesssim 0.15 $ dex variation). As noted before, the best fit for our lowest redshift galaxy sample ($z=3$) is in excellent agreement with the \cite{delooze14}, but deviations towards lower $L_{[\ion{C}{II}]}$ are evident at $z \gtrsim 5$. Consequently, at a given SFR, $L_{[\ion{C}{II}]}$ decreases with increasing redshift. 
This is reflected in the monotonically increasing values of the intercept $b$ from $z=7$ to $z=3$ (see Table~\ref{tab:best-fits1}), which increases by a factor of $\sim 3$ (0.5 dex) in this redshift range. To conclude, we find a redshift evolution in the intercept of the [\ion{C}{II}]-SFR relation, as evident from the offset between our best-fit relation and the \cite{delooze14} relation that is the same in all panels. 


\subsection{Comparison with previous work}
\label{sec:lit_comparison}

\edit{In Fig.~\ref{fig:cii_sfr}, we also compare our best-fit [\ion{C}{II}]-SFR relation (Table~\ref{tab:best-fits1}) with previous numerical work in the literature, briefly described in the following. 
\cite{lagache18} used a semi-analytical galaxy-formation model coupled to the photoionisation code \textsc{Cloudy} \citep{ferland92} to obtain the [\ion{C}{II}] luminosity for a statistical sample of galaxies at redshifts $4 \leq z \leq 8$. From their full sample of galaxies, they obtain a [\ion{C}{II}]-SFR relation of the form: $\mathrm{log} (L_{[\ion{C}{II}]} /\mathrm{L_{\odot}}) = (1.4 - 0.07 
z) \mathrm{log (SFR / M_{\odot} \, yr^{-1})} + 7.1 - 0.07 z$, which is shown as solid lime line in the figure. \cite{vallini15} calculated the [\ion{C}{II}] emission from a single  $z \sim 7$ galaxy from a high-resolution ($\approx 60 \, \rm pc$) SPH simulation \citep{pallottini14} assuming optically thin emission. To do so, they adopt a log-normal sub-grid density distribution with a variable Mach number. In this regard, their approach is similar to ours except that we adopt a log-normal+power-law PDF in regions where self-gravity cannot be neglected \citep[see][for details]{khatri24}. }
They investigate how the total [\ion{C}{II}] luminosity changes with metallicity, assuming a uniform metal distribution. To derive the [\ion{C}{II}]-SFR relation, they scale the [\ion{C}{II}] luminosity with the molecular gas mass ($L_{[\ion{C}{II}]} \propto M_{\mathrm{H_2}}$), which is assumed to scale with the SFR in accordance with the Kennicutt-Schmidt relation: $\Sigma_{\mathrm{SFR}} \propto \Sigma_{\mathrm{H_2}}^{N}$. 
In Fig.~\ref{fig:cii_sfr}, we show their results for $N=1$ and $N=2$. For both cases, the dotted line is for the median gas-metallicity ($Z_{\mathrm{gas}}$) of our galaxies at a given redshift.

\edit{Previously, \cite{lagache18} found that their [\ion{C}{II}]-SFR agrees well with the one from \cite{vallini15} for $N=2$. 
We also include the results from \cite{leung20} and \cite{vizgan22}, both of which were obtained by post-processing $z\sim 6$ galaxies from the \textsc{Simba} simulations using different versions of the emission line tool \textsc{S\'igame} \citep{olsen15, olsen17}. \textsc{S\'igame} includes a multi-phase ISM model and accounts for the contribution of different phases (molecular gas phase, cold neutral medium, and \ion{H}{II} regions) to the [\ion{C}{II}] emission of a galaxy and employs \textsc{Cloudy} to obtain the chemical abundances in the different phases. The relation from \cite{casavecchia24a} is obtained by post-processing $z\geq6 $ galaxies from the COLDSIM simulations.}

Overall, we see that the variation among the different models increases with redshift, and can span up to an order of magnitude at $\mathrm{SFR} \sim 1-10  \, \mathrm{M_{\odot} \, yr^{-1}} $. This is enhanced towards lower and higher SFRs. We find that at all redshifts our [\ion{C}{II}]-SFR relation differs significantly from \cite{vallini15}. In this regard, while their approach for calculating the [\ion{C}{II}] emission accounting for the sub-grid densities is similar to ours (albeit with different sub-grid density PDFs), their method to derive the [\ion{C}{II}]-SFR relation relies on scaling relations between $L_{[\rm CII]}$ and $M_{\mathrm{H_2}}$ and $\Sigma_{\mathrm{SFR}}-\Sigma_{\mathrm{H_2}}$, while we follow the dynamical evolution of these quantities in our simulations. Moreover, as shown by our PCA analysis (Sect.~\ref{sec:pca}), the $L_{[\ion{C}{II}]}-M_{\mathrm{mol}}$ relation exhibits a secondary dependence on the SFR that evolves with redshift.  
This provides a natural explanation for the different results from the two studies.
At $5 \leq z \leq 7$,  our results, as well as those from \cite{lagache18}, show slightly steeper slopes compared to those from \cite{leung20} and \cite{vizgan22}. However, these differences remain within the scatter of $\sim 0.3 - 0.6$. Notably, at these redshifts, the numerical predictions fall slightly below the empirical \cite{delooze14} relation. At $z=3$, our best-fit shows an excellent agreement with \cite{delooze14} and exhibits a slightly higher offset compared to others except \cite{lagache18}.


\subsection{Spatially resolved $[\ion{C}{II}]-\mathrm{SFR}$ relation}
\label{sec:sigma_cii_sfr}
Resolved observations of $[\ion{C}{II}]$ and SF in high-redshift galaxies have found that these galaxies exhibit a ``$[\ion{C}{II}]$ deficit'' with respect to the local $\Sigma_{[\ion{C}{II}]}-\Sigma_{\mathrm{SFR}}$ relation. For example, \cite{carniani18} found that the galaxy-integrated $[\ion{C}{II}]$-SFR relation at $5 \lesssim z \lesssim 7$ is similar to the local one but in the $\Sigma_{[\ion{C}{II}]}-\Sigma_{\mathrm{SFR}}$ plane, the galaxies have a substantially lower $\Sigma_{[\ion{C}{II}]}$ compared to the local galaxies at any given $\Sigma_{\mathrm{SFR}}$. They attributed the offset mainly to the lower metallicity of these galaxies and stressed that the more extended $[\ion{C}{II}]$ emission in high-$z$ galaxies could also play a role. 

Here we examine the (spatially resolved) $\Sigma_{[\ion{C}{II}]}-\Sigma_{\mathrm{SFR}}$ relation in our simulated galaxies. For simplification, we present results based on $z=4$ galaxies only. In each galaxy, the [\ion{C}{II}] surface brightness and SFR surface density are obtained from a face-on projection of a cube centred on the galaxy and of side length equal to twice the size of the galaxy. The spatial resolution (minimum grid cell size; see Table~\ref{tab:sims}) of our simulations is better than that achieved in current high-$z$ $[\ion{C}{II}]$ observations that can resolve kpc-scale regions within $z \gtrsim 4$ galaxies \citep[e.g.,][]{posses24}.
Therefore, we applied a  2D Gaussian smoothing to our simulated surface brightness and surface density maps to mimic observations. 
For this analysis, we adopted the beam sizes (in terms of the full-width at half-maximum, FWHM) for the SFR surface density and the $[\ion{C}{II}]$ surface brightness measurements from \cite{posses24}, which are $0.2\arcsec$ and $0.17\arcsec$, respectively. At $z=4$, these correspond to $\sim 1.4$ kpc and $\sim 1.2$ kpc. 

In the left panel of Fig.~\ref{fig:sigma_cii_sfr}, we show the $\Sigma_{[\ion{C}{II}]} - \Sigma_{\mathrm{SFR}}$ relation for our simulated galaxies. We split our simulated galaxies into three groups according to their stellar mass ($M_*$), with the number of galaxies in each group indicated in the legend. For each group, solid lines show the median $\Sigma_{[\ion{C}{II}]}$ as a function of $\Sigma_{\mathrm{SFR}}$ and the shaded area represents the interquartile range. The median $\Sigma_{[\ion{C}{II}]}$ shows slight variations among the different $M_*$ bins. Most notably, towards higher $\Sigma_{\mathrm{SFR}}$, lower $M_*$ galaxies exhibit a lower $\Sigma_{[\ion{C}{II}]}$, likely because of their (relatively) low metallicity. The median $\Sigma_{[\ion{C}{II}]}$ in all $M_*$ bins is lower than the local relation and similar to the relation from \citet[][same as in Figure 12 of \citealt{posses24}]{schaerer20}. 

At $\Sigma_{\mathrm{SFR}} \lesssim 1 \, \mathrm{M_{\odot} \, yr^{-1} \, kpc^{-2}}$, our medians for all $M_*$ bins are consistent with the \cite{schaerer20} relation, but at higher $\Sigma_{\mathrm{SFR}}$, they start to deviate towards lower $\Sigma_{[\ion{C}{II}]}$ values and the shaded area overlaps with the \cite{carniani18}. This trend continues towards higher $\Sigma_{\mathrm{SFR}}$, and for $\Sigma_{\mathrm{SFR}} \gtrsim 10 \, \mathrm{M_{\odot} \, yr^{-1} \, kpc^{-2}}$, the median $\Sigma_{[\ion{C}{II}]}$ from our simulations is lower than the \cite{schaerer20} and \cite{carniani18} relations by a factor of $\approx 2.5$. 

For the same galaxies, we also show the $\Sigma_{[\ion{C}{II}]}-\Sigma_{\mathrm{gas}}$ and $\Sigma_{[\ion{C}{II}]}-\Sigma_{\mathrm{H_2}}$ in Fig.~\ref{fig:sigma_cii_sfr}. The gas surface densities are smoothed with a Gaussian beam of FWHM of \edit{$\sim 1.2 \, \rm kpc$} (same as for $\Sigma_{[\ion{C}{II}]}$).

Once again, there are no strong variations among the different $M_*$ bins except at the highest gas surface densities. In the $\Sigma_{[\ion{C}{II}]}-\Sigma_{\mathrm{H_2}}$ plane, a similar trend is observed although with larger differences, similar to those in the $\Sigma_{[\ion{C}{II}]}-\Sigma_{\mathrm{SFR}}$ plane. We also observe that in all panels, the slope decreases towards higher surface densities.

\subsection{Possible $[\ion{C}{II}]$-deficit?}
\label{sec:cii_deficit}
At $\Sigma_{\mathrm{SFR}} \gtrsim 1 \, \mathrm{M_{\odot} \, yr^{-1} \, kpc^{-2}}$, our $\Sigma_{[\ion{C}{II}]}-\Sigma_{\mathrm{SFR}}$ relation deviates from the \cite{schaerer20} relation towards lower $\Sigma_{[\ion{C}{II}]}$ and exhibit values similar to the \cite{carniani18} relation, albeit with a shallower slope. As regions of high SF are expected to be rich in dust and thereby bright in FIR emission, this trend is similar to the observed `$[\ion{C}{II}]$-deficit', that is to say, the decrease in the ratio of the $[\ion{C}{II}]$ to the FIR luminosity in luminous infrared galaxies. 

We investigated the underlying cause of this deficit by examining the surface density of $\mathrm{C^+}$, and CO as a function of $\Sigma_{\mathrm{SFR}}$, $\Sigma_{\mathrm{gas}}$, and the surface density of metals ($\Sigma_{\mathrm{metals}}$) in our 
galaxies in Appendix~\ref{sec:appB}. We see that $\Sigma_{\mathrm{C^+}}$ plateaus towards high $\Sigma_{\mathrm{SFR}}$ and $\Sigma_{\mathrm{gas}}$, while $\Sigma_{\mathrm{CO}}$ continues to grow. This happens for two reasons: firstly, a high $\Sigma_{\mathrm{gas}}$ leads to a higher $\mathrm{H_2}$ abundance. Second, a higher $\Sigma_{\mathrm{SFR}}$ is associated with a higher dust abundance; together these provide better shielding of CO against the dissociating UV radiation in dense environments. Therefore, in regions with a high SFR surface density, the bulk of the carbon is locked up in CO, leading to a flattening of $\Sigma_{\mathrm{C^+}}$ towards high $\Sigma_{\mathrm{SFR}}$, and consequently a slower increase of the $[\ion{C}{II}]$ surface brightness towards high SFR and gas surface densities. Further note that our simulations do not account for the UV radiation from active galactic nuclei (AGN), that can contribute to the first ionisation of carbon (and subsequently to [\ion{C}{II}] emission). While some studies suggest that AGN are not a significant source of [\ion{C}{II}] emission \citep[see for example,][]{deBreuck22}, their contribution has not yet been estimated using a large sample of galaxies at high redshifts. 
However, the harder ionizing radiation from AGN can also ionize $\mathrm{C^+}$, resulting in a decrease in the [\ion{C}{II}] emission \citep{langer15}. In this case, the [\ion{C}{II}] luminosity could be even lower than predicted by our model.

A similar decline in $[\ion{C}{II}]$ was also reported by \cite{delooze14} (see their Figure 2) who found that the slope of the $\Sigma_{\mathrm{SFR}}-\Sigma_{[\ion{C}{II}]}$ relation steepens for $\Sigma_{[\ion{C}{II}]} \gtrsim $ few times $10^6 \, \rm L_{\odot} \, kpc^{-2}$ indicating that the $[\ion{C}{II}]$ line is not the dominant coolant in intensely star-forming regions. Such a deficit would manifest as a shallower slope of the $\Sigma_{[\ion{C}{II}]}-\Sigma_{\mathrm{SFR}}$ relation towards higher $\Sigma_{[\ion{C}{II}]}$ (and higher $\Sigma_{\mathrm{SFR}}$). 

Overall we find that for our simulated galaxies at $z=4$, the median $\Sigma_{[\ion{C}{II}]}$ in a given $\Sigma_{\mathrm{SFR}}$ bin shows an excellent agreement with the best-fit to ALPINE galaxies at $4.4 \leq z \leq 5.9$ for $\Sigma_{\mathrm{SFR}} \lesssim 1 \, \mathrm{M_{\odot} \, yr^{-1} \, kpc^{-2}}$. Towards higher $\Sigma_{\mathrm{SFR}} $, however, the median shows a flattening. This is driven by the increased abundance of CO at the expense of $\mathrm{C^+}$ along high surface density lines of sight. 
\section{$[\ion{C}{II}]$ emission as a molecular gas tracer}
\label{sec:cii_mmol}

\begin{figure*}
    \centering
     \includegraphics[width=0.78 \textwidth, trim={1cm 0 1cm 1cm},clip]{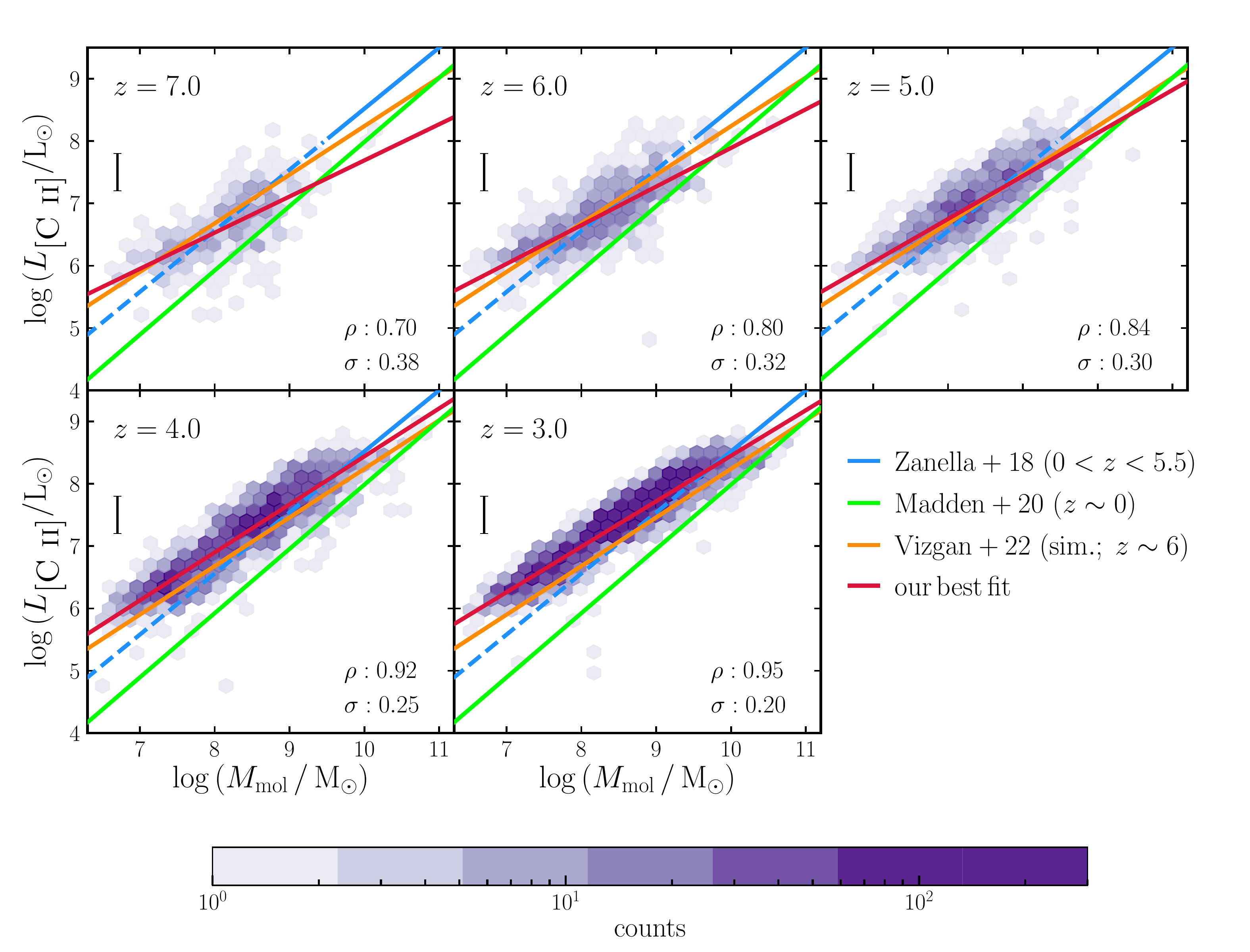}
    \caption{$[\ion{C}{II}]- M_{\mathrm{mol}}$ relation from our simulations compared with observations. The simulated galaxies are represented by purple hexbins, where the colour indicates the number of galaxies in each bin. The solid red line gives the ordinary least squares linear fit to these galaxies, with the fit parameters listed in Table~\ref{tab:best-fits1}. In each panel, we report the Spearman's rank correlation coefficient ($\rho$) and the $1 \sigma$ scatter around the best-fit relation. The best fit to the observed galaxy sample at $z=0-5.5$ by \cite{zanella18} is shown in blue and the fit to the $z \sim 0$ dwarf galaxies \citep{madden20} is shown in lime. The relation from \textsc{Simba} simulations at $z=6$ \citep{vizgan22} is shown in orange. As in Fig.~\ref{fig:cii_sfr}, the extrapolated \cite{zanella18} relation is shown as a dashed line of the same colour. }
    \label{fig:cii_mmol}
\end{figure*}

\cite{zanella18} identified a strong correlation between the luminosity of the $[\ion{C}{II}]$ line and the molecular gas mass based on a compilation of galaxies in the redshift range $0 \leq z \leq 5.5$. Their sample includes local dwarf galaxies, main-sequence galaxies at $0 \leq z \leq 5.5$, starburst galaxies at $0 \lesssim z \lesssim 2$ and local luminous/ultra-luminous infrared galaxies. The best fit to their galaxy sample is given by:
\begin{equation}\label{eq:zanella_cii_mmol}
    \mathrm{log} \, \left ( \frac{L_{[\ion{C}{II}]}}{\rm L_{\odot}} \right ) \, = \, -1.28(\pm 0.21) \, + 0.98(\pm0.02) \, \mathrm{log} \, \left ( \frac{M_{\mathrm{mol}}}{\mathrm{M_{\odot}}} \right ) \, ,
\end{equation}
with a scatter of $\approx 0.3$ dex around the best fit. Consequently, $[\ion{C}{II}]$ emission is now routinely used as a molecular gas tracer in high-redshift galaxies and it seems to be as good as conventional tracers like CO rotational lines and dust continuum emission. For instance, \cite{dessauges-zavadsky20} found that the [\ion{C}{II}]-based molecular gas mass estimates for ALPINE galaxies were consistent with those derived from dynamical mass estimates and (rest-frame) infrared luminosities. \cite{aravena24} found similar results for REBELS galaxies at $6.5 \lesssim z \lesssim 7.5$. 


However, this relation and its redshift evolution have not been extensively explored in numerical simulations. While \cite{vizgan22} investigated the [\ion{C}{II}]-$M_{\mathrm{mol}}$ relation in \textsc{Simba} simulations at a specific redshift of $z \sim 6$, a detailed study at different epochs is still lacking. To address this, here we examine the $[\ion{C}{II}]$-molecular gas connection in the \textsc{Marigold} galaxies at different redshifts and develop a prescription for inferring the molecular gas mass of a galaxy from its [\ion{C}{II}] emission. In Fig.~\ref{fig:cii_mmol}, we compare the $L_{[\ion{C}{II}]}-M_{\mathrm{mol}}$ relation for our simulated galaxies at different redshifts against the calibrations from \cite{zanella18}, \cite{madden20}, and \cite{vizgan22}.

The molecular gas mass, $M_{\mathrm{mol}}$, was obtained by scaling up the $\mathrm{H_2}$ mass directly obtained from the simulations by 1.36 to account for the contribution of helium and heavier elements confined with $\mathrm{H_2}$. At each redshift, we performed an ordinary least squares regression to fit a linear relation of the form: $\mathrm{log}(L_{[\ion{C}{II}]}) = a \, \mathrm{log}(M_{\mathrm{mol}}) + b$ to our galaxies. The corresponding best-fit parameters and the $1 \sigma$ dispersion around the best fit are listed in Table~\ref{tab:best-fits1}. We observe that the slope increases with time from $z=7$ to $z=4$ and thereafter decreases slightly (by $\approx 0.04$ dex). The overall change of $\approx 0.2$ dex in the slope between $z=7$ and $z=3$ is similar to the change of $\approx 0.15$ dex in the slope of the $[\ion{C}{II}]-\mathrm{SFR}$ relation. At all redshifts, our best-fit slope is shallower than that of the \cite{zanella18} relation. As a result only our high-mass ($M_{\mathrm{mol}} \gtrsim 10^{9} \, \mathrm{M_{\odot}}$) galaxies follow their relation, while our low-mass galaxies exhibit higher $L_{[\ion{C}{II}]}$ than expected from extrapolation of their relation. 


We also report in In Fig.~\ref{fig:cii_mmol}, the Spearman's rank correlation coefficient ($\rho$) between $L_{[\ion{C}{II}]}$ and $M_{\mathrm{mol}}$ at each redshift. From the monotonically increasing values of $\rho$ across redshift, we see that the [\ion{C}{II}]-$M_{\mathrm{mol}}$ correlation is relatively weak at $z\gtrsim 5$ and becomes progressively stronger over time. This is in contrast with the [\ion{C}{II}]-SFR correlation that exhibits a high value ($\rho \gtrsim 0.86$) out to $z =7$. This trend is also evident from the decreasing values of the scatter ($\sigma$) around the best-fit linear relation between $L_{[\ion{C}{II}]}$ and $M_{\mathrm{mol}}$ with decreasing redshift.


\renewcommand{\arraystretch}{1.3} 
\begin{table*}
    \centering
    \caption{Best-fit scaling relations between the $[\ion{C}{II}]$ luminosity and the SFR (averaged over the last 200 Myr, $\mathrm{SFR}_{200}$) and the molecular gas mass ($M_{\mathrm{mol}}$) in our simulated galaxies at different redshifts. }
    \begin{tabular}{cc|ccc|ccc}
        \hline
        \hline
        \textit{z} & No. of galaxies
        & \multicolumn{3}{c|}{$\mathrm{log}(L_{[\ion{C}{II}]}/{\rm L_{\odot}}) = a \, \mathrm{log} ({\mathrm{SFR}_{200}/  \mathrm{M_{\odot} \, yr^{-1}}}) \,+\, b$}
        & \multicolumn{3}{c}{$\mathrm{log}(L_{[\ion{C}{II}]}/{\rm L_{\odot}}) = a \, \mathrm{log} (M_{\mathrm{mol}}/ {\mathrm{M_{\odot}}}) \,+\, b$}
        \\
        \hline
        & 
        & $a$ & $b$ & 1$\sigma$ 
        & $a$ & $b$ & 1$\sigma$  
        \\
        \hline
        7 & 330 
        & 0.947 $\pm$ 0.034 & 6.670 $\pm$ 0.016 & 0.29
        & 0.580 $\pm$ 0.034 & 1.893 $\pm$ 0.277 & 0.38
        \\
        6 & 692 
        & 0.907 $\pm$ 0.018 & 6.786 $\pm$ 0.009 & 0.24
        & 0.624 $\pm$ 0.018 & 1.690 $\pm$ 0.148 & 0.32 
        \\
        5 & 1210 
        & 0.852 $\pm$ 0.011 & 6.905 $\pm$ 0.007 & 0.22 
        & 0.695 $\pm$ 0.013 & 1.223 $\pm$ 0.108 & 0.30 
        \\
        4 & 2787 
        & 0.921 $\pm$ 0.006 & 6.985 $\pm$ 0.004 & 0.21
        & 0.772 $\pm$ 0.006 & 0.742 $\pm$ 0.052 & 0.25
        \\
        3 & 4458 
        & 0.805 $\pm$ 0.004 & 7.205 $\pm$ 0.003 & 0.18
        & 0.733 $\pm$ 0.004 & 1.160 $\pm$ 0.032 & 0.20
        \\
        \hline
    \end{tabular}
    \label{tab:best-fits1}
\end{table*}

\subsection{The conversion factor $\alpha_{[\ion{C}{II}]}$}
\label{sec:alpha}

\begin{figure}
    \centering
    \includegraphics[width=0.45\textwidth]{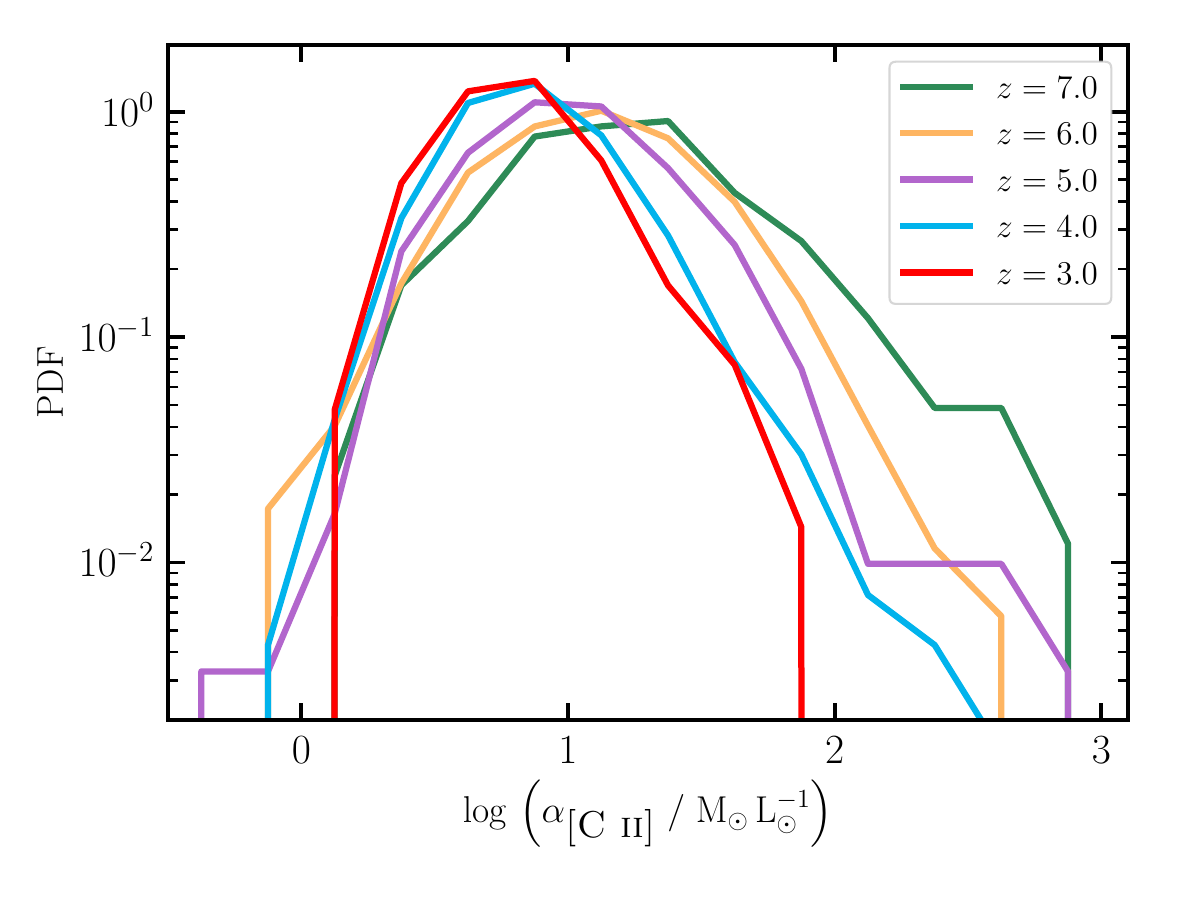}
    \caption{\edit{Probability distribution function of the conversion factor $\alpha_{[\ion{C}{II}]}$ exhibited by our simulated galaxies at different redshifts.}}
    \label{fig:alpha_hist}
\end{figure}
\begin{figure*}
    \centering    
    \includegraphics[width=0.98\textwidth, trim={0 0 0 0},clip]{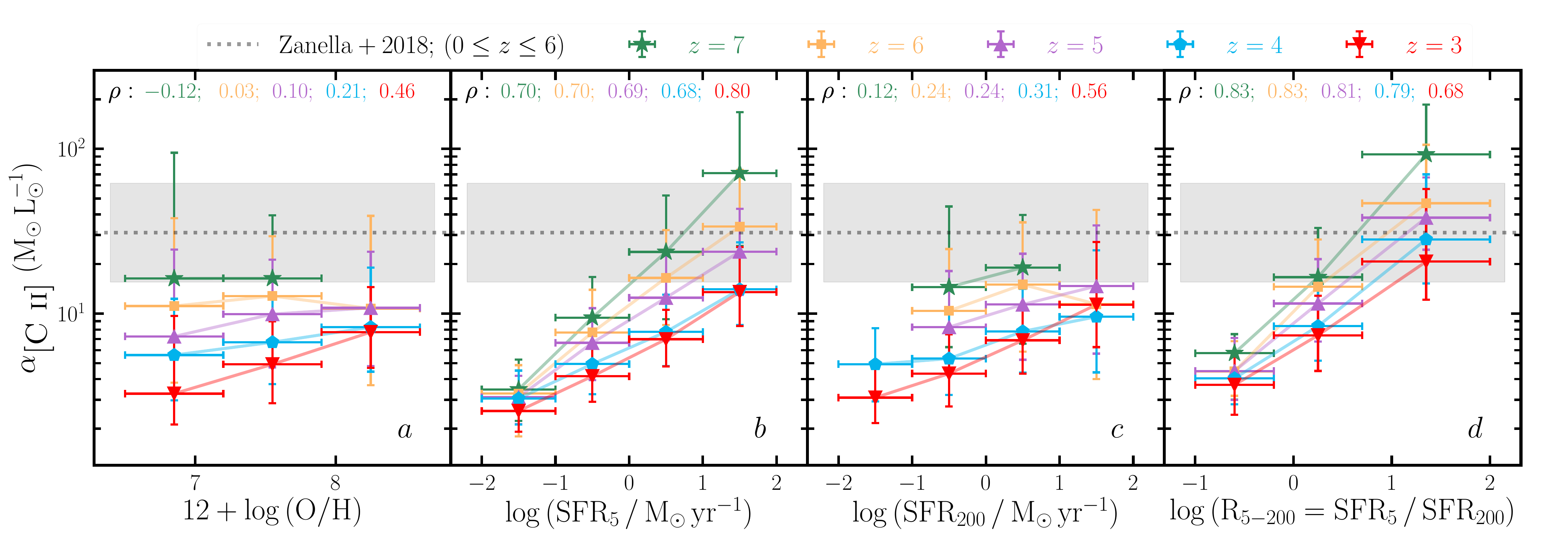}
    \caption{Conversion factor $\alpha_{\rm [\ion{C}{II}]}$ in our simulated galaxies as a function of gas metallicity (panel $a$), the SFR averaged over the last 5 Myr (panel $b$), the SFR averaged over the last 200 Myr (panel $c$), and the SFR change diagnostic $R_{5-200}$ (panel $d$, see text) at different redshifts. The coloured symbols represent the median $\alpha_{[\ion{C}{II}]}$ in each bin, while the error bars represent the 16-84 percentiles. The Spearman's rank correlation coefficient between the variables on the $y$ and $x$ axes are shown in each panel as well. The grey dotted line denotes the mean $\alpha_{[\ion{C}{II}]}$ from \cite{zanella18} and the shaded band represents the corresponding scatter. 
    }
    \label{fig:alpha_cii}
\end{figure*}

Next we computed the conversion factor between the $[\ion{C}{II}]$ luminosity and the molecular gas mass: $\alpha_{[\ion{C}{II}]} \equiv M_{\mathrm{mol}} \, / \, L_{[\ion{C}{II}]}$.
\edit{Fig.~\ref{fig:alpha_hist} shows the probability distribution function of the $\alpha_{[\ion{C}{II}]}$ exhibited by the \textsc{Marigold} galaxies. We see that at all redshifts, $\mathrm{log} \alpha_{[\ion{C}{II}]}$ exhibits a skewed distribution with an extended tail towards higher values. The tail is particularly prominent at high redshifts. However, the scatter reduces towards lower redshifts. The peak $\alpha_{[\ion{C}{II}]}$ at these redshifts lies in the range 7-24 $\mathrm{M_{\odot} \, L_{\odot}^{-1}}$ and increases with redshift.}

Fig.~\ref{fig:alpha_cii} shows the correlation between $\alpha_{[\ion{C}{II}]}$ and various galaxy properties for our simulated galaxies, namely the gas-phase metallicity, $\rm 12 + log (O/H)$, the SFR averaged over the last 5 Myr ($\mathrm{SFR}_5$), the SFR averaged over the last 200 Myr ($\mathrm{SFR}_{200}$), and the ratio of the SFRs averaged over the last 5 and 200 Myr (hereafter $R_{5-200}$). The use of the latter is motivated by \cite{lomaeva22}, who proposed an SFR change diagnostic derived from the ratio of $\mathrm{SFR}_{5}$ and $\mathrm{SFR}_{200}$ to quantify the current rate of change of the SFR. As different SFR indicators are sensitive to the SF happening on different timescales, their ratio can be used to quantify the SFR change over time. \cite{lomaeva22} proposed the ratio of $\rm H\alpha$ to FUV emission as a proxy for $\mathrm{SFR}_5/\mathrm{SFR}_{200}$. This quantity has also been used recently to quantify the bursty SF in galaxies observed with the James Webb Space Telescope \citep[see for example,][]{atek24, clarke24}. 

In each panel of Fig.~\ref{fig:alpha_cii}, we show the median $\alpha_{[\ion{C}{II}]}$ in different bins of the quantity on the $x$-axis along with the 16-84 percentile range (denoted by error bars).  Firstly, we observe that the $\alpha_{[\ion{C}{II}]}$ values span about two orders of magnitude at all redshifts, particularly at $z \geq 6$. The Spearman's rank correlation coefficient (denoted by $\rho$) for the galaxy sample at different redshifts is reported in each panel. We find a weak correlation between $\alpha_{[\ion{C}{II}]}$ and  gas metallicity at all redshifts and observe a large scatter at all values of $12+\rm log(O/H)$, in agreement with \cite{zanella18} who found little systematic variation of $\alpha_{[\ion{C}{II}]}$ with metallicity.
 
In panel $b$ of Fig.~\ref{fig:alpha_cii}, we observe that $\alpha_{[\ion{C}{II}]}$ increases with $\mathrm{SFR_5}$ at all redshifts, with the highest variation occurring at $z=7$. At SFR $\gtrsim 10 \, \mathrm{M_{\odot} \, yr^{-1}}$, our $\alpha_{[\ion{C}{II}]}$ values at all redshifts are in good agreement with the range from \cite{zanella18}, but deviate towards lower values at lower SFRs. Therefore, using a constant $\alpha_{[\ion{C}{II}]}$ (calibrated on the high-SFR galaxies alone) would lead to an overestimate of the molecular gas mass in low-SFR galaxies. 
\edit{A similar trend is observed between $\alpha_{[\ion{C}{II}]}$ and SFR$_{200}$ in panel $c$. However, the strength of the correlation between this pair of variables is much lower than the previous, as indicated by the lower values of $\rho$. This shows that the conversion factor depends strongly on the SFR averaged over a small timescale and less so on the SFR averaged over a longer timescale.}

Finally, in panel $d$, we see that $\alpha_{[\ion{C}{II}]}$ increases with $R_{5-200}$, with $\rho \in [0.68, 0.83]$. Interestingly, in this case the correlation is the strongest at high redshifts and slowly decays at lower redshifts, highlighting the increased sensitivity of $\alpha_{[\ion{C}{II}]}$ on the star-formation history of galaxies at $z \gtrsim 6$, as quantified by SFR change parameter $R_{5-200}$. 

\edit{We note however that observations of starburst galaxies at $z \sim 4$ from \cite{rizzo21} exhibit lower $\alpha_{[\ion{C}{II}]}$ than the \cite{zanella18} calibration. These galaxies have $\alpha_{[\ion{C}{II}]} \sim 5-19 \, \mathrm{M_{\odot} \, L_{\odot}^{-1}}$, despite having SFRs in excess of 100 $\mathrm{M_{\odot} \, yr^{-1}}$. This is in contrast to our finding of  $\alpha_{[\ion{C}{II}]}$ increasing with the SFR. A comparison of the molecular gas mass and depletion timescale of \cite{rizzo21} and our galaxies shows that four out of five of the \cite{rizzo21} galaxies have an order of magnitude lower $\tau_{\mathrm{depl}}$ ($\lesssim 0.02$ Gyr) compared to our galaxies with similar masses ($M_{\mathrm{mol}} \geq 10^{10} \, \mathrm{M_{\odot}}$, see Fig.~\ref{fig:ks}), indicating a faster consumption of molecular gas in these galaxies compared to the \textsc{Marigold} galaxies.}


\begin{table}
    \centering
    \caption{Results of PCA at $z=4$.}
    \begin{tabular}{ccccccc}
         \hline
         \hline
         &&&&&& \% of\\
         & $\widetilde{M_{\mathrm{mol}}}$ & $\widetilde{L_{[\ion{C}{II}]}}$ & $\widetilde{\mathrm{SFR}_5}$ & $\widetilde{\mathrm{SFR}_{200}}$ &  $\widetilde{Z}$ & variance \\
         \hline
         PC1 & 0.46 & 0.46 & 0.43 & 0.46 & 0.42 & 88.10 \\
         PC2 & -0.30 &  0.12 & -0.60 &  0.10 & 0.72 & 7.33 \\
         PC3 &  0.05 & -0.60 &  0.45 & -0.38 & 0.55 & 3.15 \\
         PC4 & -0.20 & -0.57 &  0.03 &  0.80 &-0.07 & 1.06 \\
         PC5 & -0.81 &  0.32 &  0.49 &  0.01  & 0.03 & 0.35 \\
        \hline
    \end{tabular}
    \tablefoot{The PCA is performed in the 5D parameter space of scaled variables as expressed in Eq.(~\ref{eq:pca_scaled}), where $Z$ stands for 12 + log (O/H).}
    \label{tab:pca}
\end{table}


\begin{table*}
    \centering
    \caption{Coefficients for the PCA-based prescriptions for estimating the molecular gas mass at different redshifts.} \begin{tabular}{c|ccccc|cc|cc}
        \hline
        \hline
        $z$ & $a$ & 
        $b$ & 
        $c$ & 
        $d$ & 
        $e$ & 
        \multicolumn{4}{c}{1 $\sigma$ dispersion (dex)}
        \\
        & & & & & & PCA  & $M_{\mathrm{mol}}$-$L_{[\ion{C}{II}]}$  & PCA*  & $M_{\mathrm{mol}}$-$L_{[\ion{C}{II}]}$*  \\
        \hline
        \hline
        \multicolumn{10}{c}{PCA$: \; M_{\rm mol}-L_{\textsc{[C ii]}}-{\rm SFR_{5}-SFR_{200}}-Z$}     \\    
        \hline
        %
        7 & $0.50 \pm 0.15$ & $0.79 \pm 0.08$ & $-0.16 \pm 0.31$ & $+0.04 \pm 0.26$ & $4.39 \pm 2.80$ & 0.20 & 0.45 & 0.28 & 0.47\\
        6 & $0.42 \pm 0.08$ & $0.73 \pm 0.02$ & $-0.02 \pm 0.10$ & $-0.03 \pm 0.07$ & $5.46 \pm 0.87$ & 0.17 & 0.41 & 0.24 & 0.42 \\
        5 & $0.45 \pm 0.04$ & $0.65 \pm 0.01$ & $-0.00 \pm 0.04$ & $-0.03 \pm 0.03$ & $5.27 \pm 0.35$ & 0.13 & 0.36 & 0.20 & 0.37 \\
        4 & $0.47 \pm 0.03$ & $0.59 \pm 0.01$ & $+0.01 \pm 0.03$ & $+0.09 \pm 0.02$ & $4.11 \pm 0.26$ & 0.13 & 0.30 & 0.20 & 0.32\\
        3 & $0.57 \pm 0.02$ & $0.59 \pm 0.01$ & $-0.06 \pm 0.02$ & $+0.13 \pm 0.02$ & $3.12 \pm 0.17$ & 0.11 & 0.25 & 0.19 & 0.28 \\
        \hline
        \hline
        \multicolumn{10}{c}{PCA$: \; M_{\rm mol}-L_{\textsc{[C ii]}}-{\rm SFR_{5}}$}     \\    
        \hline
        7 & $0.41 \pm 0.03$ & $0.76 \pm 0.02$ & - & - &  $5.33 \pm 0.17$ & 0.20 & 0.45 & 0.28 & 0.47 \\
        6 & $0.40 \pm 0.02$ & $0.73 \pm 0.02$ & - & - &  $5.41 \pm 0.14$ & 0.17 & 0.41 & 0.25 & 0.42 \\  
        5 & $0.44 \pm 0.01$ & $0.65 \pm 0.01$ & - & - &  $5.13 \pm 0.08$ & 0.13 & 0.36 & 0.21 & 0.37 \\  
        4 & $0.52 \pm 0.01$ & $0.59 \pm 0.01$ & - & - &  $4.49 \pm 0.08$ & 0.13 & 0.30 & 0.20 & 0.32 \\
        3 & $0.57 \pm 0.01$ & $0.58 \pm 0.01$ & - & - &  $4.17 \pm 0.08$ & 0.11 & 0.25 & 0.19 & 0.28 \\
        \hline
        \end{tabular}
        \tablefoot{We use the general expression: 
        $\mathrm{log} (M_{\mathrm{mol}}/{\mathrm{M_{\odot}}}) =  a \, \mathrm{log} (L_{[\ion{C}{II}]}/{\rm L_{\odot}}) 
        + b \,\mathrm{log} ({\mathrm{SFR}}_{5}/{\mathrm{M_{\odot} \, yr^{-1}}})
        + c \,\mathrm{log} ({\mathrm{SFR}}_{200}/{\mathrm{M_{\odot} \, yr^{-1}}})
        + d \, [{\rm 12+log(O/H)}] 
        + e$ .
        We also include results obtained for two simpler 3-variable PCA using $M_{\mathrm{mol}}$, $L_{[\ion{C}{II}]}$, and SFR$_5$. 
        Errors on the coefficients are obtained from bootstrapping with replacement with a 1000 iterations. The seventh and eighth columns enlist the standard deviation of the offset between the true and predicted $M_{\mathrm{mol}}$ when using the PCA-based and the best fit $M_{\mathrm{mol}}$-$L_{[\ion{C}{II}]}$ relations, respectively (see Fig.~\ref{fig:pca_z4}), while the last two (denoted by a *) enlist the same when accounting for typical observational uncertainties (see text). }
    \label{tab:pca_z}
\end{table*}

\subsection{Secondary dependence of the [\ion{C}{II}]-$M_{\mathrm{mol}}$ relation}
\label{sec:pca}

\begin{figure*}
    \centering    
    \includegraphics[width=0.78\textwidth, trim={0 0 0 0},clip]{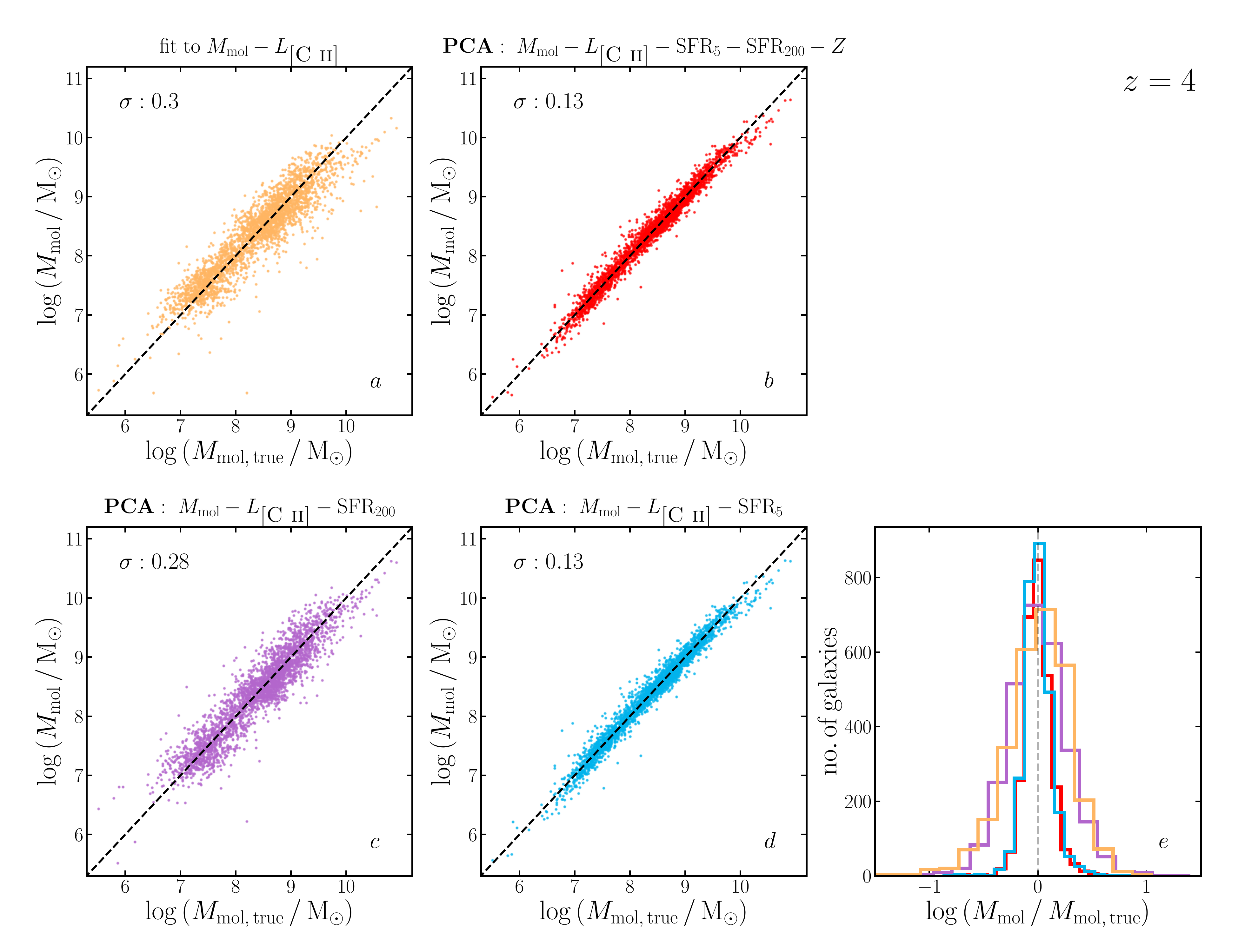}
    \caption{\edit{Comparison of the predicted molecular gas mass estimates with the true molecular gas mass using two different approaches for simulated galaxies at $z=4$. Panel $a$ shows the performance of the best-fit relation between $M_{\mathrm{mol}}$ and $L_{[\ion{C}{II}]}$, while panels $b-d$ present results from different PCA-based relations  (as indicated in the title of each panel). The corresponding relations are listed in Table~\ref{tab:pca_z}. In panels $a-d$, the top-left corner reports $\sigma$, the standard deviation of the offset between the predicted and true $M_{\mathrm{mol}}$ (both in log scale). Panel $e$ compares the distributions of these offsets from the different approaches.}
    }
    \label{fig:pca_z4}
\end{figure*}
After examining how $\alpha_{[\ion{C}{II}]}$ varies with other galaxy properties, we developed a prescription for inferring the molecular gas mass of a galaxy from its [\ion{C}{II}] emission, taking into account these secondary dependences. For this, we performed a principal component analysis (PCA) in the 5D space of parameters -- $M_{\mathrm{mol}}$, $L_{[\ion{C}{II}]}$, $\mathrm{SFR}_{5}$, $\mathrm{SFR}_{200}$ and $12+\rm log(O/H)$. PCA identifies dominant patterns and correlations between the parameters, reducing the dimensionality while approximately preserving variance. This technique has been previously used to identify secondary dependencies in the stellar-mass-metallicity relation on SFR and other parameters \citep{mannucci10, laraLopez10, hunt12, bothwell16}.


Since PCA is highly sensitive to extremes in the data, we first scaled all parameters by their respective mean and standard deviation before performing the analysis such that for a parameter $X$, 
\begin{equation}
    \label{eq:pca_scaled}
    \widetilde{X} = \frac{\mathrm{log} X - \langle \mathrm{log} X \rangle}{\sqrt{{\rm variance} (\mathrm{log} X)}}
\end{equation}
\edit{denotes the scaled variable.}

The results of this analysis at $z=4$ are shown in Table~\ref{tab:pca}. About 88\% of the variance is explained by the first principal component and $\sim$~95~\% is explained by the first two principal components. The first component contains nearly equal contributions from all variables, while PC2 is dominated by metallicity. The last principal component, PC5 is dominated by $M_{\mathrm{mol}}$ while containing only 0.35\% of the sample variance. Therefore, we set PC5 to zero \edit{and renormalize} to obtain an expression for $M_{\mathrm{mol}}$ in terms of the other quantities:

\begin{multline}\label{eq:mmol_pca}
    \mathrm{log} (M_{\mathrm{mol}}/{\mathrm{M_{\odot}}})  \,=\, 4.11 + 0.47 \, \mathrm{log} (L_{[\ion{C}{II}]}/{\rm L_{\odot}}) \\ 
    + 0.59 \,\mathrm{log} ({\mathrm{SFR}}_{5}/{\mathrm{M_{\odot} \, yr^{-1}}}) 
    + 0.01 \,\mathrm{log} ({\mathrm{SFR}}_{200}/{\mathrm{M_{\odot} \, yr^{-1}}}) \\
    + 0.09 \, [\rm 12+log(O/H)] \, .
\end{multline}

The $M_{\mathrm{mol}}$ obtained from Eq.~(\ref{eq:mmol_pca}) versus the true $M_{\mathrm{mol}}$ is shown in Fig.~\ref{fig:pca_z4}. We also contrast this with the $M_{\mathrm{mol}}$ obtained from the best-fit relation between $M_{\mathrm{mol}}$ and $L_{[\ion{C}{II}]}$\footnote{Note that this best fit is different from the one listed in Table~\ref{tab:best-fits1} as the dependent and independent variables are reversed.}. The latter shows $\approx 2.3$ times higher scatter. The $1 \sigma$ standard deviation between the true $M_{\mathrm{mol}}$ and the predicted $M_{\mathrm{mol}}$ using the PCA-based relation is $0.13$ implying that for most ($95\%$) of the galaxies, the PCA relation predicts the true molecular gas mass within a factor of $\sim 1.8$ while using the two variable linear best-fit, the molecular gas mass is predicted within a factor of 4. It is worth noting that while the linear relation systematically underpredicts $M_{\mathrm{mol}}$ for  $M_{\mathrm{mol}} \gtrsim 10^{10} \, \mathrm{M_{\odot}}$ (as the linear fit is heavily influenced by the more numerous low-mass galaxies), the PCA-based prediction does not suffer from this discrepancy.

We obtain similar results at other redshifts; 
these are listed in Table~\ref{tab:pca_z}. To estimate the uncertainties in the coefficients of the PCA-based relation, we performed a bootstrapping analysis with replacement, using 1000 iterations at each redshift. We find that the coefficient $a$, that quantifies the dependency of $M_{\mathrm{mol}}$ on $L_{[\ion{C}{II}]}$, increases from $z=6$ to $z=3$, while the coefficient $b$, that represents the dependency of $M_{\mathrm{mol}}$ on $\mathrm{SFR}_{5}$ decreases such that at $z=3$, $a \approx b$.
Interestingly, the dependence of $M_{\mathrm{mol}}$ on $\mathrm{SFR}_{200}$ and metallicity is relatively low at all times and does not evolve significantly with redshift.

Based on this finding, we inspect the efficacy of a simpler PCA-based relation namely using three variables: $M_{\mathrm{mol}}$, $L_{[\ion{C}{II}]}$, and SFR$_{200}$ as well as $M_{\mathrm{mol}}$, $L_{[\ion{C}{II}]}$, and SFR$_5$. A comparison of the $M_{\mathrm{mol}}$ obtained this way against the true molecular gas mass is presented in panels $c$ and $d$ of Fig.~\ref{fig:pca_z4}. The resulting relations at all redshifts are listed in Table~\ref{tab:pca_z}. Across redshift, we find that the $M_{\mathrm{mol}}-L_{\textsc{[C ii]}}-\mathrm{SFR}_5$ relation offers a similar level of accuracy as the 5-variable PCA-based relation, while being more convenient to use from an observational point of view. In contrast, the $M_{\mathrm{mol}}-L_{\textsc{[C ii]}}-\mathrm{SFR}_{200}$ relation does not provide a significant gain compared to the best-fit $M_{\mathrm{mol}}-L_{[\ion{C}{II}]}$ relation (panel $a$ of Fig.~\ref{fig:pca_z4}).

\subsection{Accounting for observational uncertainties}
To estimate the impact of observational uncertainties in measurements of $L_{[\ion{C}{II}]}$, $\mathrm{SFR_5}$, $\mathrm{SFR_{200}}$, and 12+log (O/H) in estimating $M_{\mathrm{mol}}$, we added a random perturbation, $\delta$, to each of the four quantities, where $\delta$ was drawn from a normal distribution with a mean of zero and a standard deviation corresponding to the typical errors reported in observations of high-$z$ galaxies. For instance, we adopted an error of 0.1 dex for $L_{[\ion{C}{II}]}$ and 0.24 dex for $\mathrm{SFR_5}$ and $\mathrm{SFR_{200}}$, both based on ALPINE galaxies \citep{bethermin20}. For $12+\mathrm{log (O/H)}$, we adopted an error of 0.05 dex \citep{sanders15}. Then we employed the PCA-based relations shown in Table~\ref{tab:pca_z} to obtain a prediction for $M_{\mathrm{mol}}$ and computed the offset between our prediction and the true $M_{\mathrm{mol}}$ (as before, the offset is computed from the log of the quantities). We then computed the standard deviation of this offset. We repeated the procedure for the best-fit $M_{\mathrm{mol}}-L_{[\ion{C}{II}]}$ relation. The resulting $\sigma$ values are reported in the last two columns of Table~\ref{tab:pca_z}. Overall, we found that even accounting for observational uncertainties, the PCA-based relation provides a significant gain in the precision/accuracy of molecular gas mass estimates and is capable of predicting the true molecular gas mass within a factor of 3 at $3\leq z \leq 6$. 



\begin{table*}
    \centering
    \caption{Best-fit scaling relations between the $[\ion{C}{II}]$ luminosity and the total gas mass ($M_{\mathrm{gas}}$) and the metal mass in the gas phase ($M_{\mathrm{metal}}$) in our simulated galaxies at different redshifts. }
    \begin{tabular}{c|ccc|ccc}
        \hline
        \hline
        \textit{z}
        & \multicolumn{3}{c|}{$\mathrm{log}(L_{[\ion{C}{II}]}/{\rm L_{\odot}}) = a \, \mathrm{log} (M_{\mathrm{gas}}/ {\mathrm{M_{\odot}}}) \,+\, b$}
        & \multicolumn{3}{c} {$\mathrm{log}(L_{[\ion{C}{II}]}/{\rm L_{\odot}}) = a \, \mathrm{log} (M_{\mathrm{metal}}/ {\mathrm{M_{\odot}}}) \,+\, b$}
        \\
        \hline
        & $a$ & $b$ & 1$\sigma$ 
        & $a$ & $b$ & 1$\sigma$  
        \\
        \hline
        7 
        & 1.080 $\pm$ 0.032 & -3.437 $\pm$ 0.298 & 0.25
        & 0.888 $\pm$ 0.016 & -0.379 $\pm$ 0.128 & 0.17
        \\
        6 
        & 1.064 $\pm$ 0.019 & -3.209 $\pm$ 0.179 & 0.22
        & 0.851 $\pm$ 0.009 & -0.044 $\pm$ 0.072 & 0.14
        \\
        5 
        & 1.116 $\pm$ 0.014 & -3.641 $\pm$ 0.132 & 0.22
        & 0.843 $\pm$ 0.005 & +0.048 $\pm$ 0.045 & 0.12
        \\
        4 
        & 1.193 $\pm$ 0.006 & -4.169 $\pm$ 0.057 & 0.16
        & 0.851 $\pm$ 0.003 & +0.059 $\pm$ 0.024 & 0.11
        \\
        3 
        & 1.158 $\pm$ 0.004 & -3.753 $\pm$ 0.040 & 0.14
        & 0.798 $\pm$ 0.002 & +0.481 $\pm$ 0.018 & 0.10
        \\
        \hline
    \end{tabular}
    \label{tab:best-fits2}
\end{table*}

\subsection{Correlation with other quantities}
In addition to the SFR and the molecular gas mass, the [\ion{C}{II}] emission at high redshifts has been shown to correlate with the metal mass in the gas-phase of high-$z$ galaxies \citep{heintz23}. In \cite{khatri24}, we found that the $\mathrm{C^+}$ distribution in our post-processed galaxy shows a grater similarity with the total gas distribution while CO and atomic carbon correlate better with the molecular gas. Inspired by these findings, we further investigated the correlation between $L_{[\ion{C}{II}]}$
and the total gas mass ($M_{\mathrm{gas}}$) and the mass in metals ($M_{\mathrm{metal}}$). At each redshift, we performed an ordinary least squares regression and the resulting parameters and the $1 \sigma$ scatter around the  best fit are listed in Table~\ref{tab:best-fits2}.  

We find that among the galaxy properties explored so far, namely the SFR, the molecular gas mass, the total gas mass, and the metal mass, the [\ion{C}{II}] emission in our simulated galaxies shows the strongest/tightest correlation with $M_{\mathrm{metal}}$ across redshifts. This is evident from the lowest scatter in this relation. In other words, we can say that \edit{the [\ion{C}{II}]
luminosity is the best predictor of the gas-phase metal mass} in any given galaxy. 
Interestingly, the scatter in the $L_{[\ion{C}{II}]}-M_{\mathrm{metal}}$ relation is lower than the scatter in the $L_{[\ion{C}{II}]}-M_{\mathrm{gas}}$ relation at all times.  A strong correlation between [\ion{C}{II}] emission and total gas mass simply highlights the multi-phase origin of the emission.

After exploring the correlation between $L_{[\ion{C}{II}]}$ and other galaxy properties in our simulations, we now turn our attention to examining the spatial extent of the [\ion{C}{II}] emission in these galaxies and how it varies across the galaxy population.

\section{Extended $[\ion{C}{II}]$ emission}
\label{sec:extended_cii}

\begin{figure*}
    \centering    
    \includegraphics[width=0.73 \textwidth, scale=0.9, trim={0 0 0 0},clip]{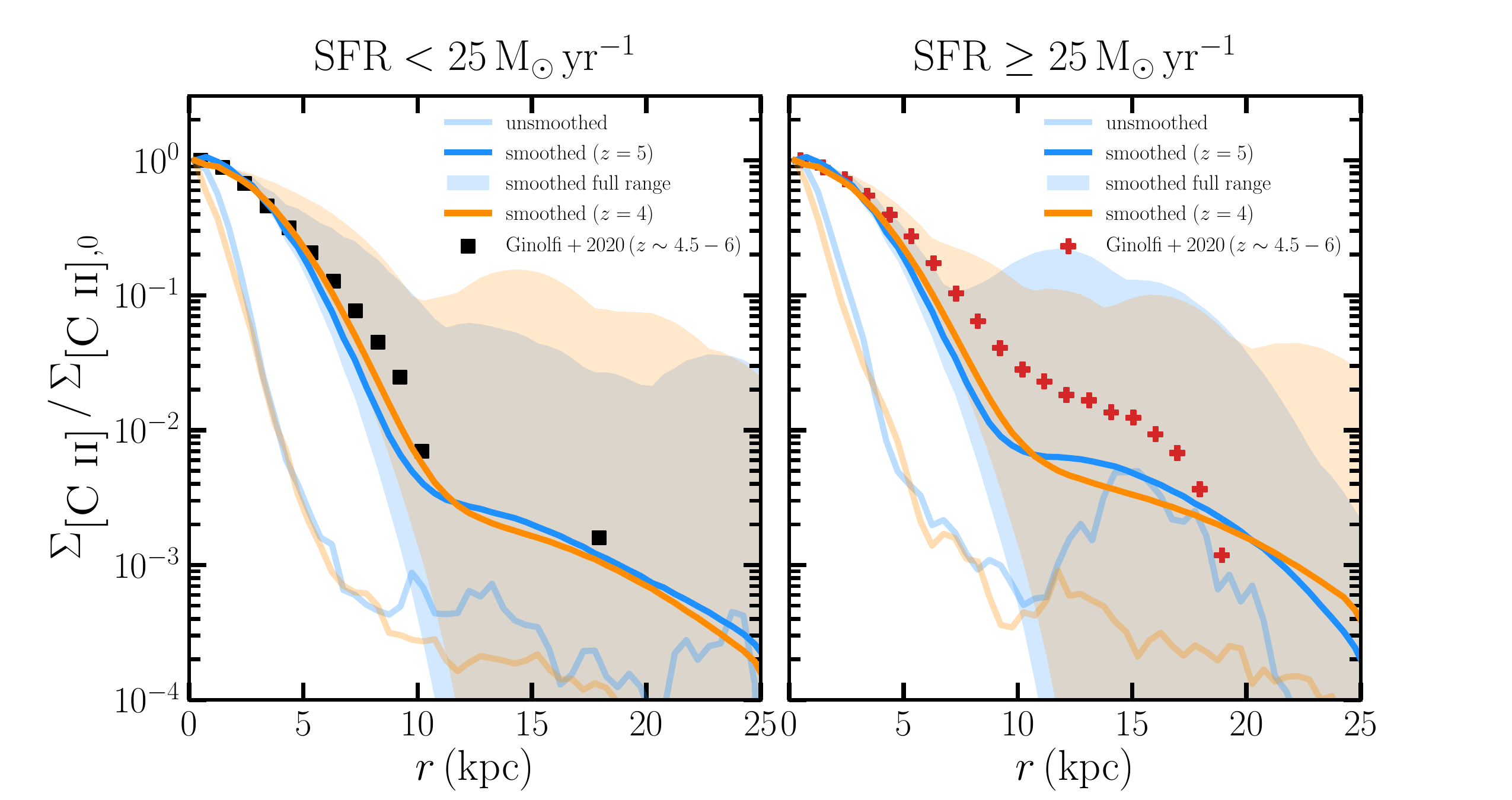}
    \caption{Comparison of the simulated and observed stacked (radial) surface brightness profiles of the [\ion{C}{II}] emission.  \edit{The left and right panels show, respectively, the stacked surface density profiles for the low-SFR ($\mathrm{SFR} < 25 \, \mathrm{M_{\odot} \, yr^{-1}}$) and high-SFR ($\mathrm{SFR} \geq 25 \, \mathrm{M_{\odot} \, yr^{-1}}$) samples as defined by \cite{ginolfi20a} based on galaxies from the ALPINE survey at $z=4.5-5.9$}. The solid lines represent the stacked profiles of (central) galaxies from the simulation at $z=5$ (in blue) and $z=4$ (in orange). The shaded areas represent the full range spanned by the individual profiles, which were constructed from the 2D projection of a 50 kpc cube centred on the galaxy along three orthogonal lines of sight.
    All profiles were smoothed with a 2D Gaussian beam of FWHM 0.9$\arcsec$ \citep[as in][]{ginolfi20a} and normalised by the peak value of the stack. For reference, the unsmoothed stacked profiles are shown in lighter shades. The observed profiles from \cite{ginolfi20a} are shown in black squares and red plusses for their low- and high- SFR samples, respectively.
    }
    \label{fig:cii_profile}
\end{figure*}

In recent years, several observations of high-$z$ ($z \gtrsim 4$) galaxies have revealed that the [\ion{C}{II}] emission extends $2-3$ times farther than the UV continuum emission. These findings come from both stacked galaxy samples \citep[e.g.,][]{fujimoto19, ginolfi20a, fudamoto22} and individual galaxies \citep[e.g.,][]{fujimoto20, lambert23, posses24}, with some studies, like \cite{posses24}, resolving emission on $\sim 1$ kpc scales within galaxies. Potential sources of this extended emission include unresolved satellites, outflows, and extended PDRs \citep[see, for example, Figure 12 in][]{fujimoto19}. In \cite{khatri24}, we examined the extent of $\mathrm{C^+}$ in a simulated galaxy by calculating the $\mathrm{H_2}$, CO, C, and $\mathrm{C^+}$ abundances in post-processing using the sub-grid model HYACINTH, and found that the $\mathrm{C^+}$ surface density profile is more extended than other components such as $\mathrm{H_2}$ and CO, and closely resembles the total gas distribution and that of young stars (with ages $\le 20$ Myr). In this paper, we extended our analysis to a statistical sample of galaxies from the \textsc{Marigold} suite and examined the extent of their [\ion{C}{II}] emission. For this, we first looked at the stacked emission from a sample of galaxies at two different redshifts (Sect.~\ref{sec:extended_cii_stacked}). Then we inspected the relative extent of the [\ion{C}{II}] emission with respect to SF in individual galaxies and investigate possible causes of an extended [\ion{C}{II}] emission (Sect.~\ref{sec:extended_cii_solo}). 


\subsection{Stacked [\ion{C}{II}] emission}
\label{sec:extended_cii_stacked}
First, we compared the stacked $[\ion{C}{II}]$ emission from our simulated galaxies with that from \cite{ginolfi20a}, based on 50 [\ion{C}{II}]-detected galaxies from ALPINE at $z = 4.5-5.9$. The stacked [\ion{C}{II}] emission from these galaxies extends out to $\sim 15 \, \rm kpc$. The authors further split their sample into low star-forming ($\mathrm{SFR} < 25 \, \mathrm{M_{\odot} \, yr^{-1}}$) and high star-forming ($\mathrm{SFR} \geq 25 \, \mathrm{M_{\odot} \, yr^{-1}}$) galaxies and reported that the extended [\ion{C}{II}] emission is more prominent in the latter.

To compare with these observations, we selected 
galaxies at redshifts $z=4$ and $z=5$ with $M_*$ and SFR in the range exhibited by the \cite{ginolfi20a} sample, that is, $10^{9.25} \, \mathrm{M_{\odot}} \leq M_* \leq 10^{11.25} \, \mathrm{M_{\odot}}$ and $5  \, \mathrm{M_{\odot} \, yr^{-1}}  \leq \mathrm{SFR} \leq 600 \, \mathrm{M_{\odot} \, yr^{-1}} $. We note that while the \cite{ginolfi20a} sample includes galaxies with SFR up to 600 $\mathrm{M_{\odot} \, yr^{-1}}$, the highest SFR in our sample is $\sim 300 \, \mathrm{M_{\odot} \, yr^{-1}} $ at $z=4$ and $\sim  170 \, \mathrm{M_{\odot} \, yr^{-1}} $ at $z=5$. Following \cite{ginolfi20a}, we further split our simulated galaxies into low-SFR (SFR$< 25  \, \mathrm{M_{\odot} \, yr^{-1}}$) and high-SFR (SFR$\geq 25  \, \mathrm{M_{\odot} \, yr^{-1}}$) samples.

For each galaxy, we obtained the [\ion{C}{II}] surface brightness map from the 2D projection of a cylinder centred on the galaxy with a line-of-sight velocity cut of $\varv_{\mathrm{los}} \in [-200,200] \, \rm km \, s^{-1}$ \citep[same as in][]{ginolfi20a} along three orthogonal lines of sights (these are taken to be the coordinate axes of the simulation box). Note that in this analysis, we adopted the stellar centre of the galaxy to be the spatial centre of the [\ion{C}{II}] emission\footnote{Note that the centres of the different baryonic components do not necessarily overlap. However, since we apply Gaussian smoothing to the surface brightness maps with a 0.9 $\arcsec$ beam, which is equivalent to 6.4 kpc (5.5 kpc) at $z=5$ ($z=4$), we do not expect this offset to significantly impact our results.}. 
We smoothed the resulting surface brightness maps with a 2D Gaussian beam of FWHM $0.9 \arcsec$ to mimic the synthesised ALMA beam in \cite{ginolfi20a}. We then stacked the individual smoothed surface brightness maps. Note that, since we do not add noise to the galaxy images, our stacking procedure is unweighted by construction and is different from the variance-weighted stacking employed in observations. 

From the stacked surface brightness map at each redshift, we extracted radial profiles in bins of size 0.5 kpc. The resulting profiles are shown as solid lines in Fig.~\ref{fig:cii_profile}. \edit{The left and right panels show the profiles for the low- and high-SFR samples, respectively.} The shaded areas represents the full range covered by the individual profiles. We also include for reference, the profiles obtained without smoothing using a faint line of the same colour. We see that while the true [\ion{C}{II}] distribution in the simulated galaxies is rather compact, smoothing extends the profiles out to larger radii. We observe that, \edit{for the low-SFR sample}, our stacked profiles show a remarkable agreement with \cite{ginolfi20a}. \edit{In our simulation, the low- and high-SFR samples exhibit similar profiles. In contrast, the \cite{ginolfi20a} profiles show some differences at larger radii.}
Nevertheless, some individual galaxies exhibit a similar extent as the high-SFR \cite{ginolfi20a} profile (as indicated by the spread of our stacked profiles).

\begin{figure*}
    \centering
    \includegraphics[width=0.90\textwidth, trim={0 0 0 0},clip]{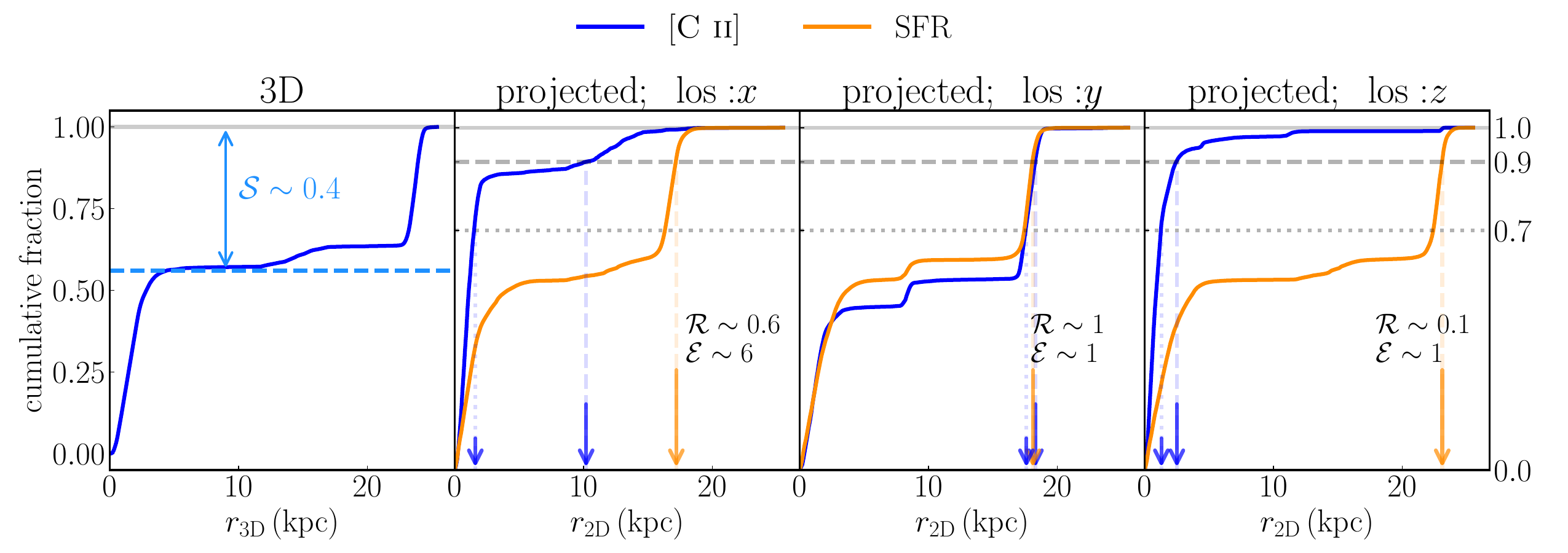}
    \caption{Example illustrating the calculation of the $\mathcal{S}$, $\mathcal{R}$, and $\mathcal{E}$ parameters in a simulated galaxy. The left panel shows the cumulative profile for the 3D distribution of [\ion{C}{II}], with the dashed line marks the contribution of the central galaxy to the total [\ion{C}{II}] emission (as evident from the flattening of the cumulative profile). The remaining fraction (denoted by $\mathcal{S}$) represents the contribution of satellites. The other panels show cumulative profile constructed from the [\ion{C}{II}] surface brightness (blue) and SFR surface density (orange). For the profiles obtained from projections, the value of the $\mathcal{R}$ and $\mathcal{E}$ parameters are indicated in each panel. In all but the leftmost panel, the dotted, dashed and solid horizontal lines denote cumulative fractions of 70\%, 90\%, and 100\%, respectively. 
    The small and large blue arrows mark $r_{70, \, [\ion{C}{II}]}$ and $r_{90, \, [\ion{C}{II}]}$, respectively and the orange arrow denotes $r_{90, \, \mathrm{SFR}}$. The parameter $\mathcal{R} \equiv r_{90, \, [\ion{C}{II}]}/r_{90, \mathrm{SFR}}$ is calculated from the ratio of the $r$ values denoted by the large blue and orange arrows, while the parameter $\mathcal{E} \equiv r_{90, \, [\ion{C}{II}]}/r_{70, \rm [\ion{C}{II}]}$ from the ratio of the large and small blue arrows.    }
    \label{fig:cartoon}
\end{figure*}
\begin{figure*}
    \centering
    \includegraphics[width=0.75\textwidth, trim={0 0 0 0},clip]{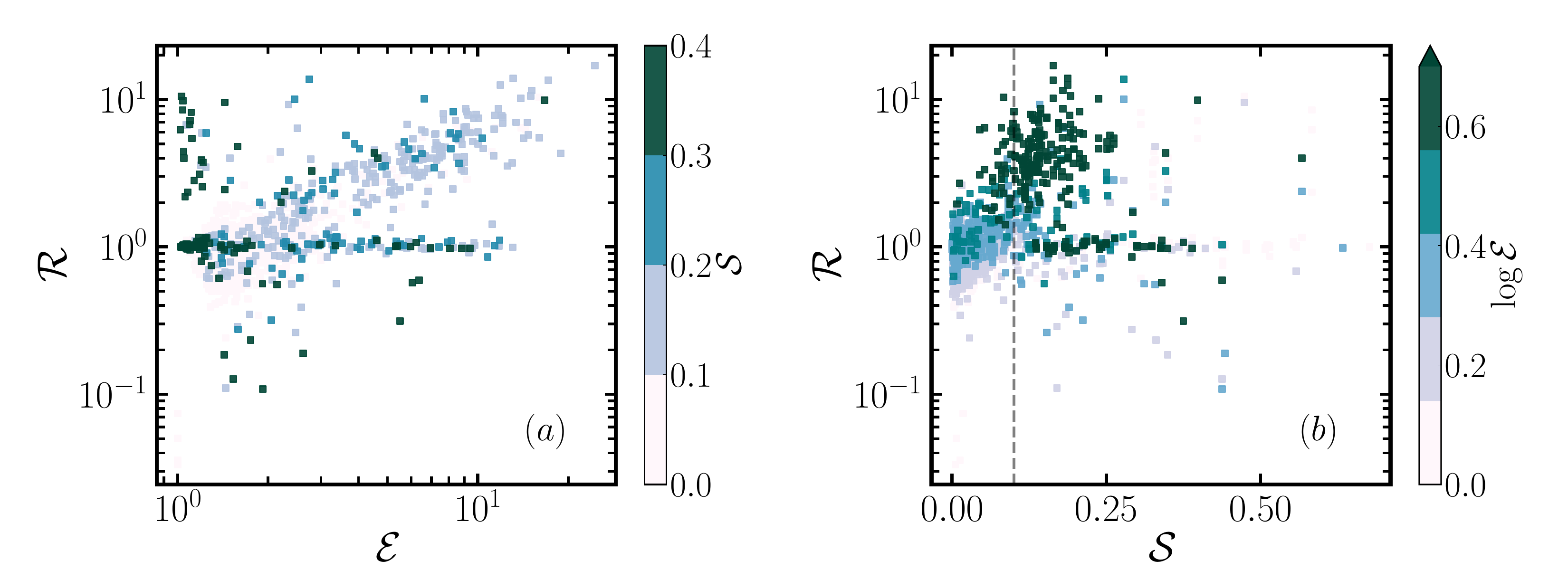}
    \caption{Distribution of our galaxies at redshift $z=4$ in the $\mathcal{R}$-$\mathcal{E}$ (\textit{left}), $\mathcal{R}$-$\mathcal{S}$ (\textit{right}) planes. The scatter points are colour-coded by the parameter $\mathcal{S}$ in the left panel and by the log of parameter $\mathcal{E}$ in the right panel. The dashed grey lines in the right panel indicates the threshold value of $\mathcal{S}$ used to separate galaxies into low and high $\mathcal{S}$. 
    }
    \label{fig:qes}
\end{figure*}
\begin{figure*}
     \centering
     \includegraphics[width=0.78\textwidth, trim={0 0.5cm 2cm 2cm},clip]{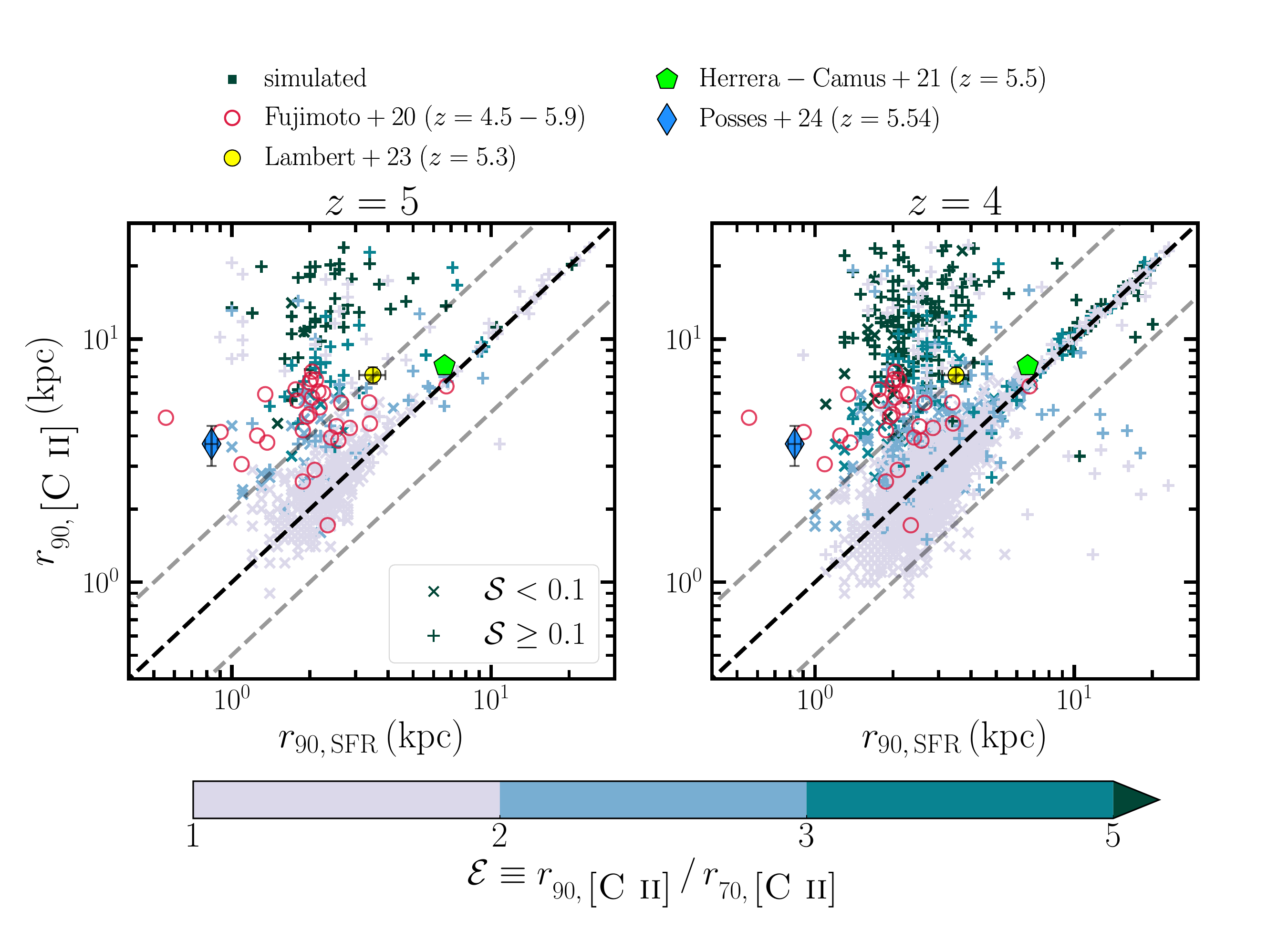}
    \caption{Comparison of $r_{90, \, \textsc{[C ii]}} $ and $ r_{90, \, \mathrm{SFR}}$ for simulated galaxies at redshifts $z=5$ (left) and  $4$ (right). The galaxies are colour-coded by their multicomponent extent parameter $\mathcal{E}$ defined as the ratio of the $r_{90}$ and $r_{70}$ values of the $[\ion{C}{II}]$ surface brightness profile. 
    The shape of the symbol reflects the $\mathcal{S}$ parameter that quantifies the satellite contribution to the total [\ion{C}{II}] emission (see text for details). We use `low $\mathcal{S}$' and `high $\mathcal{S}$', respectively to denote galaxies with $< 10 \%$ and $\geq 10\%$ satellite contribution.
    The $r_{90, \, \textsc{[C ii]}} $ versus $ r_{90, \, \mathrm{SFR}}$ values of observed galaxies are shown as red open circles \citep{fujimoto20}, a yellow plus \citep{lambert23}, a green pentagon \citep{herrera-camus21}, and a blue diamond \citep{posses24}. The black dashed line indicates a 1:1 relation, while the top and bottom grey dashed lines indicate 2:1 and 1:2 relations, respectively. Note that the error bars are not shown for \cite{fujimoto20} galaxies for the sake of clarity.}
    \label{fig:q_vs_e}
    \end{figure*}    

\subsection{Extent of [\ion{C}{II}] emission in individual objects}
\label{sec:extended_cii_solo}

Next we quantified the extent of the [\ion{C}{II}] emission relative to the SF activity in our simulated galaxies. For this analysis, we included all galaxies that meet the following criteria: (i) $M_* \geq 10^{8.5} \, \mathrm{M_{\odot}}$ and (ii) $\mathrm{SFR} \geq 3  \, \mathrm{M_{\odot} \, yr^{-1}}$. These were chosen to match the range of $M_*$ and SFR values in ALPINE galaxies \citep{lefevre20}. For each galaxy in our sample, we took a sphere of radius 25 kpc centred on the galaxy and obtain projections of the same along three orthogonal lines of sights (we take these to be the coordinate axes of the simulation box) to obtain the [\ion{C}{II}] surface brightness and SFR surface density maps. Then we computed the (cumulative) radial surface brightness profiles for each projection, from which we derived the radius enclosing 70\% and 90\% of the total [\ion{C}{II}] emission -- these are denoted as $r_{70, [\ion{C}{II}]}$ and $r_{90, [\ion{C}{II}]}$, respectively. Similarly, from the (cumulative) SFR surface density profile, we derived $r_{90, \, \mathrm{SFR}}$ (as in Sect.~\ref{sec:cii_sfr_all}, we use the SFR averaged over the last 200 Myr). For each galaxy, we also computed the cumulative [\ion{C}{II}] luminosity profile from the full 3D distribution of [\ion{C}{II}] within the region. 

Based on these, we calculated the following parameters for each galaxy for the three orthogonal projections. As an example, we show in Fig.~\ref{fig:cartoon}, cumulative profiles for a simulated galaxy and how these are used for calculating the three parameters and to ease their interpretation.
\begin{enumerate}
    \item The multicomponent extent parameter $\mathcal{E} \equiv r_{90, \, [\ion{C}{II}]} \, / \, r_{70, \, [\ion{C}{II}]}$ that measures the spread in the [\ion{C}{II}] emission. A higher $\mathcal{E}$ indicates a relatively large extent of the diffuse [\ion{C}{II}] component. This would occur in galaxies where $70\%$ of the emission is relatively confined, while the remaining $10-30\%$ is more spread out, likely due to the presence of satellite galaxies. Thus, a higher $\mathcal{E}$ denotes more extended emission relative to the bulk of the emission.
    \item The parameter $\mathcal{R} \equiv r_{90, \, [\ion{C}{II}]} \, / \, r_{90, \, \mathrm{SFR}}$ that quantifies the relative extent of the [\ion{C}{II}] emission compared to SF. In this analysis, we were particularly interested in galaxies where the [\ion{C}{II}] emission is at least twice as extended as the SF activity (i.e., $\mathcal{R} \geq 2$). 
    \item The parameter $\mathcal{S}$ that quantifies the fraction of the total [\ion{C}{II}] emission that arises from outside the central galaxy and represents the contribution from satellites. Unlike $\mathcal{R}$ and $\mathcal{E}$, this parameter is computed from the true 3D distribution of [\ion{C}{II}] emission and is therefore agnostic to the projection axis by construction.
\end{enumerate}
Note that all three parameters are agnostic to whether the [\ion{C}{II}] emission at a given location arises from a $\mathrm{C^+}$ ion formed in situ or transported there, for example, via outflows/inflows.

From Fig.~\ref{fig:cartoon}, it is evident that the parameters $\mathcal{R}$ and $\mathcal{E}$ are sensitive to the orientation of the galaxy. 
For instance, the $y$-projection of the galaxy has $\mathcal{R} \sim 1$, indicating equal extent of the [\ion{C}{II}] emission and SFR. In contrast, in the $x$- and $z$- projections, the SFR profile is more extended than the [\ion{C}{II}] profile, resulting in $\mathcal{R} < 1$. Note that in this case, the $\mathcal{E}$ value is relatively high, meaning that $r_{70, \, [\ion{C}{II}]}$ and $r_{90, \, [\ion{C}{II}]}$  are well separated, but owing to a more extended SFR profile, the $\mathcal{R}$ value is low. 

To better understand how the three parameters are correlated, we show in Fig.~\ref{fig:qes}, how $\mathcal{R}$ varies with $\mathcal{E}$ and $\mathcal{S}$ for our galaxy sample at $z=4$. Each scatter point represents one of the orthogonal projections of a galaxy and is colour-coded by value of the third parameter. Firstly, in panel (a), we see that the parameter $\mathcal{R}$ increases in general with $\mathcal{E}$. However, some galaxies\footnote{Note that while $\mathcal{R}$ and $\mathcal{E}$ are properties of the projection of a given galaxy, for simplicity, we associate these parameters directly with the galaxy itself.} with $\mathcal{R} \gtrsim 2$ exhibit a low $\mathcal{E}$ ($\lesssim 2$). In contrast, in panel (b), we see that all galaxies with a high $\mathcal{R}$ generally have a high $\mathcal{S}$ as well, although the reverse is not always true -- some galaxies have a high $\mathcal{S}$ but $\mathcal{R} \sim 1$. In these galaxies, satellites contribute similarly to the [\ion{C}{II}] emission and the SF, driving the $r_{90}$ of both quantities to high values, and thereby reducing the $\mathcal{R}$ (the $y$-projection in Fig.~\ref{fig:cartoon} is an example of this). 

In Fig.~\ref{fig:q_vs_e}, we compare the extent of the $r_{90}$ values of the [\ion{C}{II}] emission and the SFR. The galaxies are colour-coded according to their $\mathcal{E}$ parameter while the shape of the symbol represents the value of the $\mathcal{S}$ parameter: we split the galaxies into `low $\mathcal{S}$' ($\mathcal{S}<0.1$) and `high $\mathcal{S}$' ($\mathcal{S} \geq 0.1$) subsamples. The threshold of 0.1 or 10\% was inspired by \cite{springel08}, who found that $\approx 11\%$ of the mass fraction in virialised halos is present in substructures.
For reference, we also show the $r_{90}$ values for [\ion{C}{II}] and SFR from observations of individual galaxies: to obtain these, we used the half-light radii $r_{\mathrm{e}}$ for the [\ion{C}{II}] emission and UV continuum emission, reported in the respective observations. In all observations shown in Fig.~\ref{fig:q_vs_e}, $r_{\rm e, \, [\ion{C}{II}]}$ and $r_{\rm e, \, UV}$ are obtained from fitting exponential disk-profiles to the [\ion{C}{II}] emission and UV continuum emission, respectively. We scaled these $r_{\rm e}$ values by 2.318 to obtain the respective $r_{90}$ values\footnote{We remind the reader that for an exponential disk-profile $\Sigma = C \, \exp{(-r/h)}$, where $h$ is the scale length of the disk, the half-light radius $r_{\rm e} \approx 1.678 \, h$ and $r_{90} \approx 3.890 \, h$. Therefore $r_{90} \approx 2.318 \, r_{\rm e}$. }. Since the bulk of the UV emission from galaxies arises from stars younger than $\approx 100 $ Myr \citep{kennicutt12}, we further assumed that the $r_{\rm 90, \, SFR}$ in the observed galaxies can be approximated by the $r_{90, \, \rm UV}$ values. A possible caveat is that the UV sizes might be larger than the corresponding SFR sizes \citep[e.g., see][ for a comparison of the half-light and half-mass radii of $0.5 < z <2$ galaxies.]{szomoru13}.





From Fig.~\ref{fig:q_vs_e}, we see that the [\ion{C}{II}] emission in many of our simulated galaxies has a similar extent as the observed galaxies at $4.5 \lesssim z \lesssim 5.9$ from \cite{fujimoto20}, \cite{herrera-camus21}, and \cite{lambert23}. However, our simulated galaxies occupy a larger area of the parameter space compared to observations. Our galaxies where the [\ion{C}{II}] emission is at least twice as extended as the SF activity (i.e., above the top grey dashed line, $\mathcal{R} \geq 2$) exhibit preferentially higher $\mathcal{E}$ values (darker colours of the symbol) and a higher contribution from satellites, compared to galaxies lying along the diagonal (i.e. $\mathcal{R} \sim 1$ ). This is also evident from the higher median of the $\mathcal{E}$ and $\mathcal{S}$ parameters for the galaxies with $\mathcal{R} \geq 2$ (see Table~\ref{tab:median_qes}).


To summarize, we find that the inferred detection of an extended [\ion{C}{II}] emission in a given galaxy is sensitive to the orientation of the galaxy. Nevertheless, some statistical differences emerge between galaxies with extended [\ion{C}{II}] emission compared to their SF activity (i.e., with $\mathcal{R} \equiv r_{90, \, [\ion{C}{II}]} \, / \, r_{90, \, \mathrm{SFR}} \geq 2$) and those without (i.e., $\mathcal{R} < 2$). Galaxies with $\mathcal{R} \geq 2$ tend to have a higher contribution from satellites. They also exhibit higher $\mathcal{E}$ values compared to the latter, indicating that while the bulk ($\sim 70\%$) of the [\ion{C}{II}] emission is relatively concentrated, the remaining fraction can extend out to 4-5 times larger radii. Overall, about 10\% of our simulated galaxies at $z=4$ have $\mathcal{R} \geq 2$ that is, their [\ion{C}{II}] emission extends $\geq 2$ times farther than the SF activity; this fraction increases to $20\%$ at $z=5$. This is in agreement with recent observations pointing out the increased prevalence of extended [\ion{C}{II}] emission towards higher redshifts \citep{carniani18, fujimoto19, ginolfi20a, fudamoto22}.

\begin{table}
    \caption{Median values of the parameters $\mathcal{E}$ and $\mathcal{S}$ for galaxies with and without extended [\ion{C}{II}] emission at $z=4$ and $z=5$.}
    \centering
    \begin{tabular}{c|cc|cc}
    \hline\hline
    & \multicolumn{2}{c|}{$z=4$} & \multicolumn{2}{c}{$z=5$}\\
    & median $\mathcal{E}$ & median $\mathcal{S}$ &  median $\mathcal{E}$ & median $\mathcal{S}$ \\
    \hline
    $\mathcal{R} \geq 2$ &5.5 & 0.14 &  3.99 & 0.17  \\
    $\mathcal{R} < 2$ & 1.5 & 0.02 &  1.5 & 0.03 \\
    \hline
    \end{tabular}
    \label{tab:median_qes}
\end{table}

\section{Summary}
\label{sec:discussion}
Based on a sample of simulated galaxies with molecular cloud chemistry evolved on the fly and $[\ion{C}{II}]$ 158 $\mu$m line emission calculated in post-processing, we have investigated the reliability of this line as a tracer of the SF activity and molecular gas content in galaxies at redshifts $3\leq z \leq 7$. Here we briefly summarize our findings:

\begin{enumerate}
\setlength{\itemsep}{1em}
\item Redshift evolution of the $[\ion{C}{II}]$ luminosity function: Our simulations predict a strong time evolution in the number density of [\ion{C}{II}] emitters, especially at the bright end. The number density of $L_{[\ion{C}{II}]} \sim 10^9 \, \mathrm{L_{\odot}}$ galaxies increases by 600 times in the above redshift range. At all redshifts, a double power-law provides a better fit to our simulated LFs (Table~\ref{tab:cii_lf}).

\item Redshift evolution of the [\ion{C}{II}]-SFR relation:
The slope of the $L_{[\ion{C}{II}]}-\mathrm{SFR}$ relation shows little evolution in the redshift range $3 \le z \leq 7$ while the intercept increases by $\approx 0.5$ dex in this interval, indicating that the $[\ion{C}{II}]$ luminosity at given SFR increase roughly by a factor of three from $z=7$ to $z=3$. Notably, the scatter in the relation increases towards higher redshifts (Table~\ref{tab:best-fits1}).

\item The conversion factor $\alpha_{[\ion{C}{II}]}$:
The conversion factor $\alpha_{[\ion{C}{II}]}$ between the $[\ion{C}{II}]$ luminosity and the molecular gas mass in our simulated galaxies ranges from $\sim 1 - 200 \, \mathrm{M_{\odot} \, L_{\odot}^{-1}}$ and does not show a systematic dependence on metallicity in agreement with the findings from \cite{zanella18} based on a compilation of galaxies from $z =0-5.5$. Across redshifts, $\alpha_{[\ion{C}{II}]}$ shows a strong correlation with the SFR averaged over 5 Myr and with the SFR change diagnostic $R_{5-200}=\mathrm{SFR_5/SFR_{200}}$ (Fig.~\ref{fig:alpha_cii}).

\item Secondary dependences in the $L_{[\ion{C}{II}]}-M_{\mathrm{mol}}$ relation:
We performed a principal component analysis to quantify secondary dependences in the $[\ion{C}{II}]$-$M_{\mathrm{mol}}$ relation and found a strong dependence on the SFR$_5$ (the star formation rate measured on a 5 Myr timescale) that evolves with redshift and a weak dependence on metallicity across redshifts (Table~\ref{tab:pca_z}). The resulting 3-variable PCA relation predicts the true molecular gas within a factor of $1.7$ ($2.5$) at $z=3$ ($z=7$). When accounting for typical observational uncertainties on $L_{[\ion{C}{II}]}$, SFR, and gas metallicity, the PCA-based relation predicts the true molecular gas mass within a factor of $\sim 2.5$ at $3 \leq z \leq 5$.

\item What does the $[\ion{C}{II}]$ emission really trace?: 
We investigated the correlation of $L_{[\ion{C}{II}]}$ with several galaxy-integrated properties, namely the SFR, the molecular gas mass, the total gas mass, and the metal mass in gas phase ($M_{\mathrm{metal}}$). Among these, the [\ion{C}{II}] emission in our simulated galaxies shows the tightest correlation with $M_{\mathrm{metal}}$ across redshifts (Tables~\ref{tab:best-fits1} and \ref{tab:best-fits2}). 

\item Extended $[\ion{C}{II}]$ emission:
We observed that our stacked [\ion{C}{II}] surface brightness profiles show a similar extent as the low-SFR galaxy sample from \cite{ginolfi20a}, although some individual galaxies also exhibit a similar extent as their high-SFR sample. We further found that galaxies where the [\ion{C}{II}] emission extends twice or more farther than the SF activity preferentially exhibit a spatial distribution wherein the bulk ($\gtrsim 70 \%$) of the [\ion{C}{II}] emission is confined to the central galaxy, while the remaining $\lesssim 30\%$ extends out to larger distances because of the presence of satellites. The typical fractional contribution of satellites to the total [\ion{C}{II}] emission in these galaxies is $\approx 5-7$ times higher than that in galaxies without extended emission (see Table~\ref{tab:median_qes}).

\end{enumerate}


\begin{acknowledgements}
The authors thank an anonymous reviewer for their valuable feedback that helped improved this manuscript. We thank R. Teyssier for making the \textsc{Ramses} code publicly available. They gratefully acknowledge the Collaborative Research Center 1601 (SFB 1601 sub-project C5) funded by the Deutsche Forschungsgemeinschaft (DFG, German Research Foundation) – 500700252. Early stages of this work were carried out within the Collaborative Research Centre 956 (SFB 956 sub-project C4), funded by the DFG – 184018867.  The authors acknowledge the Gauss Centre for Supercomputing
e.V. (\href{www.gauss-centre.eu}{www.gauss-centre.eu}) for funding this project by providing
computing time on the GCS Supercomputer SuperMUC-NG at the Leibniz Supercomputing Centre (\href{http://www.lrz.de}{http://www.lrz.de}). They are thankful to the Max Planck Society (MPG) for providing computing time on the supercomputer Raven at the Max Planck Computing and Data Facility (\href{https://www.mpcdf.mpg.de/}{https://www.mpcdf.mpg.de}), where part of the \textsc{Marigold} simulations were performed. PK is a part of the International Max Planck Research School in Astronomy and Astrophysics, the Bonn Cologne Graduate School, and a guest at the Max Planck Institute for Radio Astronomy in Bonn.  
\end{acknowledgements}

\bibliographystyle{aa} 
\bibliography{example} 

\appendix
\section{Modelling $[\ion{C}{II}]$ emission}
\label{sec:appA}
For a two-level system such as the $\mathrm{C^+}$ fine-structure transition, the excitation temperature ($T_{\rm ex}$) captures the relative population in the upper ($u$) and lower ($l$) energy levels of the transition:
\begin{equation}
\label{eq:Tex}
    \frac{n_u}{n_l} =\frac{g_u}{g_l} \, e^{-T_* / T_{\rm ex}} \, .
\end{equation}
Here $g_u$ and $g_l$ are the statistical weights of the upper and lower levels with an energy difference of $k_{\rm B} T_* $ ($k_{\mathrm{B}}$ being the Boltzmann constant),
and $n_u + n_l = n_{\mathrm{C^+}}$. In statistical equilibrium, level populations are determined by the balance between excitation and deexcitation processes:
\begin{equation}
\label{eq:balance}
     n_l (B_{lu} U + C_{lu} ) \,=\, n_u (A_{ul} + B_{ul} U + C_{ul} ), \, 
\end{equation}
where $U$ is the energy density at the transition frequency $\nu$; $A_{ul}$, $B_{ul}$ and $B_{lu}$ are the Einstein coefficients for spontaneous decay, stimulated decay, and stimulated excitation, respectively, $C_{ul}$ and $C_{lu}$ are the collision deexcitation and excitation rates and, in the case of multiple collision partners, can be obtained from the respective collision rate coefficients and number densities as:
\begin{equation}
    C_{ul} = \sum_{i=1}^{N} R_{ul, \, i} \, n_{i} .
\end{equation}
The upward and downward collision rate coefficients are related as:
\begin{equation}
    R_{lu} = R_{ul} \frac{g_u}{g_l} e^{-T_* / T_{\rm kin}}, \, 
\end{equation}
where $T_{\rm kin}$ is the kinetic temperature related to the thermal motions of the collision partner. The collision rate coefficients are taken from \cite{goldsmith12}:
\begin{align}
    R_{ul} (e^-)   &= 8.7 \times 10^{-8}  (T_e/2000)^{-0.37} \, \rm cm^{3} \, s^{-1} ; \, \\
    R_{ul} ({\rm H})   &= 7.6 \times 10^{-10} (T_{\rm kin}/100)^{0.14} \, \rm cm^{3} \, s^{-1} ; \, \\
    R_{ul} ({\mathrm{H_2}}) &= 3.8 \times 10^{-10} (T_{\rm kin}/100)^{0.14} \, \rm cm^{3} \, s^{-1} \, ,
\end{align}
where $T_e$ is the electron temperature. In the following, we assume $T_e = T_{\rm kin}$. 
We obtain the kinetic temperature at each sub-grid density using the temperature density relation \citep[from][]{hu21} and is identical to the one adopted in \textsc{Hyacinth} \citep[see][for details]{khatri24}. The $\mathrm{C^+}$ and $\mathrm{H_2}$ number densities are obtained directly from the simulations. To obtain the number density of atomic hydrogen ($\rm H$), we assume that gas at densities  $n_{\rm H,tot} \gtrsim 0.013 \, \rm cm^{-3}$ is well-shielded and $n_{\rm H^+}= 0$ \citep{tajiri98}; 
below these densities, we assume all hydrogen to be ionised, that is, $n_{\rm H^+} =  n_{\rm H,tot}$. The electron number density follows from charge conservation that is, $n_{\rm e^-} = n_{\mathrm{C^+}} + n_{\rm H^+}$. Note that, similar to \cite{gong12} and \cite{vallini15}, we do not consider the pumping effect from the soft UV background from stars at $1330\,$\AA. 


\begin{figure}
    \centering    
    \includegraphics[scale=0.4, width=0.42\textwidth, trim={0 0 0 0 },clip]{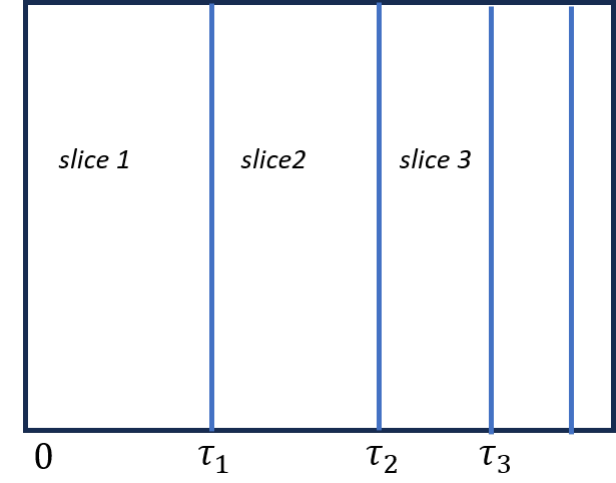}
    \caption{Schematic representation of the plane-parallel slices in a grid cell.}
    \label{fig:slice}
\end{figure}
Now, suppose that a given grid cell of sidelength $L$ can be split into $N$ plane-parallel slices as shown in Fig.~\ref{fig:slice}. In the following, we take $N=3$ for simplicity, but in practice, use 20 slices in each slice, which were sufficient to reach convergence. 
The width and density  of each slice are determined by the sub-grid density PDF (same as in \textsc{Hyacinth}) 
within the cell. If $T_{{\rm ex}, i}$ is the excitation temperature in slice $i$, then the energy density generated in the slice at the transition frequency $\nu$ can then be written as
\begin{equation}
    U_{\nu}(T_{{\rm ex},i}) \, = \, \frac{4 \pi }{c} \, B_{\nu}(T_{{\rm ex},i}) = \frac{8 \pi h \nu^3}{c^3} \, \frac{1}{\exp(h \nu/k_{\rm B} T_{{\rm ex},i}) -1} \, .
\end{equation}
Assuming the entire region is permeated by a background like the CMB 
at temperature $T_{\rm bg}$ with energy density $U_{\nu}(T_{\rm bg})$, the total energy density in a slice will have three contributions:
\begin{enumerate}
    \item the energy density generated in the slice weighted by the fraction of the energy that does not escape the slice (self-absorption, denoted by $\kappa_{ii}$);
    \item the energy density from all other slices where the energy density of the emitting slice $i$ is weighted by the fraction that is absorbed by the absorbing slice $j$ (denoted by $\kappa_{ij}$);
    \item the energy density from the background (CMB in this case).
\end{enumerate}
Following \cite{goldsmith12}, the total energy density in the slice at the $[\textsc{C ii}]$ frequency can be written as:
\begin{multline}\label{eq:U1}
    U_1 \,=\, (1 - \,\kappa_{11})\, U_{\nu}(T_{\rm bg}) +  \kappa_{11} \, U_{\nu}(T_{\rm ex,1}) \\ +\kappa_{21} \, U_{\nu}(T_{\rm ex,2}) + \kappa_{31}\, U_{\nu}(T_{\rm ex,3}) \, .
\end{multline}
Similarly, 
\begin{multline}\label{eq:U2}
    U_2 \,=\, (1 - \,\kappa_{22}) \, U_{\nu}(T_{\rm bg}) +  \kappa_{12} \, U_{\nu}(T_{\rm ex,1}) \\ + \kappa_{22} \, U_{\nu}(T_{\rm ex,2}) + \kappa_{32}\, U_{\nu}(T_{\rm ex,3}) \, ;
\end{multline}
and
\begin{multline}\label{eq:U3}
    U_3 \,=\, (1 - \,\kappa_{33}) \, U_{\nu}(T_{\rm bg}) + \kappa_{13} \, U_{\nu}(T_{\rm ex,1}) \\ + \kappa_{23} \, U_{\nu}(T_{\rm ex,2}) + \kappa_{33} \, U_{\nu}(T_{\rm ex,3}) \, .
\end{multline}

\noindent If $U_1$, $U_2$, $U_3$ are known, they can be used to evaluate the level population in each slice by balancing the upward and downward transitions 
as (e.g., for slice 1):
\begin{equation}
\label{eq:nu1}
    n_{u,1} \left( A_{ul} + B_{ul} U_1 + C_{ul} \right) = n_{l,1} \left( B_{lu} U_1 + C_{lu} \right)  \, ,
\end{equation}
and the excitation temperature $T_{\rm ex,1}$ can be obtained using Eq. (\ref{eq:Tex}). Following \cite{goldsmith12}, the optical depth for slice 1 can be calculated from the excitation temperature $T_{\rm ex,1}$ as:
\begin{equation}\label{eq:tau}
\begin{split}
    \tau_{1}
    &= \frac{h B_{lu} N(\mathrm{C^+})}{\delta v} \, \frac{1 \,-\, e^{-h \nu/ k_{\rm B} T_{{\rm ex},1}}}{1+ (g_u/g_l) \, e^{-h \nu/ k_{\rm B} T_{{\rm ex},1}}} \, \\
    &= \frac{g_u}{g_l}\, \frac{A_{ul} \lambda_{ul}^3 N(\mathrm{C^+})}{8 \pi \, \sqrt{8 \, ln(2)} \, \sigma_{\varv}} \, \frac{1 \,-\, e^{-h \nu/ k_{\rm B} T_{{\rm ex},1}}}{1+ (g_u/g_l) \, e^{-h \nu/ k_{\rm B} T_{{\rm ex},1}}} \, ,
\end{split}
\end{equation}
where $\lambda_{ul}$ is the wavelength of the [\ion{C}{II}] line and $N(\mathrm{C^+})$ denotes the column density of $\mathrm{C^+}$ from the edge of the cell to the slice $i$. The above expression approximates the line profile function $\phi_{\nu}$ at line centre by $\delta \varv^{-1}$. where $\delta \varv$ is the line width ($\delta_{\varv}=\sqrt{8 \, \ln 2} \, \sigma_{\varv}$ for a Gaussian velocity distribution with 1-D velocity dispersion $\sigma_{\varv}$). 
We obtain the 1-D velocity dispersion $\sigma_{\varv}$ for the cells in our simulations following the approach of \cite{olsen15}: the velocity dispersion in a giant molecular cloud (GMC) of mass $M_{\mathrm{GMC}}$ and radius $R_{\mathrm{GMC}}$ is given as
\begin{equation}
    \sigma_{\varv} = 1.2 {\rm km \, s^{-1}} \, \left( \frac{R_{\rm GMC}}{\rm pc}\right)^{-1/2} \, \left( \frac{M_{\rm GMC}}{290 \, \rm M_{\odot}}\right)^{1/2} \, ,
\end{equation}
where we approximate $R_{\rm GMC} \approx \frac{1}{2}\Delta L $ and $M_{\rm GMC} = M_{\mathrm{gas, \, cell}} $.  

The system of equations (\ref{eq:Tex}), (\ref{eq:U1})-(\ref{eq:tau}) can be solved iteratively. We start with the optically thin case where all emitted radiation freely escapes (i.e., $\kappa_{ij}=0 \,\,\, \forall \, \{i, j\}$), and obtain an initial estimate of $n_{u, i}$ using Eq. (\ref{eq:nu1}). These estimates are then used to obtain a first estimate of $T_{\rm ex,1}$, $T_{\rm ex,2}$, $T_{\rm ex,3}$ using Eq. (\ref{eq:Tex}). These determine the optical depths $\tau_1$, $\tau_2$, $\tau_3$ using Eq. (\ref{eq:tau}). Using these, the $\kappa$ values can be updated and the entire procedure is repeated until convergence\footnote{Here convergence is defined as the point when the difference between the norm of $\kappa$ matrices of successive iterations is $\leq 10^{-4}$.}.  

Once we have a set of self-consistent $T_{\rm ex}$ and $\kappa$ values, we can calculate the emissivity of each slice. Following \cite{goldsmith12}, the emissivity in slice 1 can be written as  
\begin{equation}
    \epsilon_1 \,=\, n_{u,1} \, A_{ul} \, \gamma_1 \, h \, \nu \, \left[ 1- \frac{e^{( h \nu /k_{\rm B} T_{\rm ex,1})}-1}{e^{( h \nu /k_{\rm B} T_{\rm bg})}-1} \right] \, ,
\end{equation}
where 
\begin{equation}
    \label{eq:gamma}
    \gamma_i = 1.0 - \sum_{j=1}^{3} \, \kappa_{ij}
\end{equation}
denotes the final escape fraction
for slice $i$, that is the fraction of photons emitted in slice $i$ that manage to escape the cell and accounts for absorption from all intervening slices.

The total luminosity from the cell can be written as:
\begin{equation}
    L_{\rm [\ion{C}{II}]} \, = \, \sum_{i=1}^{N} \gamma_i \, \epsilon_{i} \, \Delta V_{i} \, ,
    \label{eq:lcii}
\end{equation}
where $\Delta V_i = \mathcal{P}_{V}(n_i) \, \Delta n_{i} \, V_{\rm cell}$ is the volume of each slice, and $N$ is the number of slices in the cell. The galaxy luminosity is obtained by summing over the $[\textsc{C ii}]$ emission from all cells within the galaxy. We note that our method assumes that the $[\textsc{C ii}]$ emission from the different grid cells are radiatively decoupled and the total $[\textsc{C ii}]$ emission from a galaxy is the sum of the emission from each cell.

\section{Model validation}
\label{sec:cloudy}
\begin{figure*}
    \centering
    \includegraphics[width=0.97\linewidth]{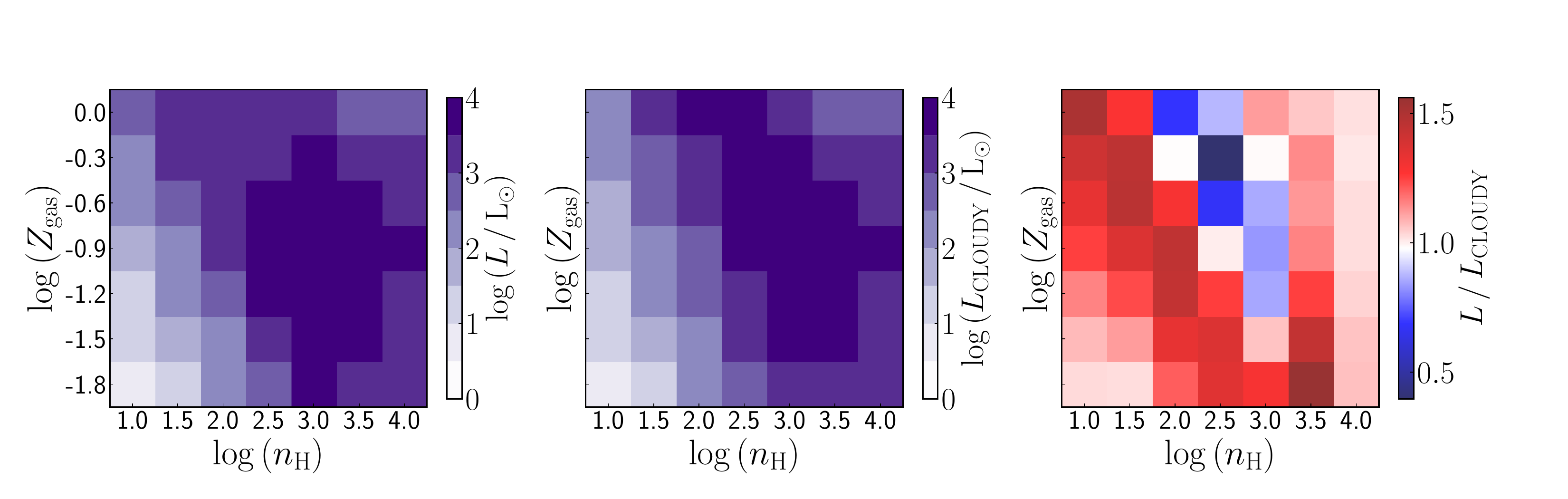}
    \caption{Tests of [\ion{C}{II}] emission model with \textsc{Cloudy}. The left and middle panels show, respectively, the luminosity from our model and \textsc{Cloudy}, while the right panel shows the ratio of the two. For each set of $(n_{\rm H}, \, Z_{\mathrm{gas}})$ values, the temperature is obtained using the temperature-density relation from \cite{hu21}. 
    }
    \label{fig:cloudy}
\end{figure*}

We validate our [\ion{C}{II}] emission model by comparing its output with the photoionisation code \textsc{Cloudy} \citep[version 17.02;][]{ferland92}. To do this, we compute the luminosity emerging from a plane-parallel slab with a side length of $\Delta x = 38 \rm pc$, corresponding to the minimum spatial resolution achieved in our {\tt M25} run at $z=4$. The CMB is included for $z=4$. 

To span the range of gas densities and metallicities exhibited by PDRs and molecular clouds, we compute the luminosity for a 2D grid with hydrogen number density $n_{\rm H} \in [10^0, 10^4] \, \rm cm^{-3}$ and gas metallicity $Z_{\mathrm{gas}} \in [10^{-1.8}, 10^{0}] \, \rm Z_{\odot}$. The slabs have a uniform density, metallicity and temperature. 
We assume that the kinetic temperature is uniform throughout the slab. For each set of ($n_{\rm H}, \, Z_{\mathrm{gas}}$) values, we obtain the temperature using the metallicity-dependent temperature density relation from \citet[][same as in HYACINTH]{hu21}. 
We assume that the elemental abundance of carbon $f_{\rm C,tot}$ scales as $f_{\rm C,tot} = 2.9 \times 10^{-4} Z_{\mathrm{gas}}/\rm Z_{\odot}$ \citep{asplund09}. For each of these models, \textsc{Cloudy} computes the abundances of different chemical species as a function of the depth into the slab. The $\mathrm{C^+}$, $\mathrm{H_2}$, and \ion{H}{I} abundances are used as inputs to our model compute the emergent luminosity from the model cells. We further assume $n_{e^-} = n_{\mathrm{C^+}}$. 

The results of the comparison are shown in Fig.~\ref{fig:cloudy}. In the following, we denote the [\ion{C}{II}] luminosity predicted by our model as $L$ and the one from \textsc{Cloudy} as $L_{\textsc{Cloudy}}$.
At all values of $T_{\rm kin}$, the distribution of $L$ is very similar to $L_{\textsc{Cloudy}}$, with $L$ from both approaches increasing with density at a fixed metallicity. Conversely, at fixed densities $n_{\rm H}\lesssim 10^{3} \, \rm cm^{-3}$, $L$ increases with metallicity and the variation with metallicity is minimal at higher densities. At ${\rm log} (Z_{\mathrm{gas}} \, / \, \rm Z_{\odot}) \gtrsim 10^{-0.6}$ (i.e., $Z_{\mathrm{gas}} \gtrsim 0.25 \, \rm Z_{\odot}$), our model overpredicts $L_{[\ion{C}{II}]}$ at low densities ($n_{\rm H} \lesssim 10^{2} \, \rm cm^{-3}$) and underpredicts at intermediate densities ($10^{2} \, \rm cm^{-3} \lesssim n_{\rm H} \lesssim 10^{3} \, \rm cm^{-3}$). 
\edit{Across the parameter space explored here, we find that the deviation between our model and \textsc{Cloudy} can be as high as $\pm 50 \%$}. However, in most ( 33/49 grid points in the right panel of Fig.~\ref{fig:cloudy}) of the parameter space there is a $ \leq 30\%$ deviation between the $L_{[\ion{C}{II}]}$ from the two approaches, particularly at high densities ($n_{\rm H} \gtrsim 10^{3} \, \rm cm^{-3}$).
\edit{We note that a deviation of $\lesssim 30-50 \%$ is much lower than the typical systematic uncertainties associated with observational measurements of the [\ion{C}{II}] luminosity in high-redshift galaxies \citep[see for example,][for the ALPINE survey]{bethermin20, schaerer20} that typically have errors/uncertainties up to a factor of 2, implying that a difference of 30-50\% is much lower than the expected observational uncertainties.} 


\section{Fit to the simulated luminosity function}
\label{sec:schechter}
\begin{figure}
    \centering
    \includegraphics[width=0.97\linewidth]{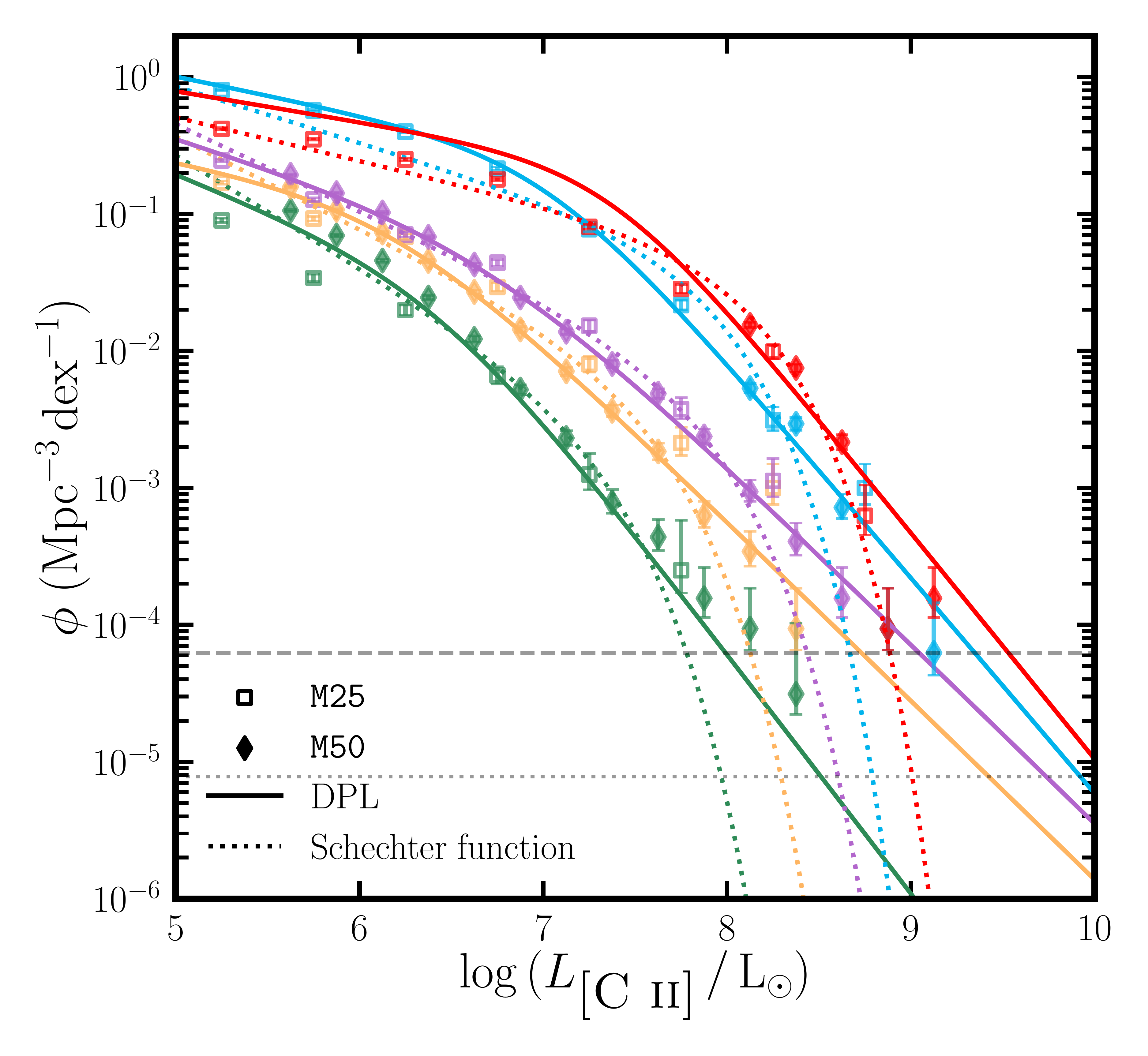}
    \caption{Comparison of the Schechter function (dotted lines) and DPL (solid lines) fits to our simulated LFs from the two simulations at different redshifts (from top to bottom, $z$ increases from 3 to 7). The open squares and diamonds represent the separate LFs from the  simulations that are used to obtain the fit parameters using an MCMC analysis. The error bars represent the 16 CL upper and lower Poisson uncertainties on number counts \citep{gehrels86}. The dashed and dotted horizontal lines represent a number count of 1 per dex in the entire simulation volume of {\tt M25} and {\tt M50}, respectively.}
    \label{fig:compare_schechter_pl}
\end{figure}

Fig.~\ref{fig:compare_schechter_pl} shows a comparison between the Schechter function and double power-law fits to the predicted [\ion{C}{II}] LFs from our simulations at $3\leq z \leq 7$.

\begin{figure*}
    \centering
    \includegraphics[width=0.81\textwidth, trim={1cm 0 0 1cm},clip]{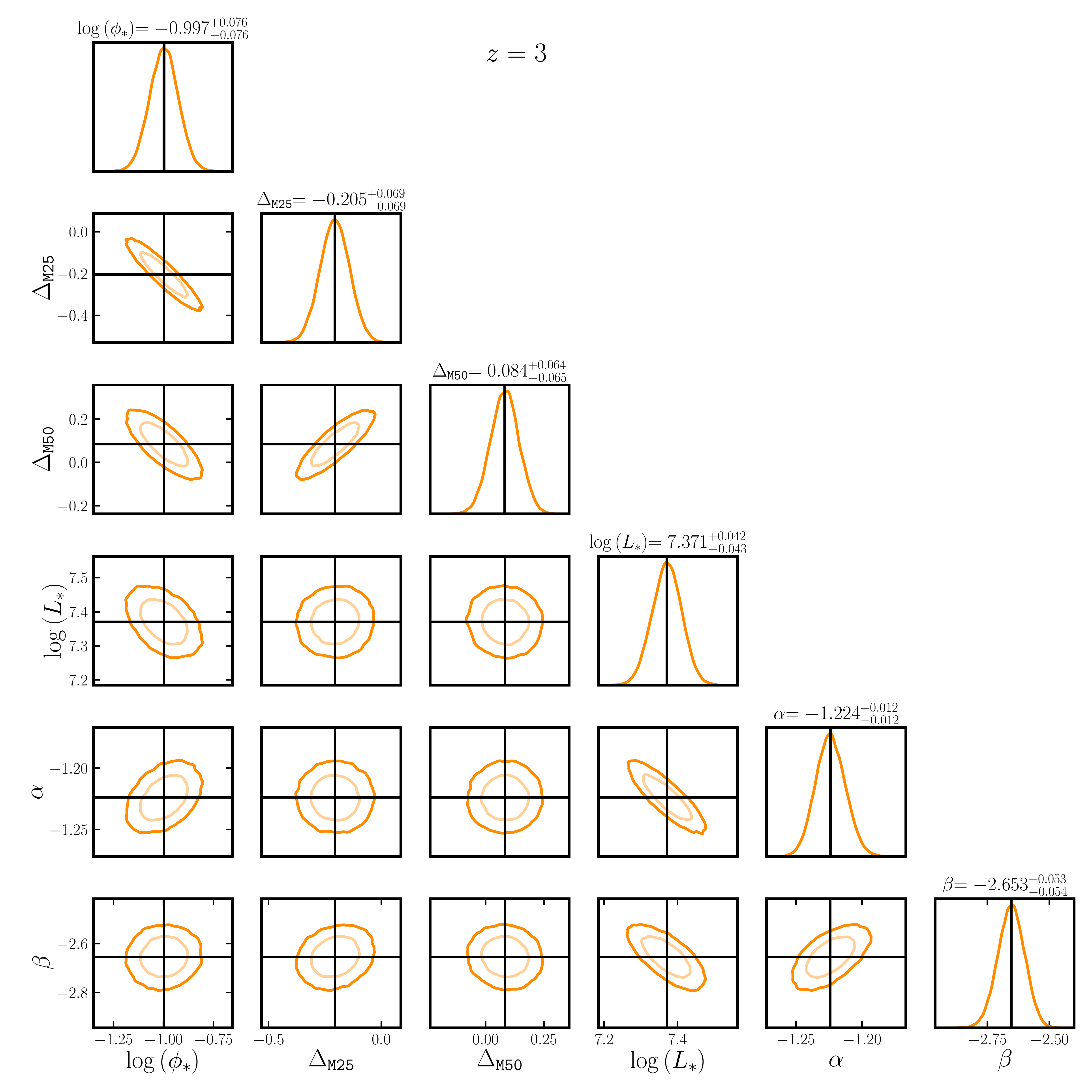}
    \caption{Covariance distributions and PDFs of the double power-law function (Eq.~\ref{eq:double_pl}) parameters: $\phi_*$, $L_*$, $\alpha$, $\beta$, and two additional parameters: $\Delta_{\tt M25}$ and $\Delta_{\tt M50}$, from MCMC runs using the python package {\tt emcee} for simulated LF at $z=3$.}
    \label{fig:corner}
\end{figure*}

In Fig.~\ref{fig:corner}, we show the covariance distributions obtained from the Markov Chain Monte Carlo (MCMC) fitting of our predicted [\ion{C}{II}] luminosity function at $z=4$ with a DPL given in Eq.~\ref{eq:double_pl}. The fitting was performed using the python package {\tt emcee} and includes two parameters -- $\Delta_{\tt M25}$ and $\Delta_{\tt M50}$ that represent variation of the $\mathrm{log} \, \phi$ of the two simulation volumes from the cosmic $\mathrm{log} \,\phi*$, because of sample variance. We see that the posterior distributions of the all parameters are well-behaved and smooth, and that the three parameters are highly correlated. We obtain similar results at other redshifts. 

\section{Surface densities}
\label{sec:appB}
\begin{figure*}
    \centering    
    \includegraphics[width=0.81\linewidth, trim={1cm 0 0 0},clip]{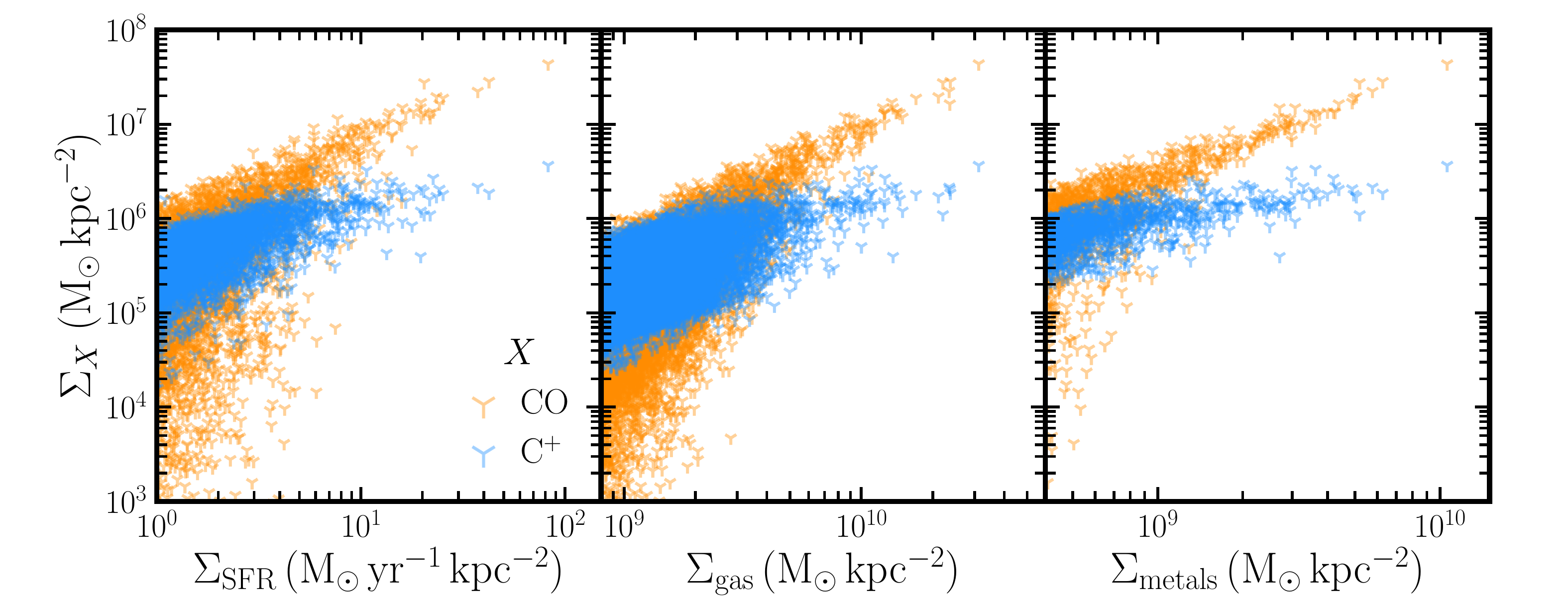}
    \caption{The surface density of CO and $\mathrm{C^+}$ as a function of the SFR surface density (left), the total gas surface density (middle), and the total metal surface density (right), for galaxies used in Fig.~\ref{fig:sigma_cii_sfr}. }
    \label{fig:sigma_cp_co_sfr}
\end{figure*}

In Fig.~\ref{fig:sigma_cp_co_sfr}, we show a scatter plot of the surface densities of CO and $\mathrm{C^+}$ as a function of the SFR (left), gas (middle), and metal (right) surface density. In all panels, we find that while $\Sigma_{\mathrm{CO}}$ continues to increase at high surface densities, $\Sigma_{\mathrm{C^+}}$ shows a plateau. This results in a decrease in the slope of the ${\rm log}\,\Sigma_{[\ion{C}{II}]}$ versus ${\rm log}\,\Sigma_{\mathrm{SFR}}$ curve at high $\Sigma_{\mathrm{SFR}}$.


\end{document}